# Going beyond richness: Modelling the BEF relationship using species identity, evenness, richness and species interactions via the `DImodels` R package

Rafael A. Moral[1], Rishabh Vishwakarma[2], John Connolly[3], Laura Byrne[2], Catherine Hurley[1], John A. Finn[4], Caroline Brophy[2]

[1]Department of Mathematics and Statistics, Maynooth University, Co Kildare, Ireland

[2]School of Computer Science and Statistics, Trinity College Dublin, Dublin 2, Ireland

[3]School of Mathematics and Statistics, University College Dublin, Dublin 4, Ireland

[4]Department of Environment, Soils and Land Use, Teagasc, Johnstown Castle, Wexford, Ireland

**CORRESPONDING AUTHOR**

Rafael De Andrade Moral

Email: Rafael.DeAndradeMoral@mu.ie

Address: Department of Mathematics and Statistics, Maynooth University, Co Kildare, Ireland

**RUNNING HEADLINE**

Modelling BEF data with the `DImodels` R package

**Abstract**

1. Biodiversity and ecosystem function (BEF) studies aim to understand how ecosystems respond to a gradient of species diversity. Generalised Diversity-Interactions (DI) models are suitable for analysing the BEF relationship. These models relate an ecosystem function response of a community to the identity of the species in the community, their evenness (proportions) and interactions. The number of species in the community (richness) is included implicitly in a DI model.

2. It is common in BEF studies to model an ecosystem function as a function of richness; while this can uncover trends in the BEF relationship, by definition, species diversity is broader than richness alone, and important patterns in the BEF relationship may remain hidden. Here, we introduce the `DImodels` R package for implementing DI models. We show how richness is mathematically equivalent to a simplified DI model under certain conditions, and illustrate how using the DI multi-dimensional definition of species diversity can provide deeper insight to the BEF relationship compared to traditional approaches.

3. Using DI models can lead to considerably improved model fit over other methods; it does this by incorporating variation due to the multiple facets of species diversity. Predicting from a DI model is not limited to the study design points, the model can interpolate or extrapolate to predict for any species composition and proportions (assuming there is sufficient coverage of this space in the study design).

4. Expressing the BEF relationship as a function of richness alone can be useful to capture overall trends. However, collapsing the multiple dimensions of species diversity to a single dimension (such as richness) can result in valuable ecological information being lost. DI modelling provides a framework to test the multiple components of species diversity in the BEF relationship. It facilitates uncovering a deeper ecological understanding of the BEF relationship and can lead to enhanced inference. The open-source `DImodels` R package provides a user-friendly way to implement this modelling approach.

# 1. INTRODUCTION

In biodiversity and ecosystem function (BEF) studies, there is usually a range of communities that may vary in the number, identities and/or proportions of species, and in overall density. One or more ecosystem function responses are recorded for each community a period after species diversity is manipulated (an experiment) or observed (an observational study). Since communities in BEF studies are characterised by the initial richness, composition (species identities) and evenness (species proportions) of their species, statistical analyses should ideally jointly assess how ecosystem function (community-level responses) are affected by these three variables. Diversity-Interactions (DI) modelling, introduced by Kirwan et al., (2009) and Connolly et al., (2013), jointly assesses the effects of species identity, richness, evenness, community composition and interspecific species interactions on an ecosystem function in a regression modelling framework. The `DImodels` R package implements this approach (Moral et al., 2022).

DI models typically include three components in the linear predictor:

$$y = Identities + Interactions + Structures + \varepsilon \qquad (eq. 1)$$

where $y$ is a community-level response (e.g., biomass for a plant community); *Identities* are the effects of species identities and enter the model as initial individual species proportions; *Interactions* are the pairwise effects of interspecific interactions between the initial species proportions; while *Structures* include other experimental structures, such as blocks, treatments or environmental gradients. A possible DI model (excluding structures) is:

$$y = \sum_{i=1}^{S} \beta_i p_i + \sum_{\substack{i,j=1 \\ i<j}}^{S} \delta_{ij} (p_i p_j)^\theta + \varepsilon, \qquad (eq.\ 2)$$

Where $p_i$ is the initial proportion of species *i*, $\beta_i$ is the identity effect of species *i*, $\delta_{ij}$ is the interaction parameter for species *i* and *j*, and $\varepsilon$ is a normally distributed random error term, with constant variance. The non-linear exponent parameter $\theta$ on each $p_i p_j$ in the *Interactions* component, determines the shape of the BEF relationship by allowing the importance and impact of interaction terms to be directly modelled (Connolly et al., 2013). Imposing different constraints in the *Interactions* term can make DI models more biologically informative and estimable (Table 1).

A major strength of the DI modelling framework is its ability to decompose the various elements of species diversity, including species identity (composition), species richness and evenness. While 'richness' does not appear explicitly in eq. 2, richness is implicitly included (see Appendix S1 in Connolly et al., 2013): when modelling data with monocultures and/or mixtures that have an equal proportion of their non-zero species (equi-proportional mixtures), the average pairwise DI model (see Table 1) with $\theta = 0.5$ *and* all *Identities* effects set equal is equivalent to the richness model (Figure 1a). A mathematical proof and illustration of this equivalency are in SI1. Under the same conditions, a $\theta = 0.72$ approximates the BEF model with predictor the square root of richness, and $\theta = 0.87$ approximates the log(richness) model (Figure 1a). Through DI models, we can go beyond richness as the main explanatory driver of the BEF relationship and additionally test for: (1) deviations from a linear richness BEF relationship (the shape of the BEF relationship changes with $\theta$: Figure 1a); (2) species-specific effects (different identity effects ($\beta_i$), Figure 1b); (3) departures from equi-proportional communities (the richness model is not sensitive to changes in species proportions at a given richness); and (4) varying interaction structures (the richness model assumes any pair of species interacts with the same strength). These benefits of DI models are illustrated in SI1 and Section 3 (Case Studies).

**TABLE 1** The 'Identities' and 'Interactions' components for a range of DI models; there are *s* species in the pool that are categorised by *T* functional groups (FGs), and $p_i$ is the initial proportion of species *i*.

| Model name | Identities | Interactions | Model tag[1] |
|---|---|---|---|
| Species identity (no interactions) | $\sum_{i=1}^{s} \beta_i p_i$ | — | `ID` |
| Average pairwise (all interactions equal) | $\sum_{i=1}^{s} \beta_i p_i$ | $\delta \sum_{\substack{i,j=1 \\ i<j}}^{s} (p_i p_j)^\theta$ | `AV` |
| Functional groups (interactions dictated by functional group membership) | $\sum_{i=1}^{s} \beta_i p_i$ | $\sum_{q=1}^{T} \omega_{qq} \sum_{\substack{i,j \in FG_q \\ i<j}} (p_i p_j)^\theta + \sum_{\substack{q,r=1 \\ q<r}}^{T} \omega_{qr} \sum_{i \in FG_q} \sum_{j \in FG_r} (p_i p_j)^\theta$ | `FG` |
| Additive species (species-specific contribution) | $\sum_{i=1}^{s} \beta_i p_i$ | $\sum_{\substack{i,j=1 \\ i<j}}^{s} (\lambda_i + \lambda_j)(p_i p_j)^\theta$ | `ADD` |
| Full pairwise (all interactions unique) | $\sum_{i=1}^{s} \beta_i p_i$ | $\sum_{\substack{i,j=1 \\ i<j}}^{s} \delta_{ij} (p_i p_j)^\theta$ | `FULL` |

[1] Each model can be implemented in the `DImodels` package and the `DImodel` argument within the `DI()` function is referred to as the "model tag" (see Section 2).

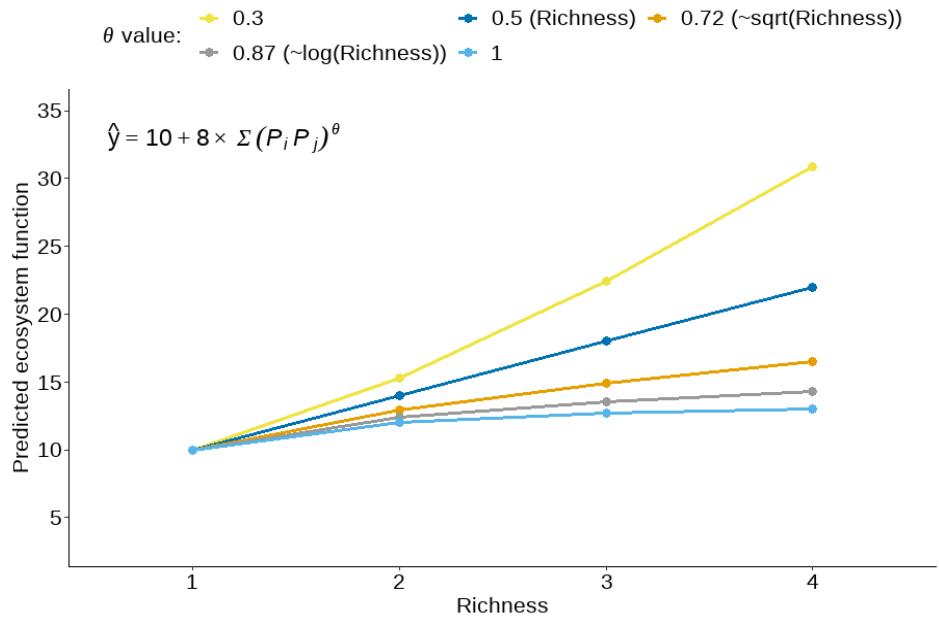

(a) Changing $\theta$ alters the shape of the BEF relationship

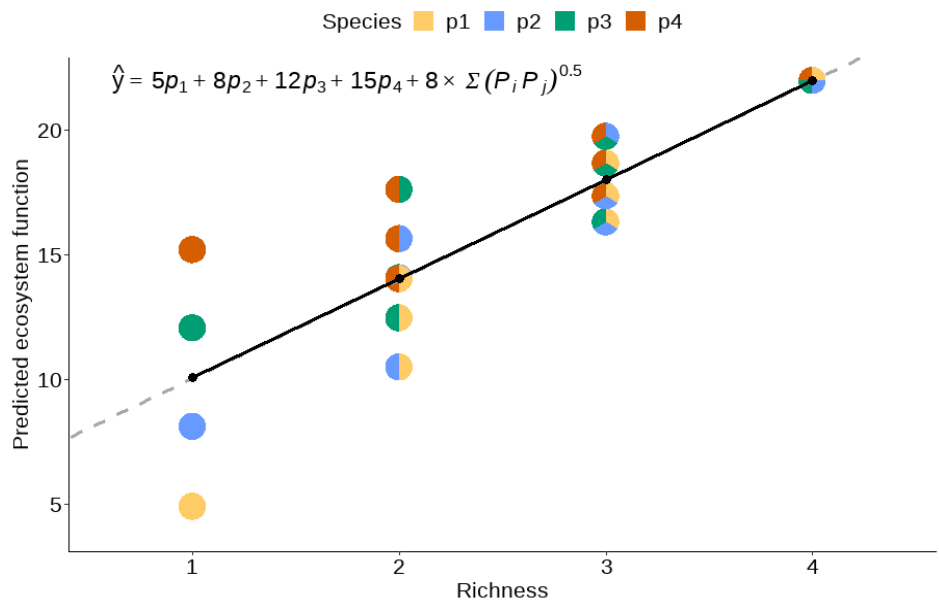

(b) Predictions from a DI model can be on average at each level of richness (black points) and for specific communities with varying combinations of species (pie-glyphs)

**FIGURE 1** Conceptual diagram to illustrate how richness is implicitly included in a DI model. It is assumed the underlying data is a balanced design with only equi-proportional communities from a pool of four species. Predicted ecosystem function versus richness for: (a) a range of DI models where all species interact in the same way (AV model in Table 1), with identity effects equal, and with $\theta$ varying, and (b) a DI model with unique identity effects for each species and an average interaction term with $\theta = 0.5$; predictions are shown for select individual communities (the pie glyphs show the initial community proportions) and on average for each richness level (black dots; computed as $\hat{y} = \frac{5+8+12+15}{4} + 8\sum(p_i p_j)^{0.5}$). The dotted grey line is the fitted richness model, it coincides with the average DI model predictions.

The DI models framework can be used to analyse community-level responses without the need to have the response broken down into the various species contributions (a requirement of partitioning models, see Loreau and Hector, 2001). This is particularly relevant for community-level responses that *cannot* be partitioned (e.g., greenhouse gas measurements; see Cummins et al., 2021), and for community-level responses that are labour intensive to partition into species contributions. DI models relate a 'total' ecosystem function to initial (e.g., sown proportions in a grassland biodiversity experiment) or previous species diversity.

Here, we present the `DImodels` R package for fitting DI models and showcase its functionality. We also compare DI modelling to other commonly used BEF modelling approaches.

## 2. THE `DImodels` R PACKAGE

The `DImodels` package provides users with the flexibility to fit and compare different types of DI models. A typical workflow consists of (1) loading and exploring the dataset; (2) performing automatic model selection using the `autoDI()` function; and (3) re-fitting and/or extending the `autoDI()` selected model using the `DI()` and (sometimes) the `DI_data()` functions (Figure 2). A dataset suitable for use with the package will contain a response variable and columns indicating the initial species proportions; it may also include block, treatment, density, and/or community explanatory variables. The package contains simulated and real example datasets for illustration.

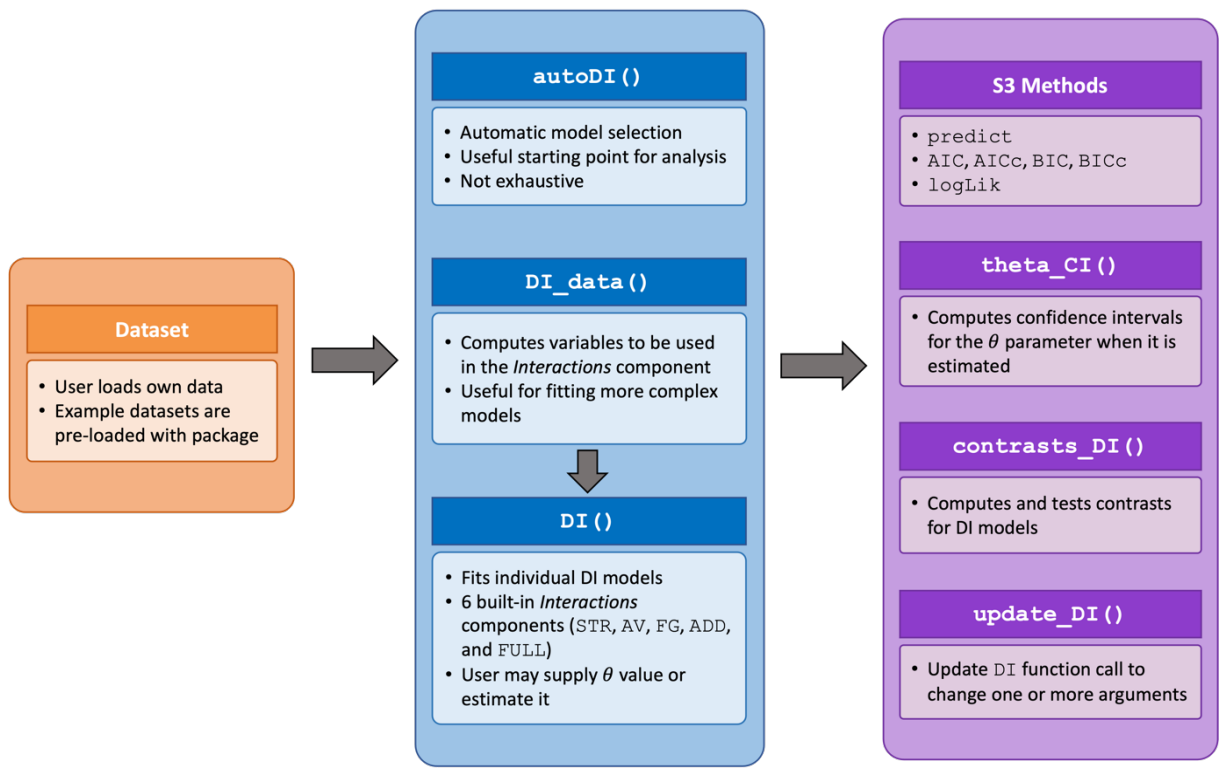

**FIGURE 2** Schematic diagram representing a typical workflow and the conceptual map of the `DImodels` package. A dataset (orange column) is passed through the core package functions (blue column) and the resulting model objects can be passed through additional functions (purple column).

## 2.1. Automatic Model Selection

The `autoDI()` function provides an automatic model selection procedure to fit a limited range of DI models. The user may choose between different criteria for model selection, such as F-tests and likelihood-based information criteria. Using `autoDI()`, a four-step selection process identifies the best DI model from those fitted (Box 1). While `autoDI()` is an excellent starting point, it fits a limited number of DI models; it is strongly recommended to assess the diagnostics of the selected model and to explore additional models as appropriate via the `DI()` function. For instance, `autoDI()` does not test for interactions of a treatment with other variables in the model.

The package provides the `richness_vs_DI()` function, a second automated model selection procedure. This function compares the richness model to a small subset of DI models (SI1).

BOX 1 Overview of `autoDI()`

The autoDI() function implements a four-step model selection procedure, plus an optional initial step (Figure 3). 'Step 0' (optional) investigates the significance of *Structures* by comparing the intercept-only model with models including the *Structures* terms. In 'Step 1', $\theta$ is estimated for the average interactions (AV) model and is tested for a difference from 1. In 'Step 2', the *Interactions* term is investigated using forward selection with five different *Interactions* structures, including the estimate of $\theta$ from 'Step 1', if it was significantly different from 1, or assuming $\theta = 1$ otherwise. In 'Step 3', the treatment (if present) is tested for inclusion in the model selected in 'Step 2'. Finally, in 'Step 4' (also optional, but conducted by default), `autoDI()` carries out a lack-of-fit test by comparing the model selected in 'Step 3' with the reference (community) model, which includes a factor variable representing each unique combination of species proportions, as the linear predictor.

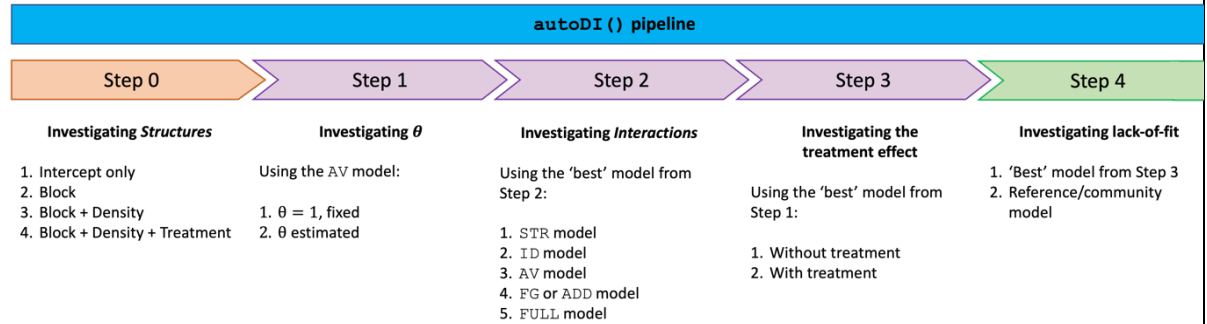

FIGURE 3 A schematic representation of the `autoDI()` pipeline.

## 2.2. Fitting Individual DI Models

The syntax for the `DI()` function is:

`DI(y, prop, DImodel, data, block, density, treat, FG, extra_formula, custom_formula, estimate_theta, theta)`

The `data` argument specifies the dataset for analysis. The arguments `y`, `prop`, `block`, `density`,

and `treat` specify the corresponding column name (or number) in the dataset. The arguments `block`, `density`, and `treat` constitute the *Structures* component of the DI model. The argument `DImodel` takes specific model tags (Table 1) and defines the corresponding *Identities* and *Interactions* components. When fitting a functional groups model (FG tag in Table 1), the argument `FG` is used to declare the functional group names. Additional terms may be added by specifying a formula in the argument `extra_formula`, e.g. interactions between species proportions and treatment effects. By default, a DI model fitted using `DI()` will assume $\theta = 1$, however, it can be set to a numeric value with the `theta` argument or can be estimated using `estimate_theta = TRUE`.

Ordinary least squares estimation is used when $\theta = 1$. The main challenge lies in generating the interaction terms of the model matrix. The `DImodels` package uses the species proportions input to internally prepare the appropriate model matrices depending on the specifications of the *Identities* and *Interactions* components. When $\theta$ is estimated, we optimise the log-likelihood $L(\theta)$ of the model using Brent's (1973) method.

## 2.3. Working with Fitted DI Models

Both `autoDI()` and `DI()` return objects of class `DI`, for which different S3 methods are available to calculate predictions and associated standard errors, information criteria and log-likelihood values, and compare nested models using F tests (Figure 2). Specific contrasts and their standard errors are provided by `contrasts_DI()` (see example in SI2.9). Updated calls to the `DI()` function are obtained via `update_DI()`. Confidence intervals for the $\theta$ parameter can be computed via the `theta_CI()` function (see example in SI4.5). We obtain $100(1-\alpha)\%$ confidence intervals for $\theta$ by profiling $L(\theta)$ over a pre-defined grid and obtaining the values of $\theta$ that correspond to the solutions of the equation $L(\theta) - L(\hat{\theta}) = \frac{1}{2}\chi^2_{1;\alpha}$, where $L(\hat{\theta})$ is the log-likelihood evaluated at the maximum likelihood estimate $\hat{\theta}$ and $\chi^2_{1;\alpha}$ is the $100\alpha\%$ percentile of the chi-squared distribution with 1 degree of freedom.

Data manipulation steps are carried out automatically within the `autoDI()` and `DI()` functions. However, the user can compute and store interaction variables using the `DI_data()` function and, for instance, may wish to use them in them in the `extra_formula` argument within the `DI()` function (see SI4 for an example), or moving beyond the `DImodels` package to other types of extended modelling approaches (e.g. Bayesian inference via MCMC samplers, or inclusion of random block effects). Moreover, users may extract the full model matrix of a fitted DI model using the `model.matrix()` S3 method for `glm` objects, or by using the `extract()` S3 method for `DI` objects, which extracts specific parts of the model matrix.

## 3. CASE STUDIES

We present two case studies to highlight the benefits of DI models and the features of our package. SI2 (simulated data) and SI3 (Bell dataset on a bacterial experiment) reproduce the analyses in the case studies. SI4 presents an additional case study on data from a grassland biodiversity experiment. All three datasets are available in the `DImodels` package.

**Simulated data case study**

We simulated a BEF dataset, assuming a pool of three species, with 16 unique communities each characterised by the proportions of species 1, 2 and 3, respectively: $p_1$, $p_2$, and $p_3$. The design (Figure 4b and SI2.2) consisted of:

- monocultures of each species (for example, 1:0:0),
- binary communities: equi-proportional (e.g., 0.5:0.5:0) and unbalanced (e.g., 0.8:0.2:0), and
- three-species communities: equi-proportional (0.333:0.333:0.333) and unbalanced (e.g.,

0.6:0.2:0.2).

A response (ecosystem function) was simulated for four replicates of each community, giving 64 experimental units (this 'sim0' dataset is available in the `DImodels` package). The responses were simulated from:

$$y = \sum_{i=1}^{3} \beta_i p_i + \sum_{\substack{i,j=1 \\ i<j}}^{3} \delta_{ij} p_i p_j + \varepsilon, \qquad (eq. 3)$$

with values: $\beta_1 = 25$, $\beta_2 = 20$, $\beta_3 = 15$, $\delta_{12} = 30$, $\delta_{13} = 20$, and $\delta_{23} = 40$, with $\epsilon \sim N(0, \sigma^2 = 4)$. A positive interaction means that combining the interacting species will result in a higher expected mean than the weighted average expected identity effects; for example, in the absence of any interaction between species 1 and 2, the expected response for a 40:60 initial two-species mix of species 1 and 2 would be $(25)(0.4) + (20)(0.6) = 22$, however, given the positive interaction term, the expected response from eq. 3 is $(25)(0.4) + (20)(0.6) + (30)(0.4)(0.6) = 29.2$.

We analysed the simulated data by fitting: (1) the richness model, (2) a full pairwise DI model, and (3) a 1-way ANOVA model (a factor with a level for each of the 16 unique communities). We found:

**The richness model:** As richness increased, the response also increased (Figure 4a, p<0.001). All variation around the line is assumed to be residual error variation (MSE=11.24, 62 residual df). At each level of richness, some variation is likely due to changing composition and/or changes in the proportions of species.

**The full pairwise Diversity-Interactions model**: Species identities and interactions strongly influenced the response, as evidenced by the 'dome' shape in the ternary diagram (Figure 4b; MSE=3.54, 58 residual df). The DI model can predict for any combination of species proportions around the three-dimensional simplex space (Figure 4b). Figure 4c also illustrates DI model predictions and highlights the increasing trend in predictions as richness increases; variation in predictions due to species composition and proportions are additionally represented by the pie-glyphs (Vishwakarma et al., 2023).

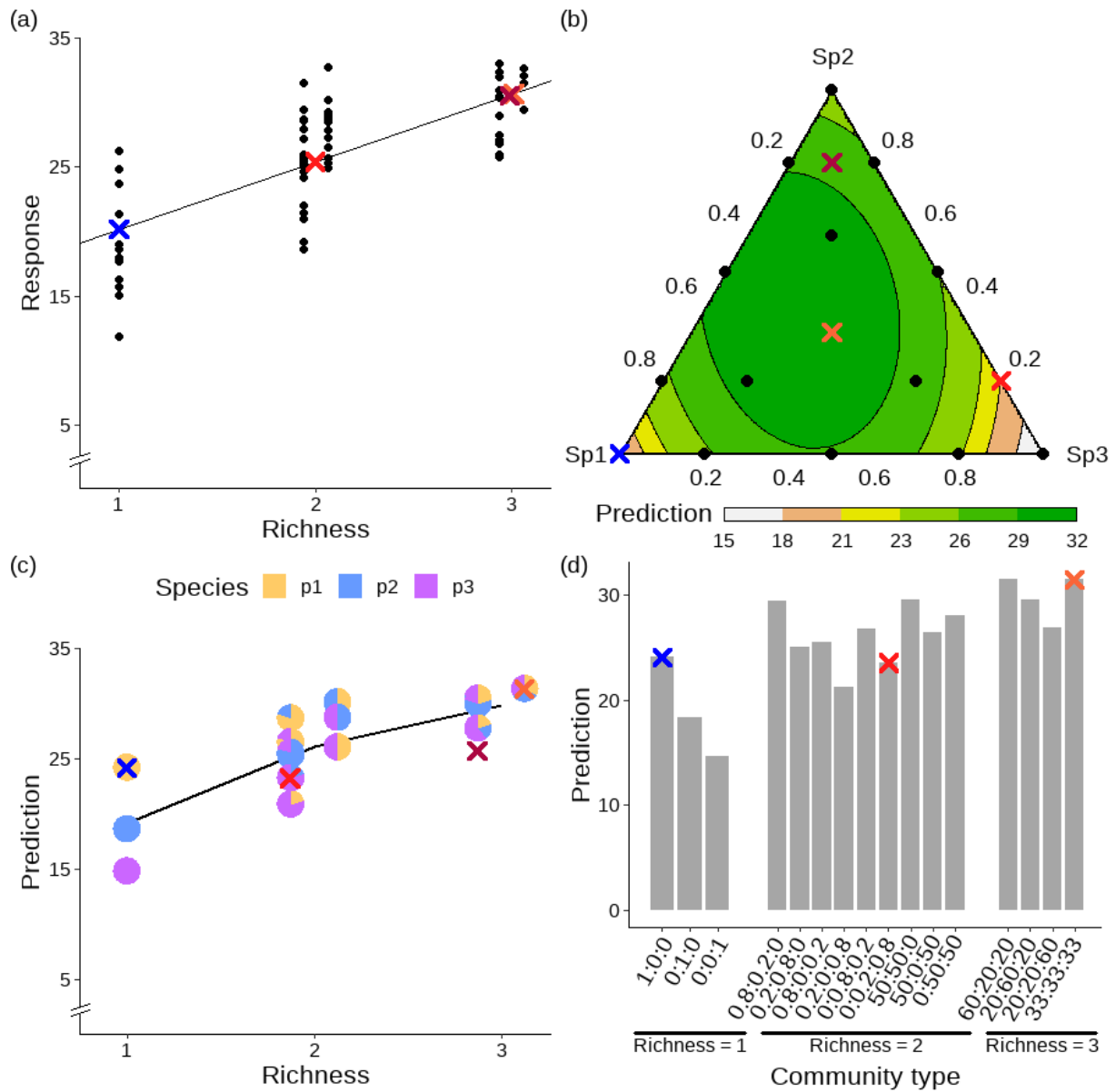

**FIGURE 4** (a) the fitted richness model with raw data; (b) the predicted response values across the three-dimensional simplex space, predicted from fitting a DI model (with design points marked with a dot or X, the wine coloured X highlights a prediction that is not a design point); (c) the predicted response for each design point from fitting a DI model, with the pie-glyphs (jittered) illustrating the initial proportions and trendline connecting the average prediction at each level of richness; (d) predicted response for each community from a 1-way ANOVA model. The coloured X's in each panel highlight select predictions for each analysis: blue for a monoculture of species 1, red for 20% of species 2 and 80% of species 3, orange for an equi-proportional 3-species mixture and wine for 10% of species 1 and 3 and 80% of species 2.

**The 1-way ANOVA model**: There was a strong effect of changing species diversity (Figure 4d, p<0.001; MSE=3.75, 48 residual df). Comparisons can be made between any two communities in the design space but there are limitations to generalise patterns across species diversity. The main strength in the ANOVA model is its ability to act as a reference or a test for lack of fit.

To illustrate the varying inferential ability of the three approaches, we highlight the predictions from each model for select communities; see the coloured X's in Figure 4. The ANOVA and DI models give a tailored prediction for the three communities in the design space (blue, red and orange X's). The richness model only gives an average prediction at any given level of richness. Only the DI model can predict uniquely for any set of species proportions. Thus, DI models can predict across levels of richness (as with the richness model), predict for any community in the design (as with the ANOVA model) and predict for any combination of $p_1$, $p_2$, and $p_3$ by interpolating and extrapolating around the simplex space, assuming there is reasonable coverage of the design space.

The MSE was highest for the richness model (11.24; AIC=340.45), while the DI model (3.54; AIC=270.26) and ANOVA model (3.75; AIC=281.76) had similar MSE values. This supports that the lower numbers of parameters in the DI model explained as much variation as the ANOVA (reference) model and was an improvement over the richness model.

**The Bell dataset case study**

This case study comes from a 72-species bacterial biodiversity experiment (Bell et al., 2005). The bacterial ecosystems used were from semi-permanent rain pool depressions near the base of European beech trees (*Fagus sylvatica*). Microcosms were inoculated with combinations of bacterial species isolated from these ecosystems. A total of 1,374 microcosms were constructed at richness levels of 1, 2, 3, 4, 6, 8, 9, 12, 18, 24, 36 and 72 species. The daily respiration rate of the bacteria in each microcosm averaged over three time intervals (days 0-7, 7-14 and 14-28) was analysed.

With 72 species, the Bell dataset presents some challenges for DI models: a full pairwise interactions model would have 72 identity and 2,556 interaction effects. Fitting this model is neither possible (there being only 1,374 experimental units) nor desirable since estimating so many interaction terms is unlikely to provide valuable biological information. We fitted the three models in Table 2. The average pairwise DI model with $\theta$ set to 1 gives a higher AIC than the richness model, however, when $\theta$ was estimated, the AIC was much lower (Table 2; $\hat{\theta} = 0.79 < 1$, more importance is given to interactions when proportions are lower). Raw data and the predictions from each model are shown in Figure 5; four $\theta$ values are shown to illustrate the flexibility of the DI modelling framework in modelling the shape of the BEF relationship. Of the three models fitted, the third model (AV + $\theta$ estimated) is the best according to AIC (Table 2; Connolly et al., 2013), although the model with all identity effects assumed equal (but keeping AV+$\theta$ estimated) gives a comparable AIC value (SI3.2). It is clear that for this data, assuming a BEF relationship to be linear with richness would be inappropriate, as would assuming $\theta = 1$.

**TABLE 2** Models fitted to the Bell dataset, with their AIC values.

| Model description | Equation(s) | AIC |
|---|---|---|
| Average interactions model, identity effects equal, $\theta$ set to 0.5 (equivalent to the richness model, see SI1). | $y = \alpha + \delta \sum_{\substack{i,j=1 \\ i<j}}^{72} (p_i p_j)^{0.5} + \epsilon$ | 6789.95 |
| Average interactions model, $\theta$ set to 1. | $y = \sum_{i=1}^{72} \beta_i p_i + \delta \sum_{\substack{i,j=1 \\ i<j}}^{72} p_i p_j + \epsilon$ | 6814.61 |
| Average interactions model, $\theta$ estimated. | $y = \sum_{i=1}^{72} \beta_i p_i + \delta \sum_{\substack{i,j=1 \\ i<j}}^{72} (p_i p_j)^{\theta} + \epsilon$ | 6751.12 |

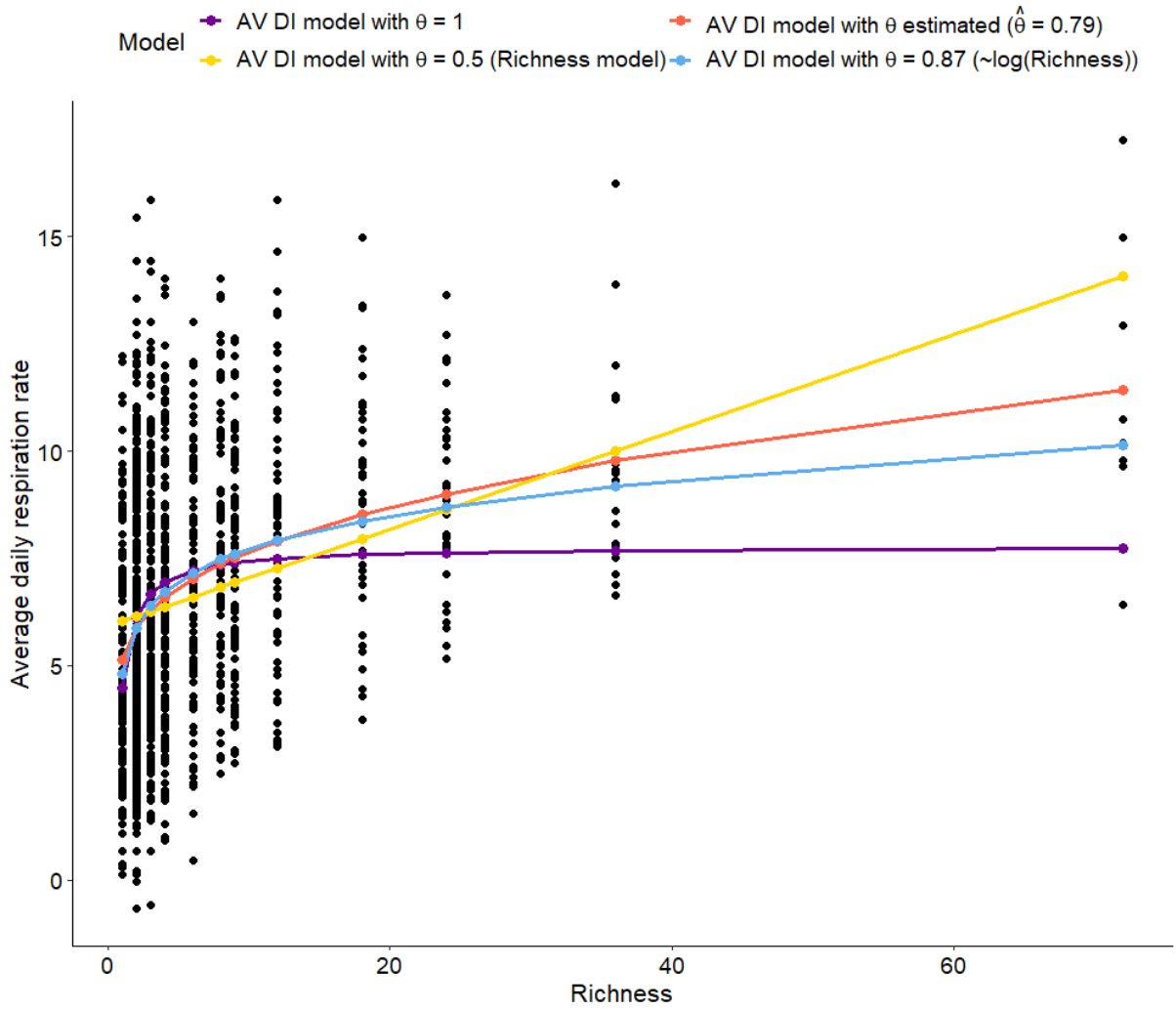

**FIGURE 5** Average daily respiration rate versus richness for the raw data (black dots) and the fitted predictions from the average interactions models with $\theta = 1$ (purple line), $\theta = 0.5$ (yellow line), $\theta = 0.87$ (blue line) and with $\theta$ estimated ($\hat{\theta} = 0.79$). Predictions for each model are computed as the average of all design communities (black dots) at each level of richness, with a smoothed line imposed across richness levels. The model with $\theta = 0.5$ depicts the BEF relationship with a linear richness shape, while the model with $\theta = 0.87$ approximates the BEF relationship with a log(richness) shape.

## 4. FINAL REMARKS

This `DImodels` package provides a user-friendly means to fit a range of models. The `autoDI()` function provides an automated preliminary analysis of a BEF dataset and is particularly beneficial to users new to the modelling approach. Only a subset of possible models that can be fitted to a given dataset are tested in `autoDI()`, and users should supplement the `autoDI()` analysis with further analysis using the `DI()` function.

Use of the full pairwise interactions DI model is typically constrained to datasets with a smaller number of species than (for example) the Bell data as it can have limited biological value (too many parameters) or may not be estimable (e.g. in studies where a pair of species do not occur together). However, alternatives involving the DI modelling framework are available to deal with datasets where communities are species rich. One solution is to use random pairwise interactions (Brophy et al., 2017, McDonnell et al., 2023); in this approach the pairwise interactions are assumed to be latent realisations of normal random variables, and it is only necessary to estimate their mean and variance rather than all possible combinations. Another solution is to specify *Interactions* components that utilise fewer degrees of freedom, such as the average pairwise interaction structure (Table 1). Indeed, even the average pairwise interaction model alone can have considerable value by allowing estimation of the shape of the BEF relationship rather than imposing a linear richness relationship, for example. Moreover, in studies where a specific pair of species is not observed in the same observational/experimental unit, their pairwise interaction coefficient cannot be estimated by a full pairwise DI model, and hence the alternative strategies outlined above can be useful.

When designing an experiment suitable for analysis using DI modelling, replicates of each community are not required. This is because DI modelling is analogous to a response surface analysis and replication is across the continuous simplex space. Of course, it may be beneficial to have replicates in some parts of the simplex design space to provide additional power; in particular, it is recommended (but not required) to have replication at the extremities of the design at the monoculture points. DI modelling is applicable to BEF data where species identities and proportions have been manipulated across an evenness and richness gradient (e.g., the Switzerland Case Study, see SI4), *and* also applicable for BEF data where only equi-proportional mixtures are available (provided that a range of richness levels are included in the design, e.g., the Bell Case Study). DI models can jointly assess the effects of species identity, richness, evenness and species interactions on ecosystem function responses in a wide range of BEF studies.

Estimating $\theta$ in a DI model allows the data to identify the best shape of the BEF relationship (rather than the user imposing, for example, a log or linear richness relationship). Frequently richness alone will not capture all variation due to species diversity and DI modelling provides a versatile framework to model the multiple dimensions of species diversity effects in the BEF relationship.

## ACKNOWLEDGEMENTS

All authors were supported by the Science Foundation Ireland Frontiers for the Future programme, grant number 19/FFP/6888 award to CB.

## AUTHORS' CONTRIBUTIONS

RAM, CB and JC developed the `DImodels` R package, RV contributed to the package and case study materials. RV and CH developed the novel visualisation approaches. RAM, CB and JC wrote the initial draft of the paper. RV, LB, CH and JF each contributed to various conceptual developments and to writing the paper.

**Supporting Information 1: Richness is implicitly included a DI model**

**SI1.1 The 'richness' model equates to a simplified version of a generalised Diversity-Interactions (DI) model when the data only contains communities with an equal proportion (equi-proportional) of their non-zero species.**

The 'richness' model is
$$y = \beta_0 + \beta_1 \text{richness} + \epsilon$$
Letting richness = $r$, this becomes
$$y = \beta_0 + \beta_1 r + \epsilon$$

Here is a simplified 'DI' model with all identity effects set equal, and the average pairwise interaction structure with $\theta$ = 0.5:
$$y = \alpha + \delta \sum_{\substack{i,j=1 \\ i<j}}^{s} (p_i p_j)^{0.5} + \epsilon$$

This DI model is equivalent to the richness model if the underlying data to which the two models are fitted contains monocultures and/or mixtures that have an equal proportion of their non-zero species (equi-proportional mixtures).

**Proof**:

Assume a pool of $s$ species, and at a given level of richness, let richness = $r$ ($\leq s$).

For a given level of richness ($r$), and initially equi-proportional communities:

$$\sum_{\substack{i,j=1 \\ i<j}}^{s} (p_i p_j)^{0.5} = \sum_{\substack{i,j=1 \\ i<j}}^{s} \sqrt{p_i p_j} = \binom{r}{2}\sqrt{\frac{1}{r} \cdot \frac{1}{r}} = \binom{r}{2}\frac{1}{r}$$

$$= \frac{r!}{2!(r-2)!} \times \frac{1}{r} = \frac{r(r-1)(r-2)(r-3)\ldots 1}{2 \times 1 \times (r-2)(r-3)\ldots 1} \times \frac{1}{r} = \frac{r(r-1)}{2} \times \frac{1}{r} = \frac{r-1}{2}$$

I.e., for an initially equi-proportional community at a given level of richness = $r$,

$$\sum_{\substack{i,j=1 \\ i<j}}^{s} (p_i p_j)^{0.5} = \frac{r-1}{2}$$

Equivalently,

$$r = 2\sum_{\substack{i,j=1 \\ i<j}}^{s} (p_i p_j)^{0.5} + 1$$

Thus,

$$\beta_0 + \beta_1 r = \beta_0 + \beta_1 \times \left( 2 \sum_{\substack{i,j=1 \\ i<j}}^{S} (p_i p_j)^{0.5} + 1 \right)$$

$$= (\beta_0 + \beta_1) + 2\beta_1 \sum_{\substack{i,j=1 \\ i<j}}^{S} (p_i p_j)^{0.5} = \alpha + \delta \sum_{\substack{i,j=1 \\ i<j}}^{S} (p_i p_j)^{0.5}$$

Where $\alpha = \beta_0 + \beta_1$, and $\delta = 2\beta_1$.

Note, this only holds for datasets where there are monocultures and/or mixtures that have an equal proportion of their non-zero species.

**SI1.2 Example of how the richness model equates to a simplified DI model**

Here is a dataset containing monocultures and mixtures that are all initially equi-proportional:

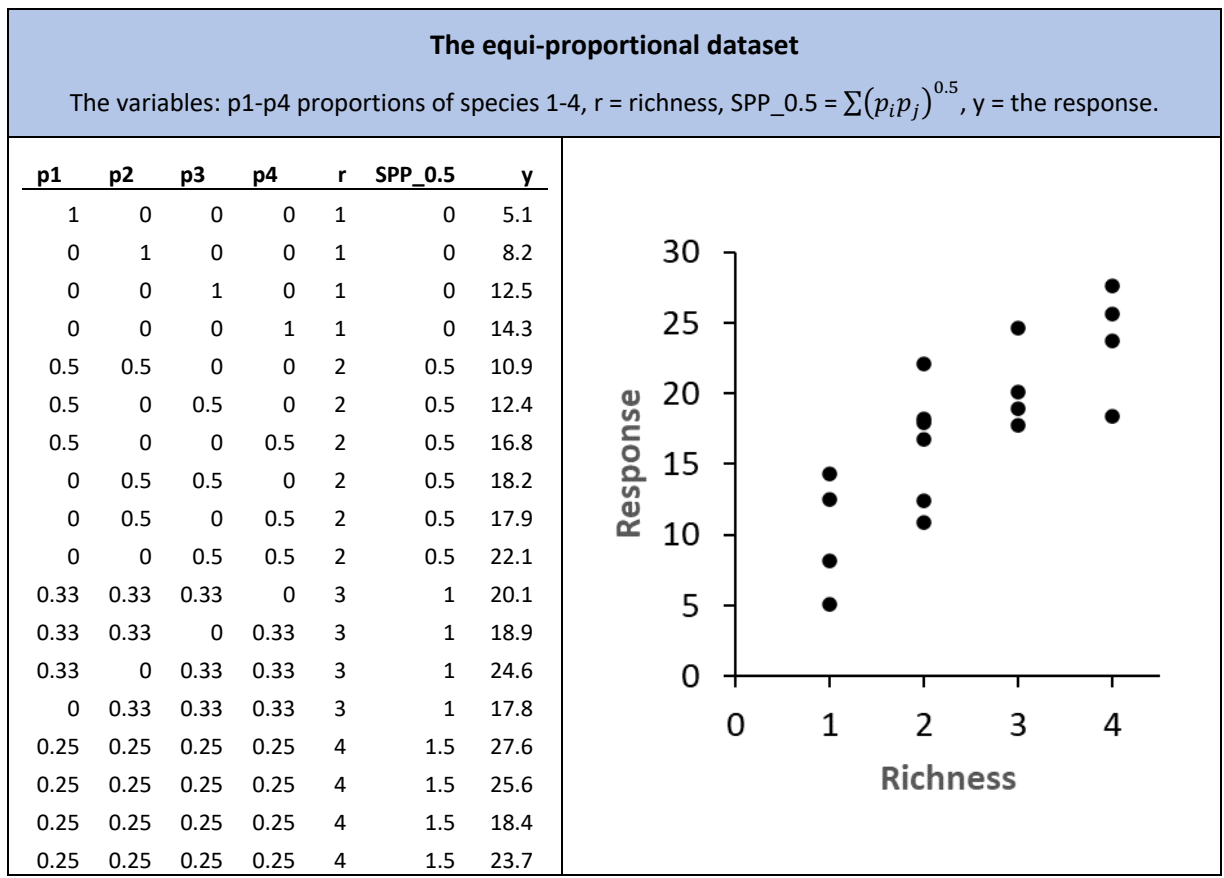

**The equi-proportional dataset**

The variables: p1-p4 proportions of species 1-4, r = richness, SPP_0.5 = $\sum (p_i p_j)^{0.5}$, y = the response.

| p1 | p2 | p3 | p4 | r | SPP_0.5 | y |
|---|---|---|---|---|---|---|
| 1 | 0 | 0 | 0 | 1 | 0 | 5.1 |
| 0 | 1 | 0 | 0 | 1 | 0 | 8.2 |
| 0 | 0 | 1 | 0 | 1 | 0 | 12.5 |
| 0 | 0 | 0 | 1 | 1 | 0 | 14.3 |
| 0.5 | 0.5 | 0 | 0 | 2 | 0.5 | 10.9 |
| 0.5 | 0 | 0.5 | 0 | 2 | 0.5 | 12.4 |
| 0.5 | 0 | 0 | 0.5 | 2 | 0.5 | 16.8 |
| 0 | 0.5 | 0.5 | 0 | 2 | 0.5 | 18.2 |
| 0 | 0.5 | 0 | 0.5 | 2 | 0.5 | 17.9 |
| 0 | 0 | 0.5 | 0.5 | 2 | 0.5 | 22.1 |
| 0.33 | 0.33 | 0.33 | 0 | 3 | 1 | 20.1 |
| 0.33 | 0.33 | 0 | 0.33 | 3 | 1 | 18.9 |
| 0.33 | 0 | 0.33 | 0.33 | 3 | 1 | 24.6 |
| 0 | 0.33 | 0.33 | 0.33 | 3 | 1 | 17.8 |
| 0.25 | 0.25 | 0.25 | 0.25 | 4 | 1.5 | 27.6 |
| 0.25 | 0.25 | 0.25 | 0.25 | 4 | 1.5 | 25.6 |
| 0.25 | 0.25 | 0.25 | 0.25 | 4 | 1.5 | 18.4 |
| 0.25 | 0.25 | 0.25 | 0.25 | 4 | 1.5 | 23.7 |

We fitted the richness model and its equivalent DI model to this dataset, with the following results.

## Comparison of the richness and simplified DI models – model specification

**The richness model:**

$$y = \beta_0 + \beta_1 r + \epsilon$$

**The simplified DI model:**

$$y = \alpha + \delta \sum_{\substack{i,j=1 \\ i<j}}^{S} (p_i p_j)^{0.5} + \epsilon$$

## Estimated models
These are equivalent models but re-parameterisations of each other.

| Parameter | Est | SE | p-value |
|---|---|---|---|
| Intercept | 6.523 | 2.1906 | 0.0089 |
| r | 4.493 | 0.8215 | <0.001 |

| Parameter | Est | SE | p-value |
|---|---|---|---|
| Intercept | 11.016 | 1.475 | <0.001 |
| SPP_0.5 | 8.986 | 1.643 | <0.001 |

## Predicting from the models for monocultures and equi-proportional communities

Data for communities 1, 5 and 20 (Comm = 1, 5, 20; values extracted from the full dataset):

| Comm | p1 | p2 | p3 | p4 | r | SPP_0.5 | y |
|---|---|---|---|---|---|---|---|
| 1 | 1 | 0 | 0 | 0 | 1 | 0 | 5.1 |
| 5 | 0.5 | 0.5 | 0 | 0 | 2 | 0.5 | 10.9 |
| 20 | 0.25 | 0.25 | 0.25 | 0.25 | 4 | 1.5 | 23.7 |

$\hat{y}_1 = 6.523 + 4.493 \times 1 = 11.0$

$\hat{y}_5 = 6.523 + 4.493 \times 2 = 15.5$

$\hat{y}_{20} = 6.523 + 4.493 \times 4 = 24.5$

$\hat{y}_1 = 11.016 + 8.986 \times 0 = 11.0$

$\hat{y}_5 = 11.016 + 8.986 \times 0.5 = 15.5$

$\hat{y}_{20} = 11.016 + 8.986 \times 1.5 = 24.5$

## ANOVA tables
The models are equivalent; thus the ANOVA tables are the same.

| | Df | SS | MS | F | p |
|---|---|---|---|---|---|
| r | 1 | 412.70 | 412.7 | 29.9 | <0.001 |
| Error | 16 | 220.75 | 13.8 | | |
| Total | 17 | 633.45 | | | |

| | Df | SS | MS | F | p |
|---|---|---|---|---|---|
| SPP_0.5 | 1 | 412.70 | 412.7 | 29.9 | <0.001 |
| Error | 16 | 220.75 | 13.8 | | |
| Total | 17 | 633.45 | | | |

## Direct mathematical association between the models

There is a direct mathematical association between the parameter estimates in the two models when fitted to datasets with equi-proportional communities:
- The estimated intercept from the simplified DI model is equal to the sum of the estimates of the intercept and slope from the richness model.
- The estimate of the slope from the simplified DI model is equal to twice the slope from the richness model.
- This association arises from the fact that SPP_0.5 = (richness-1)/2 (see proof in SI1.1).

As an example from this dataset:
- 11.016 = 6.523 + 4.493
- 8.986 = 2 × 4.493.

**NB**: This equivalency only holds because the dataset being analysed contains monocultures and only equi-proportional initial mixtures. For data where mixtures deviate from all species having equal initial proportions, this equivalency would not hold.

## SI1.3 The advantages of the DI modelling framework

If the simplified DI model and the richness model are the same for initially equi-proportional datasets, what are the advantages of using the Diversity-Interactions / DI framework?

The richness model is limited in that it assumes no unique species-specific contributions to the BEF relationship, it assumes that the relationship is linear with richness, and it assumes that any pair of species interacts with the same strength.

Testing a range of DI models can address the following questions:

- Do the species have species specific contributions (identity effects) to the BEF relationship?
- Are deviations from linear richness needed to model the BEF relationship?
- Is assuming that 'all interactions have the same strength' reasonable or do other interaction structures fit the data better?

A range of DI models were fitted to the data from the previous section with the following AIC results:

| # | Model | Fitted using (see code in SI1.4) | AIC |
|---|---|---|---|
| A1 | Richness | The `lm()` function and the `richness_vs_DI()` function in the `DImodels` package | 102.20 |
| A2 | GDI: equal identity terms, average interaction term (AV), and $\theta$ = 0.5 | The `lm()` function and the `richness_vs_DI()` function | 102.20 |
| A3 | GDI: separate identity terms, average interaction term (AV), and $\theta$ = 0.5 | `richness_vs_DI()` function | 97.36 |
| A4 | GDI: separate identity terms, average interaction term (AV), and $\theta$ estimated | `richness_vs_DI()` function and `DI()` function | 95.42 |
| A5 | GDI: separate identity terms, separate pairwise interaction terms (FULL), and $\theta$ estimated. | `DI()` function | 103.00 |

For the dataset in this supporting information document, A1 and A2 are equivalent because the data contains only monocultures and initially equi-proportional mixtures. Extending the DI model, it is shown that the species-specific identity terms are contributing to the BEF explanation (A3 AIC = 97.36 versus A2 AIC = 102.20). There is an indication of a deviation from linear richness: the AIC from the model with $\theta$ estimated has a lower AIC than when $\theta$ is set to equal 0.5 (A3 AIC = 97.36 versus A4 AIC 95.42, difference ≈ 2). There is no need for a unique interaction for each pair of species (A4 AIC = 95.42 versus A5 AIC = 103). The model with separate identity effects and average interaction term with $\theta$ estimated (A4) has the lowest AIC (= 95.42).

## SI1.4 Code to reproduce the analyses in this Supporting Information document

```r
# Load the DImodels package
library(DImodels)

# Store the variables in series of vectors
p1 <- c(1,0,0,0,0.5,0.5,0.5,0,0,0,0.333333333,0.333333333,0.333333333,0,0,0.25,0.25,0.25,0.25)
p2 <- c(0,1,0,0,0.5,0,0,0.5,0.5,0,0.333333333,0.333333333,0,0.333333333,0.25,0.25,0.25,0.25)
p3 <- c(0,0,1,0,0,0.5,0,0.5,0,0.5,0.333333333,0,0.333333333,0.333333333,0.25,0.25,0.25,0.25)
p4 <- c(0,0,0,1,0,0,0.5,0,0.5,0.5,0,0.333333333,0.333333333,0.333333333,0.25,0.25,0.25,0.25)
r <- c(1,1,1,1,2,2,2,2,2,2,3,3,3,3,4,4,4,4)
y <- c(5.1,8.2,12.5,14.3,10.9,12.4,16.8,18.2,17.9,22.1,20.1,18.9,24.6,17.8,27.6,25.6,18.4,23.7)

# Merge the vectors into a dataframe
data1 <- data.frame(p1, p2, p3, p4, r, y)

# Create the AV pairwise interaction term with theta = 0.5
data1$SPP_0.5 <- DI_data(prop = c("p1", "p2", "p3", "p4"),
                         data = data1, what = "AV", theta = 0.5)

# Plot the response versus richness
plot(data1$r, data1$y, xlab = "Richness", ylab = "Response", pch = 16)

# Fit the richness model and examine the estimates, predictions and ANOVA table
m1_richness <- lm(y ~ r, data = data1)
summary(m1_richness)
predict(m1_richness)
anova(m1_richness)
plot(m1_richness)

# Fit the simplified GDI model (i.e. the AV model with ID effects equal and theta set to 0.5)
# and examine the estimates, predictions and ANOVA table
m2_DI <- lm(y ~ SPP_0.5, data = data1)
summary(m2_DI)
predict(m2_DI)
anova(m2_DI)
plot(m2_DI)

# Examine the AIC values for the richness model and a small number of GDI models
richness_vs_DI(y = "y", prop = c("p1", "p2", "p3", "p4"), data = data1)

# Fit the AV pairwise model and allow theta to be estimated
m3_DI <- DI(y = "y", prop = c("p1", "p2", "p3", "p4"), data = data1, DImodel = "AV", estimate_theta = TRUE)
summary(m3_DI)
AIC(m3_DI)
plot(m3_DI)

# Fit the FULL pairwise model and allow theta to be estimated
m4_DI <- DI(y = "y", prop = c("p1", "p2", "p3", "p4"), data = data1, DImodel = "FULL", estimate_theta = TRUE)
summary(m4_DI)
AIC(m4_DI)
plot(m4_DI)
```

# Supporting Information 2
# Vignette: The analysis of the `sim0` dataset.

This vignette reproduces the analyses provided the 'Simulated data case study' in Section 3 of the main text and provides some additional supporting material for the analyses.

## SI2.1 DImodels installation and load

The `DImodels` package is installed from CRAN and loaded in the typical way.

```
#install.packages("DImodels")
library(DImodels)
```

Loading other packages necessary for this vignette

```
library(tidyverse) # For data manipulation
library(PieGlyph)  # For replacing points with pie-chart glyphs
library(ggtern)    # For creating ternary diagram
library(knitr)     # For creating pretty tables
library(lsmeans)   # For calculating contrasts
library(grid)      # For creating grid graphics
library(ggplotify) # For converting grid graphic plots to ggplots
```

## SI2.2 Simulating and exploring the `sim0` dataset

The dataset `sim0` is available in the `DImodels` package.

The dataset was simulated with the design points in Table SI2.1.

```
# Load the design dataset
data("sim0_design", package="DImodels")
kable(sim0_design, digits = 2, caption = "**Table SI2.1**: *Unique design points for the
↪ `sim0` dataset.*")
```

Table 1: **Table SI2.1**: *Unique design points for the `sim0` dataset.*

| community | richness | p1 | p2 | p3 |
|---|---|---|---|---|
| 1 | 1 | 1.00 | 0.00 | 0.00 |
| 2 | 1 | 0.00 | 1.00 | 0.00 |
| 3 | 1 | 0.00 | 0.00 | 1.00 |
| 4 | 2 | 0.80 | 0.20 | 0.00 |
| 5 | 2 | 0.20 | 0.80 | 0.00 |
| 6 | 2 | 0.80 | 0.00 | 0.20 |
| 7 | 2 | 0.20 | 0.00 | 0.80 |
| 8 | 2 | 0.00 | 0.80 | 0.20 |
| 9 | 2 | 0.00 | 0.20 | 0.80 |
| 10 | 2 | 0.50 | 0.50 | 0.00 |

| community | richness | p1 | p2 | p3 |
|---|---|---|---|---|
| 11 | 2 | 0.50 | 0.00 | 0.50 |
| 12 | 2 | 0.00 | 0.50 | 0.50 |
| 13 | 3 | 0.60 | 0.20 | 0.20 |
| 14 | 3 | 0.20 | 0.60 | 0.20 |
| 15 | 3 | 0.20 | 0.20 | 0.60 |
| 16 | 3 | 0.33 | 0.33 | 0.33 |

The design can be visualised using a ternary diagram, as shown in Figure SI2.1.

```r
# Generate a ternary diagram to illustrate the design points

# Load design
data("sim0_design")
# Set up the aesthetics for points
sim0_design2 <- sim0_design
sim0_design2$colour <- c('blue', rep('black', 7), 'red',  rep('black', 6), 'orange')

####################################
# Plot the experimental design in a ternary diagram
ggtern(data = sim0_design2, aes(x=p2, y=p1, z=p3)) +
    # Create borders of ternary diagram
    geom_polygon(data = data.frame(p1 = c(1,0,0),
                                   p2 = c(0,1,0),
                                   p3 = c(0,0,1)), aes(fill = NULL), alpha = 0, colour = 'black')+
    # Add masking to allow to add points on top of the ternary
    geom_mask()+
    # Add points on the ternary
    geom_point(size = 4, colour = sim0_design2$colour)+
    # Aesthetics of the ternary
    # Axis labels
    labs( x      = "Sp2",
          xarrow = "Proportion of species 2",
          y      = "Sp1",
          yarrow = "Proportion of species 1",
          z      = "Sp3",
          zarrow = "Proportion of species 3")+
    # Theme for figure
    ggtern::theme_bw()+
    # Add arrows for ease of reading the diagram
    theme_arrowlarge() +
    # Adjust text size of arrows and axis labels
    theme(axis.text = element_text(size = 14, colour = 'black'),
          tern.axis.text.R = element_text(hjust = .75),
          tern.axis.text.L = element_text(hjust = .65),
          tern.axis.text.T = element_text(hjust = .25),
          tern.axis.arrow = element_line(size = 3, colour = 'black'),
          tern.axis.arrow.text = element_text(colour = 'black', vjust = -.07),
          tern.axis.arrow.text.R = element_text(vjust = .9),
          panel.border = element_rect(fill = NA, colour = 'black'),
          axis.title = element_text(size = 14, colour = 'black'),
          legend.position = 'none')+
```

```r
    # Change labels to be displayed on ternary axes
    scale_L_continuous(limits = c(0,1), breaks = c(.2, .4, .6, .8), labels =  c(.2, .4,
    ↪  .6, .8))+
    scale_R_continuous(limits = c(0,1), breaks = c(.2, .4, .6, .8), labels =  c(.2, .4,
    ↪  .6, .8))+
    scale_T_continuous(limits = c(0,1), breaks = c(.2, .4, .6, .8), labels =  c(.2, .4,
    ↪  .6, .8))
#> Warning: The `size` argument of `element_line()` is deprecated as of ggplot2 3.4.0.
#> i Please use the `linewidth` argument instead.
#> This warning is displayed once every 8 hours.
#> Call `lifecycle::last_lifecycle_warnings()` to see where this warning was
#> generated.
```

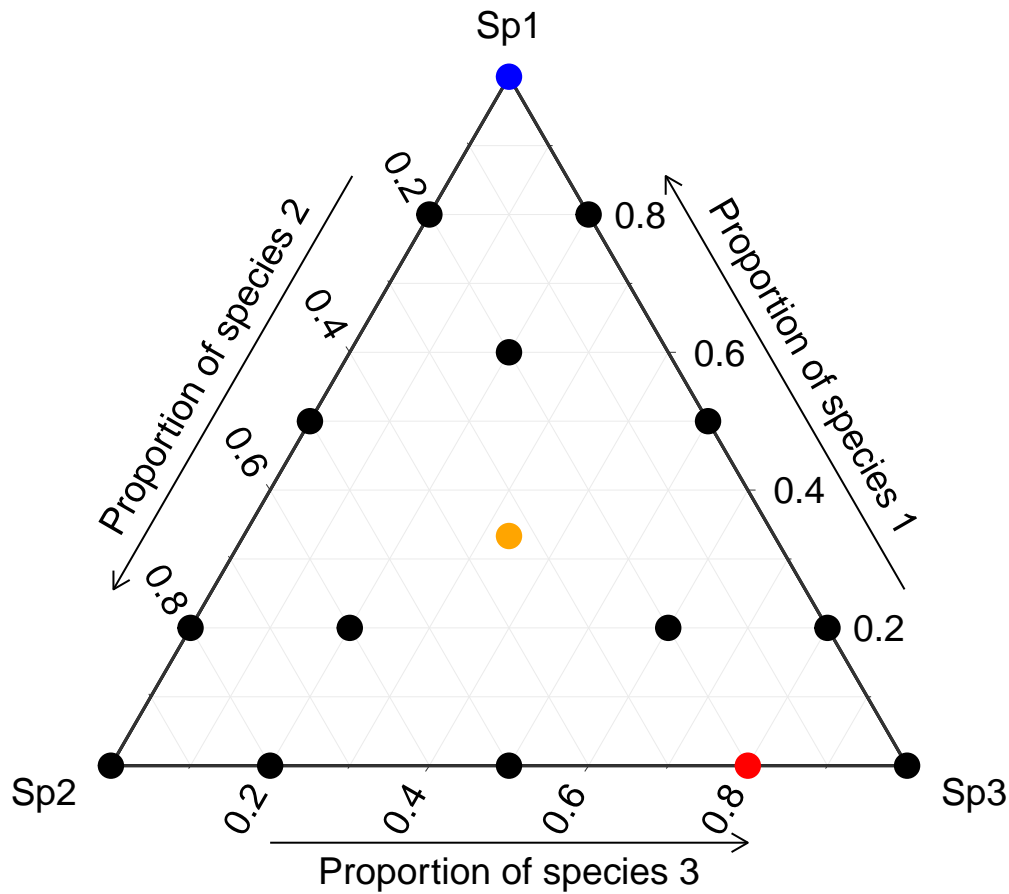

**Figure SI2.1**: *Ternary diagram showing the 16 design points from which the* `sim0` *dataset was simulated. Each vertex represents a monoculture of that species (e.g. the blue point represents a monoculture of species 1, 1:0:0); the sides of the ternary diagram represent two-species communities, and the closer a point is to one vertex, the more dominant that species is (e.g. the red point represents a community with 20% of species 2 and 80% of species 3, 0:0.2:0.8); all interior points represent a three-species mixture (e.g. the orange point is the community with an equal proportion of all three species, 0.333:0.333:0.333), and the closer a point is to a vertex, the higher the proportion of that species present.*

Using the proportions $(p1, p2, p3)$ in the design dataset as inputs, the `sim0` dataset was simulated with four replicates of each design point using the model:

$$y = \sum_{i=1}^{3} \beta_i p_i + \sum_{\substack{i,j=1 \\ i<j}}^{3} \delta_{ij} p_i p_j + \epsilon$$

where $\epsilon \sim N(0, \sigma^2)$. The values for the model parameters used to simulate from are: $\beta_1 = 25$, $\beta_2 = 20$, $\beta_3 = 15$, $\delta_{12} = 30$, $\delta_{13} = 20$, $\delta_{23} = 40$ and $\sigma^2 = 4$.

The code to simulate the `sim0` dataset is as follows (but where the data is called `box1_data` instead of `sim0`):

```r
# Reading experimental design
data("sim0_design", package="DImodels")
design <- sim0_design

# Replicate the design for four values of each community
box1_data <- design[rep(seq_len(nrow(design)), times = 4), ]
row.names(box1_data) <- NULL

# To simulate the response, first create a matrix of predictors that includes
# p1-p3, and all pairwise interaction variables
X <- model.matrix(~ p1 + p2 + p3 + (p1 + p2 + p3)^2 -1, data = box1_data)

## Create a vector of 'known' parameter values for simulating the response.
## The first three are the p1-p3 parameters,
## and the second set of 15 are the interaction parameters.
coeffs <- c(25,20,15,    30,20,40)

## Create response and add normally distributed error
box1_data$response <- as.numeric(X %*% coeffs)
set.seed(86914)
r <- rnorm(n = nrow(box1_data), mean = 0, sd = 2)
box1_data$response <- round(box1_data$response + r, digits = 3)

# Note that box1_data is exactly the same as sim0, the code is shown here for
# any user that wishes to reproduce the simulation. The different name is to
# avoid a user creating a local dataset with the same name as a package dataset.

# For users wishing to only use the sim0 dataset, it can be loaded directly
# from the DImodels package as in the next chunk of code.
```

To examine the `sim0` dataset we view the first six rows (Table SI2.2) and generate a histogram of the response variable (Figure SI2.2).

```r
# Load the dataset directly from the DImodels package
data(sim0)

# View the first five entries
kable(head(sim0), caption = "**Table SI2.2**: *The first six rows of the `sim0`
  dataset.*")
```

Table 2: **Table SI2.2**: *The first six rows of the `sim0` dataset.*

| community | richness | p1 | p2 | p3 | response |
|---|---|---|---|---|---|
| 1 | 1 | 1.0 | 0.0 | 0.0 | 24.855 |

| community | richness | p1 | p2 | p3 | response |
|---|---|---|---|---|---|
| 2 | 1 | 0.0 | 1.0 | 0.0 | 19.049 |
| 3 | 1 | 0.0 | 0.0 | 1.0 | 16.292 |
| 4 | 2 | 0.8 | 0.2 | 0.0 | 31.529 |
| 5 | 2 | 0.2 | 0.8 | 0.0 | 25.102 |
| 6 | 2 | 0.8 | 0.0 | 0.2 | 24.615 |

```
# Summary statistics
summary(sim0$response)
#>    Min. 1st Qu.  Median    Mean 3rd Qu.    Max.
#>   11.82   24.50   25.95   25.74   29.04   33.04

# Visualisation
ggplot(data = sim0)+
  geom_histogram(aes(x = response), bins = 10, fill='light green', colour = 'black')+
  labs(x = 'Response', y = 'Frequency')+
  theme_bw()+
  theme(axis.text = element_text(hjust=0.5, size=14),
        axis.title = element_text(hjust = 0.5, size = 16))
```

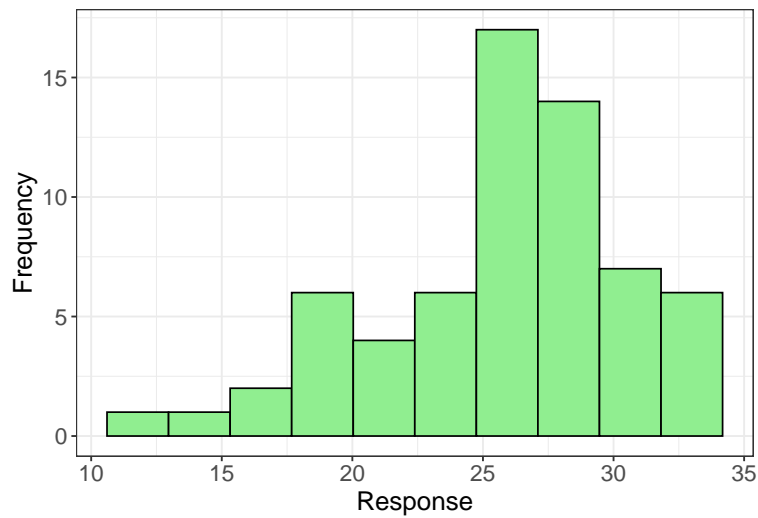

**Figure SI2.2**: *Histogram of the response variable in the `sim0` dataset.*

In the following subsections, we will fit and compare three modelling approaches: 1) a simple linear regression model with richness only as a predictor (referred to as the richness model), 2) a Diversity-Interactions (DI) model, and 3) a 1-way ANOVA model.

## SI2.3 Analysing `sim0` using the richness model

We fit the simple linear regression model with only richness as a predictor. The model is:

$$y = \beta_0 + \beta_1 Richness + \epsilon$$

where $\epsilon \sim N(0, \sigma^2)$.

The code to fit the regression model is:

```r
# Fit the simple linear regression model with Richness as a predictor
m1 <- lm(response ~ richness, data = sim0)
summary(m1)
#>
#> Call:
#> lm(formula = response ~ richness, data = sim0)
#>
#> Residuals:
#>     Min      1Q  Median      3Q     Max
#> -8.3363 -1.8101 -0.1053  2.4001  7.3317
#>
#> Coefficients:
#>             Estimate Std. Error t value Pr(>|t|)
#> (Intercept)  14.9093     1.3781   10.82 6.53e-16 ***
#> richness      5.2510     0.6365    8.25 1.46e-11 ***
#> ---
#> Signif. codes:  0 '***' 0.001 '**' 0.01 '*' 0.05 '.' 0.1 ' ' 1
#>
#> Residual standard error: 3.353 on 62 degrees of freedom
#> Multiple R-squared:  0.5233, Adjusted R-squared:  0.5156
#> F-statistic: 68.06 on 1 and 62 DF,  p-value: 1.46e-11

## Find the residual mean square
MSE <- sum(m1$residuals^2) / m1$df
MSE
#> [1] 11.24257

## Display the ANOVA table
richness_model_anova <- anova(m1)
knitr::kable(richness_model_anova, digits = 2, caption = "**Table SI2.3**: *The ANOVA
  table for the richness model.*")
```

Table 3: **Table SI2.3**: *The ANOVA table for the richness model.*

|           | Df | Sum Sq | Mean Sq | F value | Pr(>F) |
|-----------|----|--------|---------|---------|--------|
| richness  | 1  | 765.16 | 765.16  | 68.06   | 0      |
| Residuals | 62 | 697.04 | 11.24   |         |        |

The output shows that as richness increased, the response also increased (estimate = 5.25, p < 0.001). All variation around the line is assumed to be residual error variation, the residual mean square was 11.24, with 62 degrees of freedom (df). At each level of richness, some variation is likely due to changing composition and/or changes in the proportions of species (a measure of evenness). This means that the residual mean square is likely not reflecting only true error variation.

The model predictions from the model are shown in Figure SI2.3; the linear trend of increasing response as richness is apparent in this graph.

```r
# Visualise the fitted regression model
# Aesthetics for points
sim0$colour <- rep(c('blue', rep('black', 7), 'green', rep('black', 6), 'orange'),4)
sim0$size <- rep(c('4', rep('3', 7), '4', rep('3', 6), '4'),4)
sim0$evenness <- (sim0$p1*sim0$p2 + sim0$p2*sim0$p3 + sim0$p1*sim0$p3)*2*3/(3-1)
```

```r
# Scatter plot of the raw data with the SLR model fitted
richness_plot <- ggplot(data = sim0) +
    # Add the raw data points
    geom_point(aes(x = as.factor(richness), y = response, group = (evenness)),
               # Dodging to avoid overlapping of points
               position = position_dodge(0.25), size = 3) +
    # Overlay points with best fit regression line from model
    geom_abline(intercept = m1$coefficients[1],
                slope = m1$coefficients[2])+
    # Theme for plot
    theme_bw()+
    # Edit axis font size
    theme(axis.text = element_text(size=12),
          axis.title = element_text(size = 14))+
    # Axis labels
    labs(x = 'Richness', y = 'Response')

# Create predictions for communities of interest
data_interest <- data.frame(richness = c(1,2,3,3) + c(0,0,.01,-.01),
                            pred = (predict(m1)[c(1,4,13,14)] + c(0,0,.01,-0.1)))

# Highlight communities of interest
richness_plot +
  geom_point(data = data_interest,
             aes(x = richness, y = pred),
             position = position_dodge(.2),
             colour = c('blue', '#ff1a1a', '#fa6538','#ab0741'),
             size = 4, shape = 4, stroke = 3)
```

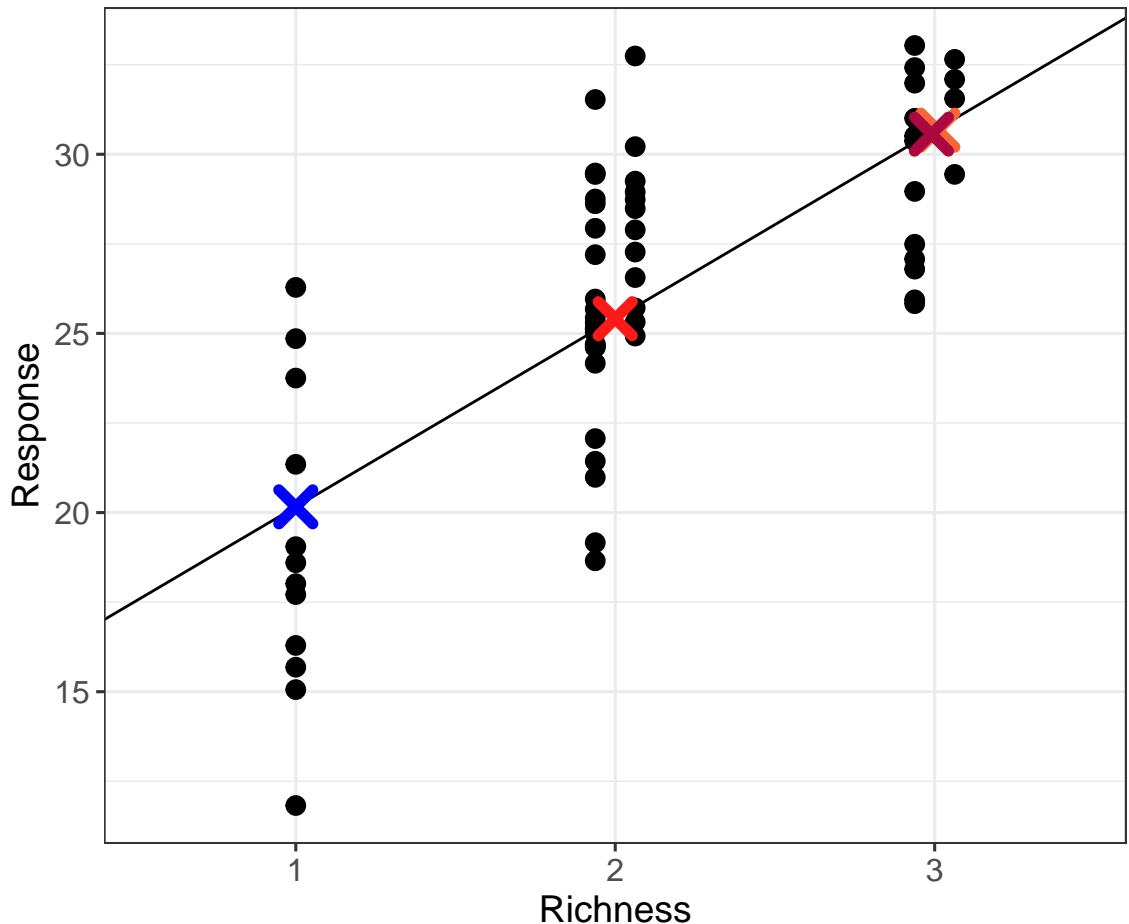

**Figure SI2.3**: *The fitted simple the fitted simple linear regression line for the response versus richness, raw data also shown. The coloured X's highlight select predictions: blue for a monoculture of species 1, red for 20% of species 2 and 80% of species 3, orange for an equi-proportional 3-species mixture and wine for 10% of species 1 and 3 and 80% of species 2. For this model, there is no difference in the prediction for the final two community types.*

## SI2.4 Comparing the richness model to a subset of DI models

When a dataset contains monocultures and only equi-proportional mixtures, the richness model is equivalent to the average pairwise DI model with all identity effects set equal and $\theta$ set to 0.5. In the `sim0` dataset, there are mixture communities that are not equi-proportional, e.g. (0.8, 0.2, 0). We can still compare this simplified DI model to the richness model using the `richness_vs_DI()` function to explore if the richness model is sufficient to model this data.

The `richness_vs_DI()` function compares the richness model to four DI models:

```
# Compare the richness model to a small subset of DI models
richness_vs_DI(y = "response", prop = c("p1", "p2", "p3"), data = sim0)
#>
#> --------------------------------------------------------------------------------
#>
#> Investigating richness model and three DI alternatives
#>      AIC      AICc     BIC       df
#> 1 340.4535 340.8535 346.9301    2
```

```
#> 2 337.2738 337.6738 343.7504 2
#> 3 330.4325 331.1104 339.0680 3
#> 4 305.3711 306.4056 318.3244 4
#> 5 288.3583 289.8320 301.3116 5
#>   Description
#> 1 Model 1: Richness only
#> 2 Model 2: Average interactions 'AV' DImodel with common identity effects and theta =
#>   0.5
#> 3 Model 3: Average interactions 'AV' DImodel with common identity effects and theta
#>   estimated
#> 4 Model 4: Average interactions 'AV' DImodel with unique identity effects and theta =
#>   0.5
#> 5 Model 5: Average interactions 'AV' DImodel with unique identity effects and theta
#>   estimated
#>
#> ---------------------------------------------------------------------------------
#> Models 1 and 2 are equivalent if all mixtures in the dataset at any given level of
#>   richness are equi-proportional.
#> richness_vs_DI is limited in terms of model selection. Only five models are explored.
#>   See ?autoDI and ?DI for more options.
#> ---------------------------------------------------------------------------------
```

For this dataset, the model with the predictor $\sum_{i<j} \delta_{ij}(p_i p_j)^{0.5}$ instead of richness (models 1 versus 2 in the output above) shows an improvement on the richness model (AIC 340.5 versus AIC 337.3). They are not equivalent models since there are unbalanced mixture communities in the `sim0` dataset.

There is evidence that the BEF relationship is not linear with richness (compare the first and third models: AIC 340.5 versus 330.4). There is also evidence for species-specific contributions (compare the first and fourth models: AIC 340.4 versus 305.4). The best model of the five is the last one (AIC = 288.4), it has unique species-specific identity effects and theta is estimated (and not forced to be 0.5).

This analysis has limitations, however, it firmly concludes that there will be considerable value in exploring the DI modelling framework, compared to using just the richness model for this dataset.

## SI2.5 Analysing `sim0` using DI modelling

We first investigate if $\theta$ different from 1 is required.

```
# Fit the DI model with full pairwise interactions and with theta = 1
theta_1 <- DI(y = "response", prop = 3:5, DImodel = "FULL", data = sim0, theta = 1)
#> Fitted model: Separate pairwise interactions 'FULL' DImodel
# Fit the DI model with full pairwise interactions and with theta = 1
theta_estimated <- DI(y = "response", prop = 3:5, DImodel = "FULL", data = sim0,
  estimate_theta = TRUE)
#> Fitted model: Separate pairwise interactions 'FULL' DImodel
#> Theta estimate: 0.9643

# Compare using AIC
AIC(theta_1)
#> [1] 270.2609
AIC(theta_estimated)
#> [1] 272.0698
```

The AIC for the model with theta equal to 1 is lower. Some users may wish to always estimate $\theta$ regardless, however, we opt to proceed with $\theta = 1$.

Here we fit the full pairwise Diversity-Interactions model:

$$y = \sum_{i=1}^{3} \beta_i p_i + \sum_{\substack{i,j=1 \\ i<j}}^{3} \delta_{ij} p_i p_j + \epsilon$$

where $\epsilon \sim N(0, \sigma^2)$.

```
# Fit the DI model with full pairwise interactions
m2 <- DI(y = "response", prop = 3:5, DImodel = "FULL", data = sim0)
#> Fitted model: Separate pairwise interactions 'FULL' DImodel

# View the summary of the fitted model
summary(m2)
#>
#> Call:
#> glm(formula = fmla, family = family, data = data)
#>
#> Deviance Residuals:
#>     Min       1Q   Median       3Q      Max
#> -2.9829  -1.1752  -0.1008   1.1086   5.0970
#>
#> Coefficients:
#>         Estimate Std. Error t value Pr(>|t|)
#> p1       24.1740     0.7698  31.403  < 2e-16 ***
#> p2       18.6155     0.7698  24.182  < 2e-16 ***
#> p3       14.8069     0.7698  19.235  < 2e-16 ***
#> `p1:p2`  34.8345     3.5694   9.759 7.60e-14 ***
#> `p1:p3`  26.1355     3.5694   7.322 8.43e-10 ***
#> `p2:p3`  47.9159     3.5694  13.424  < 2e-16 ***
#> ---
#> Signif. codes:  0 '***' 0.001 '**' 0.01 '*' 0.05 '.' 0.1 ' ' 1
#>
#> (Dispersion parameter for gaussian family taken to be 3.541817)
#>
#>     Null deviance: 43863.70  on 64  degrees of freedom
#> Residual deviance:   205.43  on 58  degrees of freedom
#> AIC: 270.26
#>
#> Number of Fisher Scoring iterations: 2
```

The identity effect estimates for species 1, 2 and 3 are 24.2, 18.6 and 14.8 respectively. The interaction parameter estimates are 34.8, 26.1 and 47.9 for species pairs 1-2, 1-3 and 2-3 respectively. All interactions are significant (p < 0.001 in each case).

The predictions from the Diversity-Interactions model are shown in the ternary diagram in Figure SI2.4.

```
# Show predictions for entire simplex space using ternary diagram

# Create ternary diagram

# Create data for plotting on ternary diagram
```

```r
trian <- expand.grid(base=seq(0,1,l=100*2*5),
                     high=seq(0,sin(pi/3),l=87*2*5))
# Subset 2-d ternary space (triangle)
trian <- subset(trian, (base*sin(pi/3)*2)>high)
trian <- subset(trian, ((1-base)*sin(pi/3)*2)>high)

# Map 2-d co-ordinates to proportions of species
trian$p1 <- trian$high*2/sqrt(3)
trian$p3 <- trian$base-trian$high/sqrt(3)
trian$p2 <- 1-trian$p3-trian$p1

# Make predictions
trian$yhat <- predict(m2, trian)

# Creating levels for the contour map
lower <- round(min(trian$yhat))
upper <- round(max(trian$yhat))
size <- 7 # Number of colours

# Breaks for the legend
breaks <- round(seq(lower,upper,length.out=size))

# Axes and labels for the ternary diagram
axis_labels <- data.frame(x1 = seq(0.2,0.8,0.2)) %>%
               mutate(y1 = c(0,0,0,0),
                      x2 = x1/2,
                      y2 = x1*sqrt(3)/2,
                      x3 = (1-x1)*0.5+x1,
                      y3 = sqrt(3)/2-x1*sqrt(3)/2,
                      label = x1,
                      rev_label = rev(label),
                      yhat = 0)

vertex_labels <- data.frame(x = c(0.5, 1 + 0.05, 0 -0.05),
                            y = c(sqrt(3)/2 + 0.05, 0,  0),
                            label = c('Sp1', 'Sp3','Sp2'),
                            yhat = 0)

# Colours for the contour
colours <- terrain.colors(n = size, rev = T)

# Creating the ternary diagram
tern_plot <- ggplot(trian, aes(x = base, y = high, z = yhat))+
    # Filled contour
    geom_raster(aes(fill = yhat))+
    # Specify colours and breaks for the contour
    scale_fill_stepsn(colours = colours, breaks = breaks,
                     labels = function(x){
                       round(x)
                     },
                     limits = c(lower, upper),
                     show.limits = F)+
    # Draw contour lines
```

```r
    geom_contour(breaks = breaks, colour = 'black') +
    # Add axis titles (vertices of triangle)
    geom_text(data = vertex_labels,
              aes(x= x, y= y, label=label),
              size = 6) +
    # Add boundary of triangle
    geom_segment(data = data.frame(x = c(0, 0, 1),
                                   y = c(0,0,0),
                                   xend = c(1, 0.5, 0.5),
                                   yend = c(0, sqrt(3)/2, sqrt(3)/2),
                                   yhat = 0),
                 aes(x=x, y=y, xend=xend, yend=yend),
                 linewidth = 1)+
    # Add labels for the three axes
    geom_text(data = axis_labels,
              aes(x=x1, y=y1, label=label, fontface = 'plain'), nudge_y=-0.055, size =
              ↪ 5)+
    geom_text(data = axis_labels,
              aes(x=x2, y=y2, label=rev_label, fontface = 'plain'),  nudge_x=-0.055,
              ↪ nudge_y=0.055, size = 5)+
    geom_text(data = axis_labels,
              aes(x=x3, y=y3, label=rev_label, fontface = 'plain'),  nudge_x=0.055,
              ↪ nudge_y=0.055, size = 5)+
    # Remove existing axes (x and y)
    theme_void()+
    # Design of the legend
    guides(fill = guide_colorsteps(frame.colour = 'black',
                                   ticks.colour = 'black',
                                   title = 'Prediction',
                                   show.limits = T))+
    # Fix the coordinate system so the triangles don't distort
    coord_fixed()+
    # Edit specific elements of theme
    theme(legend.key.size = unit(0.1, 'npc'),
          legend.key.height = unit(0.04, 'npc'),
          legend.title = element_text(size = 14, vjust = 0.75),
          legend.text = element_text(size = 12),
          legend.position = 'bottom')
#> Coordinate system already present. Adding new coordinate system, which will
#> replace the existing one.

# Add original design points on the ternary and highlight communities of interest
# Create data
design <- rbind(sim0[1:16,c(2:5,9)], c(3, .1,.8,.1, .51))
# Specify x and y coordinates on the figure for the new communities
design$x <- design$p3 + design$p1/2
design$y <- design$p1*sqrt(3)/2
# Add predictions
design$yhat <- predict(m2, design)
# Aesthetics for communities
design$colour <- fct_inorder(c(sim0$colour[1:16], 'purple'))
design$size <- c(sim0$size[1:16],'4')
design$shape <- c(sim0$size[1:16],'4')
```

```r
# Add communities on the ternary
tern_plot +
  geom_point(data = design, aes(x = x, y = y, color = colour,
                                size = size, shape = shape),
             stroke = 3)+
  # Colours of points
  scale_colour_manual(values = c('blue', 'black', '#ff1a1a', '#fa6538','#ab0741'))+
  # Enlarge points of interest
  scale_size_manual(values = c(2, 4))+
  # Marks points of interest with an X
  scale_shape_manual(values = c(16,4))+
  # Suppress legends
  guides(colour = 'none', size = 'none', shape = 'none')
```

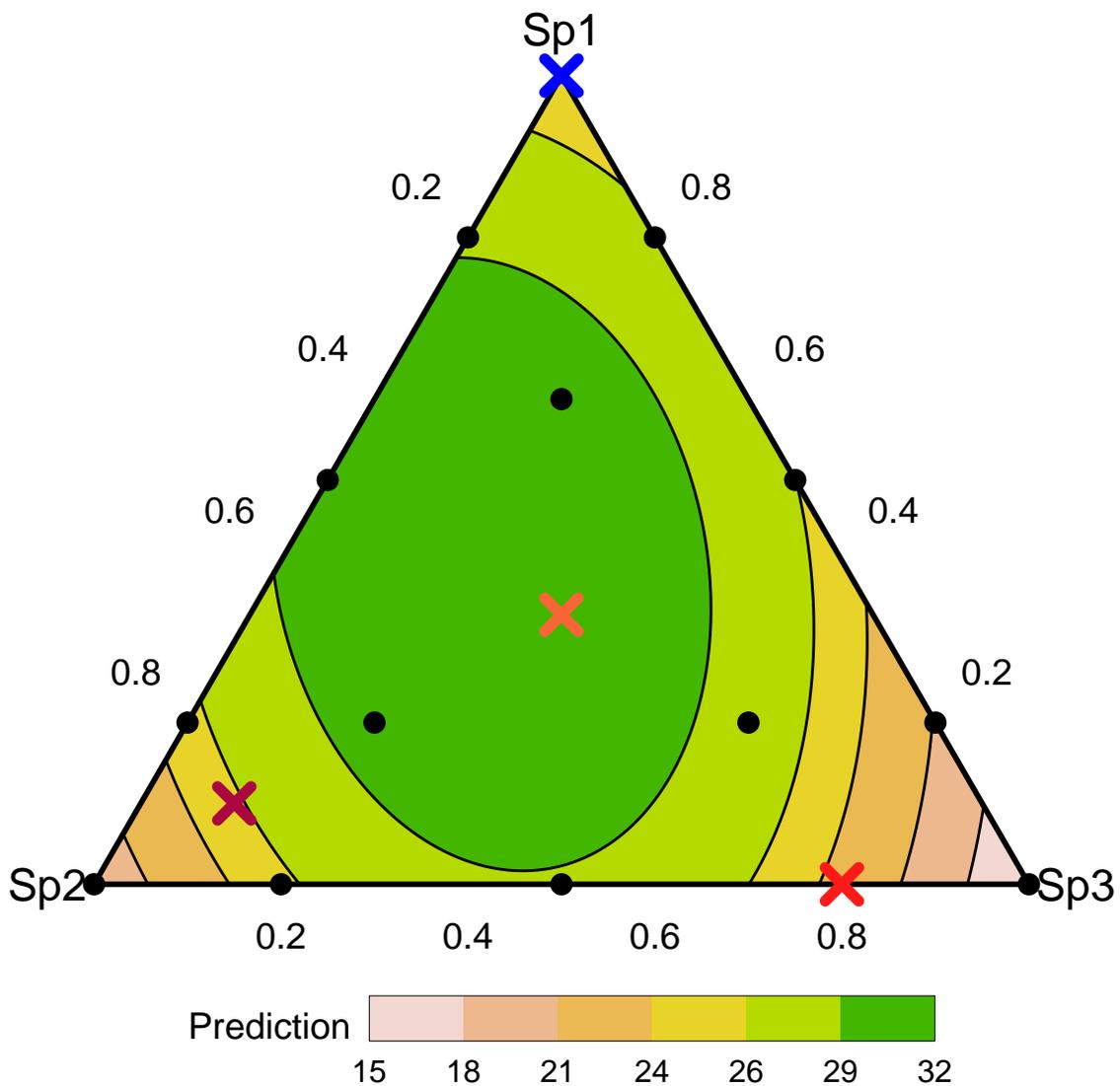

**Figure SI2.4**: *The predicted response across the three-dimensional simplex space, predicted from fitting a Diversity-Interactions model, with design points marked with a dot or X. The coloured X's highlight: blue for a monoculture of species 1, red for 20% of species 2 and 80% of species 3, orange for an equi-proportional*

*3-species mixture and wine for 10% of species 1 and 3 and 80% of species 2.*

```
# Create Deviance and ANOVA tables
DI_model_dev <- anova(m2)
knitr::kable(DI_model_dev, caption = "**Table SI2.4**: *Deviance table for the DI
↪  model.*")
```

Table 4: **Table SI2.4**: *Deviance table for the DI model.*

|       | Df | Deviance  | Resid. Df | Resid. Dev |
|-------|----|-----------|-----------|------------|
| NULL  |    |           | 64        | 43863.6987 |
| p1    | 1  | 24921.7302 | 63        | 18941.9685 |
| p2    | 1  | 12231.5403 | 62        | 6710.4282  |
| p3    | 1  | 5544.6967 | 61        | 1165.7315  |
| p1:p2 | 1  | 205.4772  | 60        | 960.2543   |
| p1:p3 | 1  | 116.5788  | 59        | 843.6755   |
| p2:p3 | 1  | 638.2501  | 58        | 205.4254   |

```
m2a <- lm(response ~ 0 + (p1 + p2 + p3)^2, data = sim0)
DI_model_anova <- anova(m2a)
knitr::kable(DI_model_anova, digits = 2, caption = "**Table SI2.5**: *ANOVA table for the
↪  DI model.*")
```

Table 5: **Table SI2.5**: *ANOVA table for the DI model.*

|           | Df | Sum Sq    | Mean Sq   | F value  | Pr(>F) |
|-----------|----|-----------|-----------|----------|--------|
| p1        | 1  | 24921.73  | 24921.73  | 7036.43  | 0      |
| p2        | 1  | 12231.54  | 12231.54  | 3453.47  | 0      |
| p3        | 1  | 5544.70   | 5544.70   | 1565.50  | 0      |
| p1:p2     | 1  | 205.48    | 205.48    | 58.01    | 0      |
| p1:p3     | 1  | 116.58    | 116.58    | 32.91    | 0      |
| p2:p3     | 1  | 638.25    | 638.25    | 180.20   | 0      |
| Residuals | 58 | 205.43    | 3.54      |          |        |

In addition to the ternary diagram, we can illustrate predictions across evenness or richness gradients. Figure SI2.5 shows the predictions versus evenness, where evenness (E) is defined as in Kirwan et al 2007:

$$E = \frac{2s}{(s-1)} \Sigma_{i<j} p_i p_j$$

Where $s = 3$, is the number of species in the pool. While Figure SI2.6 shows the predictions versus richness. In each case, the predictions vary at any given point on the x-axis, depending on the composition and proportions of species in the initial community (shown in the pie glyph).

```
# Use the PieGlyph package to create scatter plots of predictions using pie-glyphs as
↪  points on the scatter plot on the eveness scale

# Create the dataset with model predictions (from the full pairwise DI model)
```

```r
sim0$pred <- predict(m2)

# Create figure with pie-glyph showing species proportions
pie_figure <- ggplot(data = sim0[1:16,],
       aes(x = evenness, y = pred))+
    # Line connecting average response at each level of evenness
    geom_line(data = rbind(sim0 %>% group_by(evenness) %>% summarize(mean = mean(pred))),
              aes(y = mean, group = 1), size = 1) +
    # Geom that will add the pie-glyphs showing proportion of communities
    PieGlyph::geom_pie_glyph(aes(group = evenness),
                             slices = c('p1','p2','p3'),
                             data = sim0[1:16,],
                             radius = 0.375)+
    # Theme of plot
    theme_bw()+
    # Adjust colours in the legend
    scale_fill_manual(values = c('#ffcb66', '#669aff', '#cb66ff'))+
    # Axis and legend labels
    labs(x = 'Evenness', y = 'Prediction', fill = 'Species')+
    ylim(14, 32)+
    # Edit text size of legend
    theme(axis.text = element_text(size=12),
          axis.title = element_text(size = 14),
          legend.text = element_text(size=12),
          legend.title = element_text(size = 14))

# Add original design points on the plot and highlight communities of interest
# Create data of communities of interest
design <- rbind(sim0[1:16,c(2:5,9)], c(3, .1,.8,.1, .51))
# Adjust aesthetics
design$colour <- fct_inorder(c(sim0$colour[1:16], 'purple'))
design$size <- c(sim0$size[1:16],'4')
design$shape <- c(sim0$size[1:16],'4')
# Add predictions
design$pred <- predict(m2, newdata = design)

# Add points of interest
pie_figure +
  geom_point(data = design %>% filter(colour != 'black'),
             aes(x = evenness, y = pred, color = colour),
             stroke = 3, size = 4, shape = 4)+
  # colours for points
  scale_colour_manual(values = c('blue', '#ff1a1a', '#fa6538','#ab0741'))+
  # Remove legends and keep only species proportions
  guides(colour = 'none', size = 'none', shape = 'none')
```

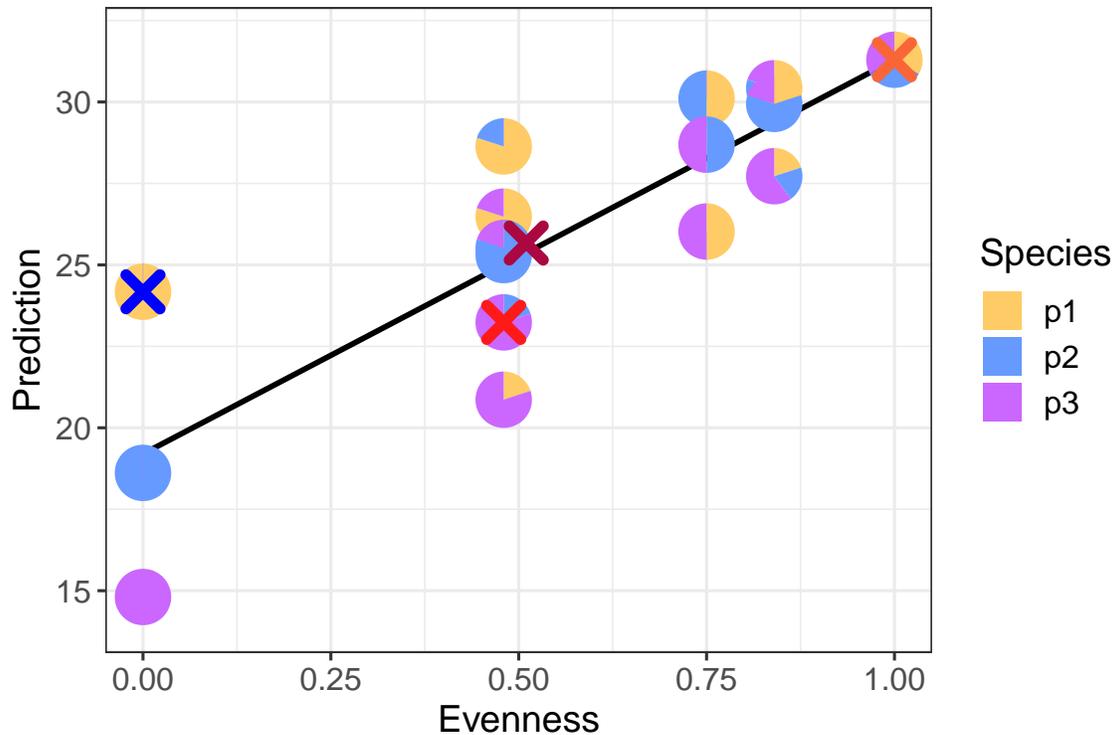

**Figure SI2.5**: *The predicted response for each design point from fitting a Diversity-Interactions model, with the pie glyphs illustrating the initial proportions and trend line connecting the average prediction at each level of evenness. The coloured X's highlight: blue for a monoculture of species 1, red for 20% of species 2 and 80% of species 3, orange for an equi-proportional 3-species mixture and wine for 10% of species 1 and 3 and 80% of species 2.*

```r
# figure with predictions using DI model, showing species proportions using pie-glyphs
#   with richness scale on the x-axis

# Create figure with pie-glyph showing species proportions
pie_figure <- ggplot(data = sim0[1:16,],
      aes(x = (richness), y = pred))+
  # Line connecting average response at each level of richness
  geom_line(data = rbind(sim0 %>% group_by(richness) %>% summarize(mean = mean(pred))),
            aes(y = mean, group = 1), size = 1) +
  # Geom that will add the pie-glyphs showing proportion of communities
  PieGlyph::geom_pie_glyph(aes(group = evenness),
                           slices = c('p1','p2','p3'),
                           data = sim0[1:16,],
                           position = position_dodge(0.5),
                           radius = 0.375)+
  # Theme of plot
  theme_bw()+
  # Adjust colours in the legend
  scale_fill_manual(values = c('#ffcb66', '#669aff', '#cb66ff'))+
  # Axis and legend labels
  labs(x = 'Richness', y = 'Prediction', fill = 'Species')+
  ylim(14, 32)+
  scale_x_continuous(breaks = 1:3)+
```

```r
    # Edit text size of legend
    theme(axis.text = element_text(size=12),
          axis.title = element_text(size = 14),
          legend.text = element_text(size=12),
          legend.title = element_text(size = 14))

# Add original design points on the plot and highlight communities of interest
# Create data of communities of interest
design <- rbind(sim0[1:16,c(2:5,9)], c(3, .1,.8,.1, .51))
# Adjust aesthetics
design$colour <- fct_inorder(c(sim0$colour[1:16], 'purple'))
design$size <- c(sim0$size[1:16],'4')
design$shape <- c(sim0$size[1:16],'4')
# Add predictions
design$pred <- predict(m2, newdata = design)

# Add points of interest
pie_figure +
  # manually adjust the dodging for points as position_dodge
  # doesn't work well due to data having only 4 points
geom_point(data = design %>% filter(colour != 'black'),
           aes(x = c(1, 1.875, 3.125, 2.875), color = colour),
           stroke = 3, size = 4, shape = 4)+
  # colours for points
  scale_colour_manual(values = c('blue', '#ff1a1a', '#fa6538','#ab0741'))+
  # remove legends and keep only species
  guides(colour = 'none', size = 'none', shape = 'none')
```

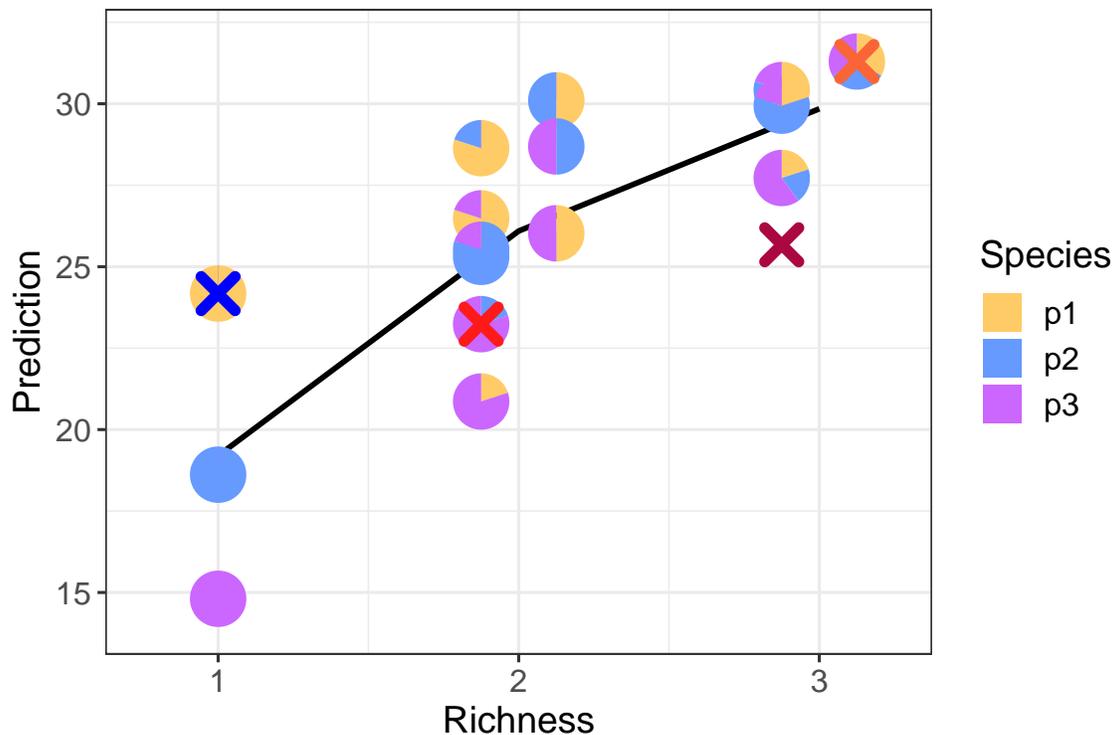

**Figure SI2.6**: *The predicted response for each design point from fitting a Diversity-Interactions model, with the pie glyphs illustrating the initial proportions and trend line connecting the average prediction at each level*

*of richness. The coloured X's highlight: blue for a monoculture of species 1, red for 20% of species 2 and 80% of species 3, orange for an equi-proportional 3-species mixture and wine for 10% of species 1 and 3 and 80% of species 2.*

## SI2.6  Analysing `sim0` using a one-way ANOVA model

Here we fit the one-way ANOVA model with community type (1-16) as the factor:

$$y = \mu + \alpha_i + \epsilon$$

for $i = 1, ..., 16$ where $\epsilon \sim \text{N}(0, \sigma^2)$.

Predictions from the one-way ANOVA model are shown in Figure SI2.7. For other approaches, predictions are shown for the 0.1:0.8:0.1 community, however, this is not possible for the one-way ANOVA model since this community was not in the design.

```
# One-way ANOVA community factor
sim0$communityF <- as.factor(sim0$community)
m3 <- lm(response ~ communityF, data = sim0)
preds <- predict(m3)
m2means <- preds[1:16]

m2_lsmeans <- lsmeans(m3, "communityF")
preds <- predict(m3)
m2means <- preds[1:16]

contrasts1 = list(Sp1vSp2 = c(1, -1, 0, 0, 0, 0, 0, 0, 0, 0, 0, 0, 0, 0, 0, 0),
                  Sp1vSp3 = c(1, 0, -1, 0, 0, 0, 0, 0, 0, 0, 0, 0, 0, 0, 0, 0),
                  Comm4vComm5 = c(0, 0, 0, 1, -1, 0, 0, 0, 0, 0, 0, 0, 0, 0, 0, 0),
                  Comm5vComm6 = c(0, 0, 0, 0, 1, -1, 0, 0, 0, 0, 0, 0, 0, 0, 0, 0),
                  Comm7vComm8 = c(0, 0, 0, 0, 0, 0, 1, -1, 0, 0, 0, 0, 0, 0, 0, 0),
                  Comm9vComm10 = c(0, 0, 0, 0, 0, 0, 0, 0, 0, 0, 0, 0, 0, 0, 1, -1))

contrast(m2_lsmeans, contrasts1)
#>  contrast      estimate   SE df t.ratio p.value
#>  Sp1vSp2          5.716 1.37 48   4.176  0.0001
#>  Sp1vSp3          9.345 1.37 48   6.827  <.0001
#>  Comm4vComm5      4.306 1.37 48   3.146  0.0028
#>  Comm5vComm6     -0.373 1.37 48  -0.273  0.7863
#>  Comm7vComm8     -5.584 1.37 48  -4.079  0.0002
#>  Comm9vComm10    -4.552 1.37 48  -3.325  0.0017
xlabs <- c("1:0:0", "0:1:0", "0:0:1",
           "0.8:0.2:0", "0.2:0.8:0", "0.8:0:0.2", "0.2:0:0.8",
           "0:0.8:0.2", "0:0.2:0.8",
           "50:50:0", "50:0:50", "0:50:50",
           "60:20:20", "20:60:20", "20:20:60", "33:33:33")

bar_data <- data.frame(y = m2means, x = factor(xlabs, levels = xlabs))
bar_data$colour <- sim0$colour[1:16]
bar_data$richness <- sim0$richness[1:16]

## Barplot with communities of interest highlighted
bar_plot <- ggplot(data = bar_data)+
    # Add bars to show predicitons from ANOVA for communities
```

```r
    geom_col(aes(x = x, y = y),
             fill = '#a6a6a6', width = 0.85, position = position_dodge(0.5))+
    # Highlight bars corresponding to communities of interest
    geom_point(data = bar_data[c(1,9,16),],
               aes(x = x, y = y),
               colour = c('blue', '#ff1a1a','#fa6538'),
               size = 4, shape = 4, stroke = 3)+
    # Edit axis labels
    labs(y = 'Prediction', x = "Community type")+
    # Choose the theme for plot
    theme_bw()+
    # Create separate panel for each level of richness
    facet_grid(~richness, scales = 'free_x', space = 'free_x',
               switch = 'x',
               labeller = as_labeller(c(`1` = 'Richness = 1', `2` = 'Richness = 2', `3` =
                 'Richness = 3')))+
    # Add some whitespace at end of facet
    scale_x_discrete(expand = expansion(add = 0.75))+
    # Modify specific elements of theme
    theme(axis.text = element_text(size=12),
          axis.title = element_text(size = 14),
          axis.text.x = element_text(angle = 60, vjust = 1.2, hjust = 1.2),
          # removes space between panels
          panel.spacing = unit(0, units = "cm"),
          strip.placement = "outside", # moves the states down
          strip.background = element_rect(colour = NA, fill = NA),
          strip.text = element_text(face = 'plain', size = 12, colour = 'black'),
          panel.border = element_blank())

# Add black line along panel labels to visually separate the panels
# Convert plot to grid objet to add lines
anova_plot <- ggplotGrob(bar_plot)
lines <- linesGrob(x=unit(c(0.05,.95),"npc"),
                   y=unit(c(0.9,.9),"npc"),
                   gp=gpar(col="black", lwd=3))

for (k in grep("strip-b",anova_plot$layout$name)) {
  anova_plot$grobs[[k]]$grobs[[1]]$children[[1]] <- lines
}

anova_plot$widths[5] <- anova_plot$widths[5]*1.25
# Convert back to ggplot object
anova_plot <- ggplotify::as.ggplot(anova_plot)
# Plot the plot
anova_plot
```

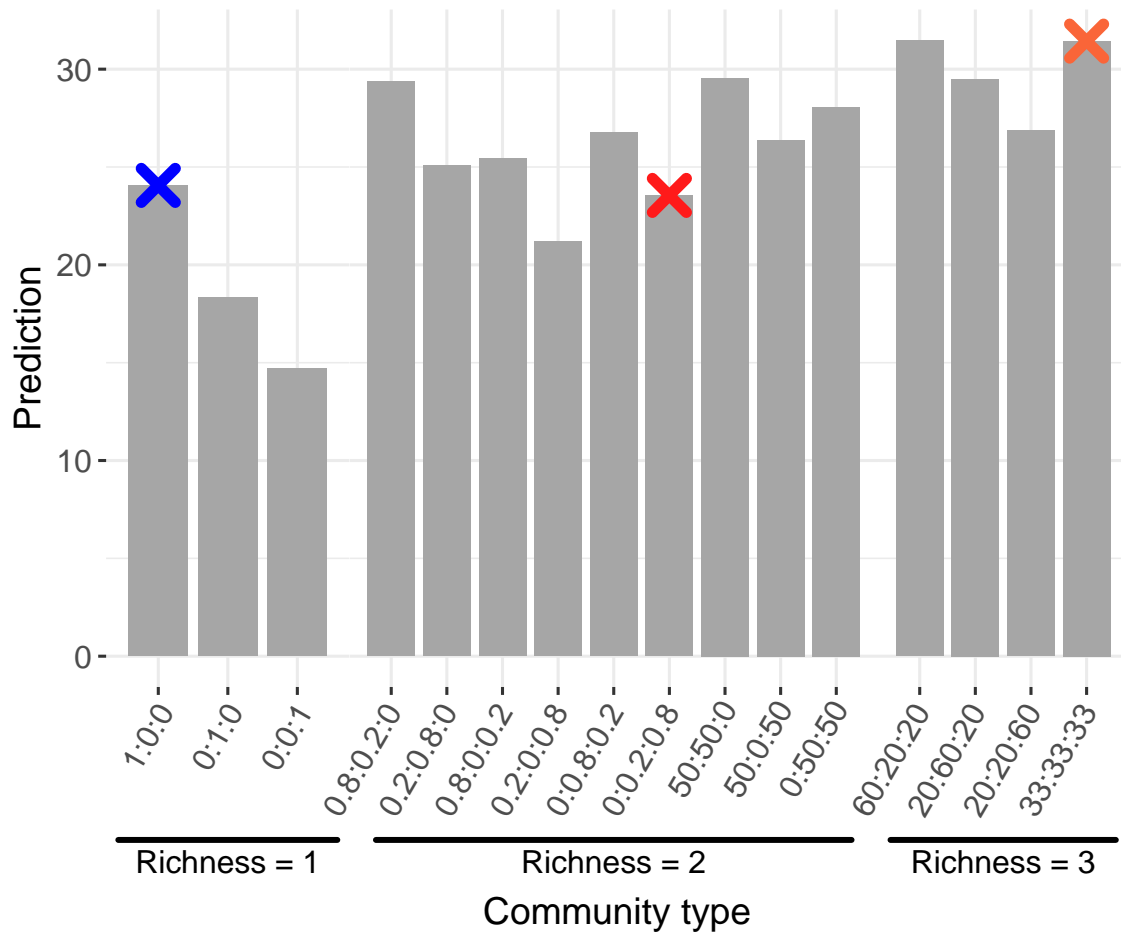

**Figure SI2.7**: *The predicted response for each community type from a 1-way ANOVA model. The coloured X's highlight select predictions: blue for a monoculture of species 1, red for 20% of species 2 and 80% of species 3, and orange for an equi-proportional 3-species mixture.*

The ANOVA table for the one-way ANOVA model is shown in Table SI2.6.

```
# ANOVA table
oneway_model_anova <- anova(m3)
knitr::kable(oneway_model_anova, digits = 2, caption = "**Table SI2.6**: *ANOVA table for
↪   the 1-way ANOVA model.*")
```

Table 6: **Table SI2.6**: *ANOVA table for the 1-way ANOVA model.*

|  | Df | Sum Sq | Mean Sq | F value | Pr(>F) |
|---|---|---|---|---|---|
| communityF | 15 | 1282.33 | 85.49 | 22.81 | 0 |
| Residuals | 48 | 179.87 | 3.75 |  |  |

## SI2.7 Comparison of the three modelling approaches

The mean squared error (MSE; estimated $\sigma^2$) was 11.2 for the richness model, 3.54 for the DI model and 3.75 for the 1-way ANOVA model. The variation due to species composition and species proportions are incorporated in the MSE in the richness model, reflected by its higher value compared to that for the ANOVA and DI models, whose two MSE values are somewhat similar.

Predicted values across the three models for a range of community types are shown in Table SI2.7.

```r
#Generate new data frame with select communities for comparison across the three models
new_comms <- data.frame(richness = c(1,   2,   3,   3),
                                 p1 = c(1,   0,   1/3, 0.1),
                                 p2 = c(0,   0.8, 1/3, 0.8),
                                 p3 = c(0,   0.2, 1/3, 0.1),
                          communityF = c("1","8","16",NA))

comm_names <- c("(1, 0 ,0)", "(0, 0.8, 0.2)", "(1/3, 1/3, 1/3)", "(0.1, 0.8, 0.1)")
richness_new_preds <- predict(m1, newdata = new_comms)
DI_new_preds <- predict(m2, newdata = new_comms)
oneway_new_preds <- predict(m3, newdata = new_comms)
all_new_preds <- data.frame(comm_names, richness_new_preds, DI_new_preds,
↪  oneway_new_preds)

# Produce a table of predictions
kable(all_new_preds,
      col.names = c("Community type", "Richness model", "DI model", "One-way ANOVA
↪  model"),
      row.names = FALSE, digits = 2, caption = "**Table SI2.7**: *A comparison of
↪  predictions for select communities from the richness model, the full pairwise
↪  interactions DI model, and the one-way ANOVA model. Community type gives the
↪  proportions for species 1, 2 and 3.*")
```

Table 7: **Table SI2.7**: *A comparison of predictions for select communities from the richness model, the full pairwise interactions DI model, and the one-way ANOVA model. Community type gives the proportions for species 1, 2 and 3.*

| Community type | Richness model | DI model | One-way ANOVA model |
|---|---:|---:|---:|
| (1, 0 ,0) | 20.16 | 24.17 | 24.06 |
| (0, 0.8, 0.2) | 25.41 | 25.52 | 26.77 |
| (1/3, 1/3, 1/3) | 30.66 | 31.30 | 31.43 |
| (0.1, 0.8, 0.1) | 30.66 | 25.67 | |

## SI2.8 Using `autoDI()` for the `sim0` dataset

In many cases, it will be helpful to start analysing a dataset with the DI models approach with the `autoDI()` function. Here is code and output from doing this.

Some points to note in the `autoDI()` output:

- In step 1, it is seen that there is no evidence that $\theta$ differs from 1 (p = 0.8162).

- In step 2, the intercept only model (since there are no structures specified), average pairwise model and full pairwise models are compared. Each of these improves significantly on the last and thus the full pairwise model is selected as the best. The additive species model is equivalent to the full pairwise model in the case of only three species, while functional groups were not specified here.

- In step 3, no treatment is tested as none was specified in the `treat` argument.

- In step 4, the selected DI model is compared to the 'reference' model, which is the one-way ANOVA model, i.e., a factor is specified for each unique community (set of proportions). There is no evidence of lack of fit (p = 0.7356).

```
# Using autoDI for the sim0 dataset
autoDI(prop = c("p1", "p2", "p3"), y = "response", data = sim0)
#> 
#> --------------------------------------------------------------------------------
#> Step 1: Investigating whether theta is equal to 1 or not for the AV model, including
↪   all available structures
#> 
#> Theta estimate: 0.9768
#> Selection using F tests
#>           Description
#> DI Model 1 Average interactions 'AV' DImodel
#> DI Model 2 Average interactions 'AV' DImodel, estimating theta
#> 
#>           DI_model treat estimate_theta Resid. Df Resid. SSq Resid. MSq Df
#> DI Model 1     AV  none          FALSE        60    281.4708     4.6912
#> DI Model 2     AV  none           TRUE        59    281.2109     4.7663  1
#>              SSq      F Pr(>F)
#> DI Model 1
#> DI Model 2 0.2599 0.0545 0.8162
#> 
#> The test concludes that theta is not significantly different from 1.
#> 
#> --------------------------------------------------------------------------------
#> Step 2: Investigating the interactions
#> Selection using F tests
#>           Description
#> DI Model 1 Structural 'STR' DImodel
#> DI Model 2 Species identity 'ID' DImodel
#> DI Model 3 Average interactions 'AV' DImodel
#> DI Model 4 Separate pairwise interactions 'FULL' DImodel
#> 
#>           DI_model treat estimate_theta Resid. Df Resid. SSq Resid. MSq Df
#> DI Model 1    STR  none          FALSE        63   1462.1967    23.2095
#> DI Model 2     ID  none          FALSE        61   1165.7315    19.1104  2
#> DI Model 3     AV  none          FALSE        60    281.4708     4.6912  1
#> DI Model 4   FULL  none          FALSE        58    205.4254     3.5418  2
#>              SSq      F Pr(>F)
#> DI Model 1
#> DI Model 2 296.4652 41.8521 <0.0001
#> DI Model 3 884.2608 249.663 <0.0001
#> DI Model 4  76.0454 10.7354   1e-04
#> 
#> Functional groups (argument 'FG') were not specified, and therefore not investigated.
#> 
#> The 'ADD' variables are only computed for > 3 species cases as the 'ADD' model is not
↪   informative for the 2 or 3 species case.
#> 
#> Selected model: Separate pairwise interactions 'FULL' DImodel
#> 
#> --------------------------------------------------------------------------------
#> Step 3: No investigation of treatment effect included, since no treatment was
↪   specified
#>         (argument 'treat' omitted)
```

```
#> 
#> ----------------------------------------------------------------------------
#> Step 4: Comparing the final selected model with the reference (community) model
#> 'community' is a factor with 16 levels, one for each unique set of proportions.
#> 
#>                model Resid. Df Resid. SSq Resid. MSq Df    SSq      F Pr(>F)
#> DI Model 1  Selected        58   205.4254     3.5418
#> DI Model 2 Reference        48   179.8704     3.7473 10 25.555  0.682 0.7356
#> 
#> ----------------------------------------------------------------------------
#> autoDI is limited in terms of model selection. Exercise caution when choosing your
↪  final model.
#> ----------------------------------------------------------------------------
#> 
#> Call:  glm(formula = fmla, family = family, data = data)
#> 
#> Coefficients:
#>     p1       p2       p3   `p1:p2`  `p1:p3`  `p2:p3`
#>  24.17    18.62    14.81    34.83    26.14    47.92
#> 
#> Degrees of Freedom: 64 Total (i.e. Null);  58 Residual
#> Null Deviance:      43860
#> Residual Deviance: 205.4     AIC: 270.3
```

## SI2.9 Comparing community performances using a DI model

It may be helpful to compare mean performances of different communities from the DI model estimates. This can be done using the `contrasts_DI` function, which takes a fitted DI model object and a list of contrasts, then calculates the contrasts and p-values associated with testing the hypothesis of whether the contrast is equal to zero.

Below we present examples comparing identity effects between themselves, then identity effects to species mixtures.

```
## Contrasts for difference between identity effects of p1, p2, and p3
con1 <- contrasts_DI(object = m2,
                     contrast = list('p1vp2 ID' = c(1, -1,  0,   0, 0, 0),
                                     'p2vp3 ID' = c(0,  1, -1,   0, 0, 0),
                                     'p1vp3 ID' = c(1,  0, -1,   0, 0, 0)))
#> Generated contrast matrix:
#>          p1 p2 p3 `p1:p2` `p1:p3` `p2:p3`
#> p1vp2 ID  1 -1  0       0       0       0
#> p2vp3 ID  0  1 -1       0       0       0
#> p1vp3 ID  1  0 -1       0       0       0
summary(con1)
#> 
#>   Simultaneous Tests for General Linear Hypotheses
#> 
#> Fit: glm(formula = fmla, family = family, data = data)
#> 
#> Linear Hypotheses:
#>              Estimate Std. Error z value Pr(>|z|)
#> p1vp2 ID == 0   5.559      1.048   5.305  < 1e-04 ***
```

```
#> p2vp3 ID == 0    3.809      1.048   3.635 0.000799 ***
#> p1vp3 ID == 0    9.367      1.048   8.941  < 1e-04 ***
#> ---
#> Signif. codes:  0 '***' 0.001 '**' 0.01 '*' 0.05 '.' 0.1 ' ' 1
#> (Adjusted p values reported -- single-step method)

## Constrast to compare ID effect of p1 to the equi-proportional three-species community
p1_ID <- c(1, 0, 0,    0, 0, 0)                    # ID of species 1
equi_mix <- c(1/3, 1/3, 1/3,    1/9, 1/9, 1/9)     # Three species equi-prop mix
p1_ID_vs_equi_mix <- p1_ID - equi_mix
print(p1_ID_vs_equi_mix)                           # Difference of two comms
#> [1]  0.6666667 -0.3333333 -0.3333333 -0.1111111 -0.1111111 -0.1111111

# Get contrast
con2 <- contrasts_DI(object = m2,
                     contrast = list('p1_ID v equi_mix' = p1_ID_vs_equi_mix))
#> Generated contrast matrix:
#>                        p1         p2         p3     `p1:p2`    `p1:p3`
#> p1_ID v equi_mix 0.6666667 -0.3333333 -0.3333333 -0.1111111 -0.1111111
#>                     `p2:p3`
#> p1_ID v equi_mix -0.1111111
summary(con2)
#>
#>    Simultaneous Tests for General Linear Hypotheses
#>
#> Fit: glm(formula = fmla, family = family, data = data)
#>
#> Linear Hypotheses:
#>                       Estimate Std. Error z value Pr(>|z|)
#> p1_ID v equi_mix == 0  -7.1232     0.9758    -7.3 2.88e-13 ***
#> ---
#> Signif. codes:  0 '***' 0.001 '**' 0.01 '*' 0.05 '.' 0.1 ' ' 1
#> (Adjusted p values reported -- single-step method)

## Constrast to compare average of ID effect of p1, p2, and p3 to the equi-proportional
↪   three-species community
avg_ID <- c(1/3, 1/3, 1/3,    0, 0, 0)             # Average ID effect of three species
equi_mix <- c(1/3, 1/3, 1/3,    1/9, 1/9, 1/9)     # Three species equi-prop mix
avg_ID_vs_equi_mix <- avg_ID - equi_mix
print(avg_ID_vs_equi_mix)                          # Difference of two comms
#> [1]  0.0000000  0.0000000  0.0000000 -0.1111111 -0.1111111 -0.1111111

# Get contrast
con3 <- contrasts_DI(object = m2,
                     contrast = list('AVG_ID v equi_mix' = avg_ID_vs_equi_mix))
#> Generated contrast matrix:
#>                   p1 p2 p3    `p1:p2`    `p1:p3`    `p2:p3`
#> AVG_ID v equi_mix  0  0  0 -0.1111111 -0.1111111 -0.1111111
summary(con3)
#>
#>    Simultaneous Tests for General Linear Hypotheses
#>
```

```
#> Fit: glm(formula = fmla, family = family, data = data)
#> 
#> Linear Hypotheses:
#>                        Estimate Std. Error z value Pr(>|z|)
#> AVG_ID v equi_mix == 0 -12.0984     0.7657   -15.8   <2e-16 ***
#> ---
#> Signif. codes:  0 '***' 0.001 '**' 0.01 '*' 0.05 '.' 0.1 ' ' 1
#> (Adjusted p values reported -- single-step method)
```

# Supporting Information 3
# Vignette: The analysis from the `Bell` Case Study.

This dataset is from a 72 species bacterial BEF experiment. With this number of species, there are 2556 possible pairwise interactions! Having so many species can provide challenges for the Diversity-Interactions (DI) modelling framework, however, the approach can still work very well for species rich datasets, as we will show with this analysis.

## SI3.1 The `Bell` dataset

This dataset comes from a bacterial biodiversity experiment (Bell et al., 2005, Nature, 436, 1157-1160). The bacterial ecosystems used were from semi-permanent rainpools that form in bark-lined depressions near the base of large European beech trees (Fagus sylvatica). Microcosms consisting of sterile beech leaf disks and 10 ml of liquid (phosphate buffer) were inoculated with random combinations of 72 bacterial species isolated from these ecosystems. A total of 1,374 microcosms were constructed at richness levels of 1, 2, 3, 4, 6, 8, 9, 12, 18, 24, 36 and 72 species. The daily respiration rate of the bacterial community in each microcosm was measured over three time intervals (days 0-7, 7-14 and 14-28) and the average over the three time intervals was recorded.

The `DImodels` package is installed from CRAN and loaded in the typical way.

```
#install.packages("DImodels")
library(DImodels)
```

Loading other packages necessary for this vignette

```
library(tidyverse)
#> -- Attaching core tidyverse packages ------------------------ tidyverse 2.0.0 --
#> v dplyr     1.1.1     v readr     2.1.4
#> v forcats   1.0.0     v stringr   1.5.0
#> v ggplot2   3.4.2     v tibble    3.2.1
#> v lubridate 1.9.2     v tidyr     1.3.0
#> v purrr     1.0.1
#> -- Conflicts ------------------------------------------ tidyverse_conflicts() --
#> x dplyr::filter() masks stats::filter()
#> x dplyr::lag()    masks stats::lag()
#> i Use the conflicted package (<http://conflicted.r-lib.org/>) to force all conflicts
#>   to become errors
```

The `Bell` dataset is available as a pre-loaded dataset in the `DImodels` package. To load it, use the following code.

```
data("Bell")
```

We can generate scatter plots of the response versus richness (Figure SI3.1) and versus log(richness) (Figure SI3.2; as presented in the Bell et al 2005 paper).

```
library(DImodels)
#>
```

```
#> Attaching package: 'DImodels'
#> The following object is masked from 'package:tidyr':
#>
#>     extract
data("Bell")
ggplot(data = Bell, aes(x = richness, y = response))+
  # Create hollow points
  geom_point(size = 2)+
  # Add labels
  labs(x = "Richness",
       y = "Average daily respiration rate")+
  # Add theme
  theme_bw()+
  # Adjust size of axis text
  theme(axis.title = element_text(face = 'plain', size = 14),
        axis.text = element_text(face = 'plain', size = 12))
```

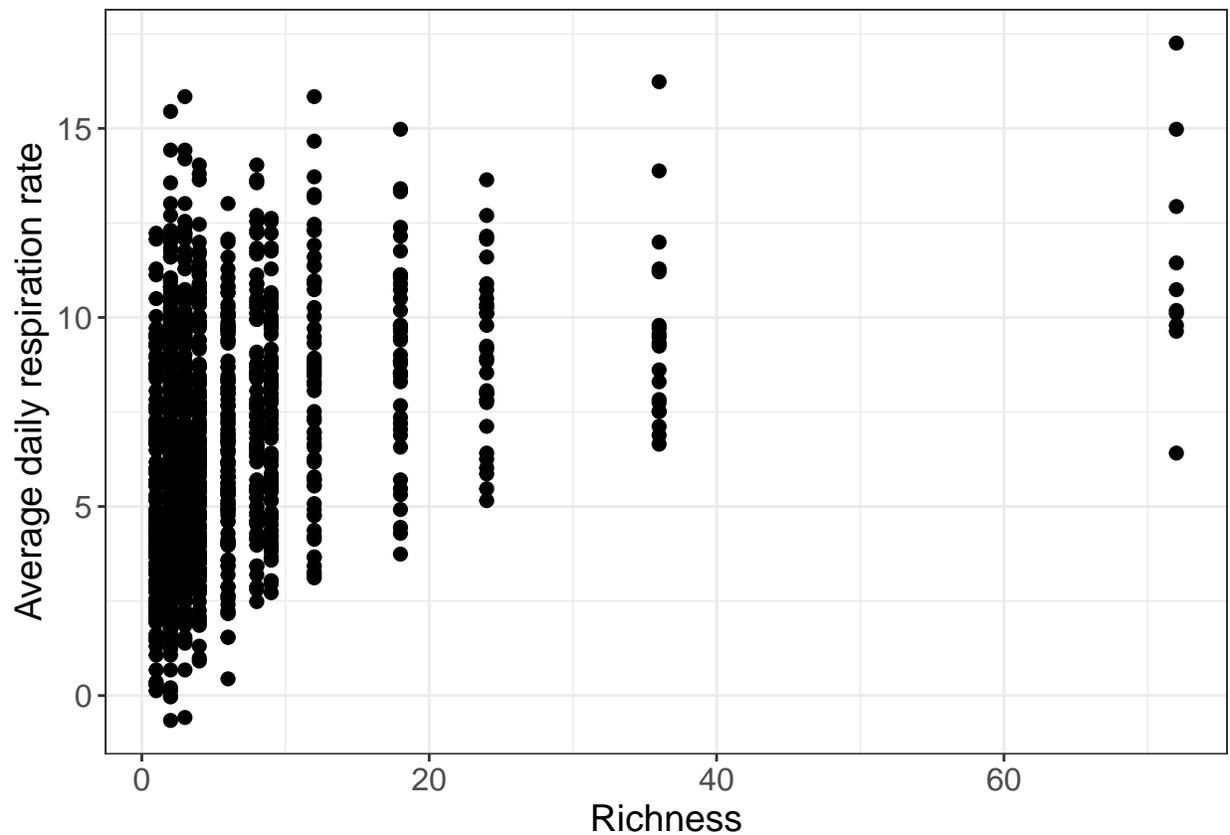

**Figure SI3.1**: *Scatter plot of the average daily respiration rate response versus richness.*

```
ggplot(data = Bell, aes(x = log(richness), y = response))+
  # Create hollow points
  geom_point(size = 2)+
  # Add labels
  labs(x = "Log Richness",
       y = "Average daily respiration rate")+
  # Add theme
  theme_bw()+
```

```r
# Adjust size of axis text
theme(axis.title = element_text(face = 'plain', size = 14),
      axis.text = element_text(face = 'plain', size = 12))
```

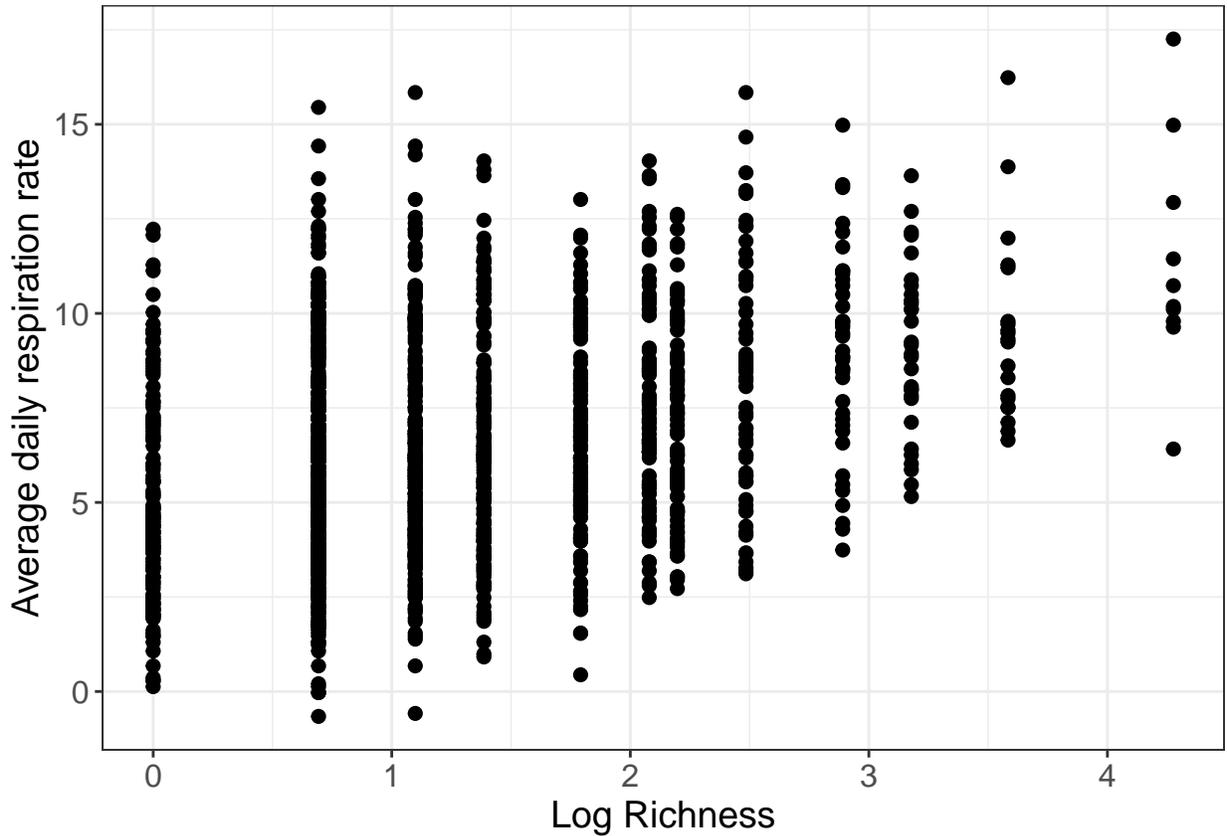

**Figure SI3.2**: *Scatter plot of the average daily respiration rate response versus log(richness).*

The dataset comprises a variable to uniquely identify each row of the dataset (id), a variable to identify each unique set of proportions (community), the number of species included in each community (richness), the proportion of each species (p1 to p72) and the community level average daily respiration rate variable (response).

Here are the first five rows (with many species proportion columns omitted):

```r
head(Bell[,c(1:13,73:76)])
#>   id community richness p1 p2 p3 p4 p5 p6 p7 p8 p9 p10 p70 p71 p72 response
#> 1  1         1        1  1  0  0  0  0  0  0  0  0   0   0   0   0 8.769750
#> 2  2         2        1  0  1  0  0  0  0  0  0  0   0   0   0   0 5.626893
#> 3  3         3        1  0  0  1  0  0  0  0  0  0   0   0   0   0 9.712607
#> 4  4         4        1  0  0  0  1  0  0  0  0  0   0   0   0   0 3.426893
#> 5  5         5        1  0  0  0  0  1  0  0  0  0   0   0   0   0 3.819750
#> 6  6         6        1  0  0  0  0  0  1  0  0  0   0   0   0   0 4.448321
```

## SI3.2  Comparing the richness model to a subset of simplified Diversity-Interactions models

For the Bell data, it is clear from SI3.1 that as richness increases, that the response is also increasing. However, is it a linear relationship with richness or would a different shape be better to model this BEF relationship (e.g. linear with log(richness)? Additionally, are there species-specific effects contributing to the BEF relationship?

We can use the `richness_vs_DI()` function for an initial evaluation of this.

```
# Compare the richness model to a small subset of DI models
richness_vs_DI(y = "response", prop = 4:75, data = Bell)
#>
#> --------------------------------------------------------------------------------
#>
#> Investigating richness model and three DI alternatives
#>     AIC      AICc     BIC      df
#> 1 6789.954 6789.971 6805.630  2
#> 2 6789.954 6789.971 6805.630  2
#> 3 6749.273 6749.302 6770.175  3
#> 4 6798.307 6806.852 7190.218 73
#> 5 6751.120 6759.903 7143.031 74
#>   Description
#> 1 Model 1: Richness only
#> 2 Model 2: Average interactions 'AV' DImodel with common identity effects and theta = 0.5
#> 3 Model 3: Average interactions 'AV' DImodel with common identity effects and theta estimated
#> 4 Model 4: Average interactions 'AV' DImodel with unique identity effects and theta = 0.5
#> 5 Model 5: Average interactions 'AV' DImodel with unique identity effects and theta estimated
#>
#> --------------------------------------------------------------------------------
#> Models 1 and 2 are equivalent if all mixtures in the dataset at any given level of richness are equi-proportional.
#> richness_vs_DI is limited in terms of model selection. Only five models are explored. See ?autoDI and ?DI for more options.
#> --------------------------------------------------------------------------------
```

Models 1 and 2 in this output are equivalent, because the Bell dataset contains monocultures and only equi-proportional mixtures (See Supporting Information 1 for further details on this). Allowing $\theta$ to be estimated gives a lower AIC than the richness model (or $\theta = 0.5$ DI model), the AIC is 6749.31, compared to 6789.95. Adding in species specific effects does give a higher AIC (6798.31). Having both unique identity effects and $\theta$ estimated gives AIC 6751.12, slightly higher than the third model (AIC 6749.27).

It is clear that further exploration of this BEF relationship using the DI modelling framework is warranted.

## SI3.3  Overview of applying DI models to the `Bell` dataset

It is not possible to fit the full pairwise interactions model to the `Bell` dataset, because there are more parameters to estimate $(72 + 2556)$ than data points $(1374)$. We will fit three DI models: the richness model, the average pairwise model with $\theta = 1$, and the average pairwise model with $\theta$ estimated. These models are fitted in the following sections.

## SI3.4 Fitting the richness model

We first fitted the richness model, equivalent to an average pairwise interactions DI model with all identity effects set equal and average interactions with $\theta$ set to 0.5.

```
m1 <- lm(response ~ richness, data = Bell)
summary(m1)
#>
#> Call:
#> lm(formula = response ~ richness, data = Bell)
#>
#> Residuals:
#>     Min      1Q  Median      3Q     Max
#> -7.6572 -2.2149 -0.3521  1.9448  9.5871
#>
#> Coefficients:
#>              Estimate Std. Error t value Pr(>|t|)
#> (Intercept)  5.914224   0.094831   62.37   <2e-16 ***
#> richness     0.113271   0.009401   12.05   <2e-16 ***
#> ---
#> Signif. codes:  0 '***' 0.001 '**' 0.01 '*' 0.05 '.' 0.1 ' ' 1
#>
#> Residual standard error: 2.859 on 1372 degrees of freedom
#> Multiple R-squared:  0.0957, Adjusted R-squared:  0.09504
#> F-statistic: 145.2 on 1 and 1372 DF,  p-value: < 2.2e-16
AIC(m1)
#> [1] 6789.954
```

## SI3.5 Fitting the average pairwise model

Secondly, we fitted the average pairwise interaction model with $\theta = 1$.

```
m2 <- DI(y = "response", prop = 4:75, DImodel = "AV", data = Bell)
#> Fitted model: Average interactions 'AV' DImodel
summary(m2)
#>
#> Call:
#> glm(formula = fmla, family = family, data = data)
#>
#> Deviance Residuals:
#>     Min       1Q   Median       3Q      Max
#> -6.4704  -2.1243  -0.3316   1.9577   9.5318
#>
#> Coefficients:
#>     Estimate Std. Error t value Pr(>|t|)
#> p1    5.7105     1.1216   5.091 4.08e-07 ***
#> p2    5.5747     1.1250   4.955 8.18e-07 ***
#> p3    3.4926     1.1193   3.120 0.001847 **
#> p4    2.3149     1.1140   2.078 0.037915 *
#> p5    3.2285     1.1153   2.895 0.003857 **
#> p6    5.5131     1.1137   4.950 8.39e-07 ***
#> p7    3.4661     1.1186   3.099 0.001985 **
#> p8    1.6133     1.1184   1.442 0.149408
#> p9    3.4184     1.1113   3.076 0.002142 **
```

```
#> p10    7.1129    1.1204    6.348 3.00e-10 ***
#> p11    3.5904    1.1152    3.219 0.001316 **
#> p12    4.9979    1.1131    4.490 7.75e-06 ***
#> p13    2.5541    1.1133    2.294 0.021936 *
#> p14    3.5794    1.1252    3.181 0.001502 **
#> p15    5.4052    1.1314    4.778 1.98e-06 ***
#> p16    4.4374    1.1178    3.970 7.59e-05 ***
#> p17    6.1111    1.1196    5.458 5.74e-08 ***
#> p18    3.2028    1.1126    2.879 0.004058 **
#> p19    3.3937    1.1136    3.047 0.002354 **
#> p20    7.8467    1.1211    6.999 4.12e-12 ***
#> p21    4.1623    1.1127    3.741 0.000192 ***
#> p22    6.4496    1.1148    5.785 9.05e-09 ***
#> p23    4.3865    1.1266    3.894 0.000104 ***
#> p24    4.7592    1.1109    4.284 1.97e-05 ***
#> p25    2.9538    1.1143    2.651 0.008125 **
#> p26    7.3707    1.1163    6.602 5.87e-11 ***
#> p27    4.6156    1.1143    4.142 3.66e-05 ***
#> p28    6.0793    1.1148    5.453 5.91e-08 ***
#> p29    4.1503    1.1174    3.714 0.000212 ***
#> p30    6.6802    1.1227    5.950 3.44e-09 ***
#> p31    5.0855    1.1163    4.555 5.72e-06 ***
#> p32    3.3478    1.1219    2.984 0.002899 **
#> p33    5.4769    1.1176    4.900 1.08e-06 ***
#> p34    3.1326    1.1184    2.801 0.005168 **
#> p35    5.4153    1.1236    4.819 1.61e-06 ***
#> p36    7.2738    1.1226    6.479 1.30e-10 ***
#> p37    3.5038    1.1286    3.104 0.001947 **
#> p38    5.6498    1.1185    5.051 5.02e-07 ***
#> p39    4.8475    1.1139    4.352 1.46e-05 ***
#> p40    4.3296    1.1168    3.877 0.000111 ***
#> p41    4.3681    1.1156    3.916 9.49e-05 ***
#> p42    4.1374    1.1155    3.709 0.000217 ***
#> p43    3.7047    1.1130    3.329 0.000898 ***
#> p44    2.3610    1.1168    2.114 0.034697 *
#> p45    2.8705    1.1156    2.573 0.010193 *
#> p46    4.0866    1.1237    3.637 0.000287 ***
#> p47    5.4158    1.1210    4.831 1.52e-06 ***
#> p48    2.9909    1.1192    2.672 0.007627 **
#> p49    4.5798    1.1132    4.114 4.13e-05 ***
#> p50    5.2774    1.1160    4.729 2.50e-06 ***
#> p51    2.8937    1.1248    2.573 0.010202 *
#> p52    5.1024    1.1307    4.513 6.98e-06 ***
#> p53    5.2599    1.1135    4.724 2.56e-06 ***
#> p54    4.0736    1.1178    3.644 0.000279 ***
#> p55    4.6093    1.1150    4.134 3.80e-05 ***
#> p56    1.5151    1.1161    1.358 0.174849
#> p57    4.9993    1.1238    4.448 9.39e-06 ***
#> p58    4.7925    1.1145    4.300 1.84e-05 ***
#> p59    3.3852    1.1172    3.030 0.002492 **
#> p60    7.9228    1.1201    7.073 2.46e-12 ***
#> p61    7.0961    1.1152    6.363 2.73e-10 ***
#> p62    4.0589    1.1224    3.616 0.000310 ***
```

```
#> p63     6.9558       1.1149    6.239 5.94e-10 ***
#> p64     3.2582       1.1304    2.882 0.004012 **
#> p65     3.4536       1.1148    3.098 0.001990 **
#> p66     5.5263       1.1170    4.947 8.51e-07 ***
#> p67     2.5118       1.1214    2.240 0.025266 *
#> p68     6.0230       1.1202    5.376 8.99e-08 ***
#> p69     4.6244       1.1307    4.090 4.58e-05 ***
#> p70     3.1420       1.1181    2.810 0.005028 **
#> p71     2.8731       1.1148    2.577 0.010071 *
#> p72     2.9895       1.1190    2.672 0.007640 **
#> AV      6.5637       0.5697   11.521  < 2e-16 ***
#> ---
#> Signif. codes:  0 '***' 0.001 '**' 0.01 '*' 0.05 '.' 0.1 ' ' 1
#>
#> (Dispersion parameter for gaussian family taken to be 7.914453)
#>
#>     Null deviance: 71873  on 1374  degrees of freedom
#> Residual deviance: 10297  on 1301  degrees of freedom
#> AIC: 6814.6
#>
#> Number of Fisher Scoring iterations: 2
AIC(m2)
#> [1] 6814.613
```

## SI3.6 Fitting the average pairwise (AV) model, with theta estimated

Thirdly, we fitted the average pairwise model with $\theta$ estimated.

```
m3 <- DI(y = "response", prop = 4:75, DImodel = "AV", data = Bell,
         estimate_theta = TRUE)
#> Fitted model: Average interactions 'AV' DImodel
#> Theta estimate: 0.7917
summary(m3)
#>
#> Call:
#> glm(formula = formula(obj), family = family, data = data_theta_E_AV)
#>
#> Deviance Residuals:
#>     Min        1Q    Median        3Q       Max
#> -6.1987   -1.9835   -0.3337    1.8725    8.6604
#>
#> Coefficients:
#>     Estimate Std. Error t value Pr(>|t|)
#> p1    6.3646     1.0859   5.861 5.82e-09 ***
#> p2    6.2287     1.0892   5.718 1.33e-08 ***
#> p3    4.1466     1.0836   3.827 0.000136 ***
#> p4    2.9689     1.0784   2.753 0.005988 **
#> p5    3.8825     1.0796   3.596 0.000335 ***
#> p6    6.1671     1.0781   5.720 1.32e-08 ***
#> p7    4.1201     1.0829   3.805 0.000149 ***
#> p8    2.2673     1.0827   2.094 0.036447 *
```

```
#> p9     4.0724     1.0758   3.786 0.000160 ***
#> p10    7.7669     1.0847   7.160 1.34e-12 ***
#> p11    4.2444     1.0796   3.932 8.88e-05 ***
#> p12    5.6520     1.0775   5.245 1.82e-07 ***
#> p13    3.2082     1.0777   2.977 0.002966 **
#> p14    4.2334     1.0894   3.886 0.000107 ***
#> p15    6.0592     1.0955   5.531 3.84e-08 ***
#> p16    5.0915     1.0822   4.705 2.81e-06 ***
#> p17    6.7652     1.0839   6.242 5.85e-10 ***
#> p18    3.8568     1.0770   3.581 0.000355 ***
#> p19    4.0477     1.0780   3.755 0.000181 ***
#> p20    8.5007     1.0854   7.832 9.93e-15 ***
#> p21    4.8163     1.0771   4.472 8.44e-06 ***
#> p22    7.1037     1.0792   6.583 6.69e-11 ***
#> p23    5.0405     1.0908   4.621 4.20e-06 ***
#> p24    5.4132     1.0753   5.034 5.47e-07 ***
#> p25    3.6078     1.0786   3.345 0.000847 ***
#> p26    8.0247     1.0807   7.426 2.02e-13 ***
#> p27    5.2696     1.0787   4.885 1.16e-06 ***
#> p28    6.7333     1.0791   6.240 5.92e-10 ***
#> p29    4.8043     1.0817   4.441 9.70e-06 ***
#> p30    7.3342     1.0869   6.748 2.26e-11 ***
#> p31    5.7395     1.0807   5.311 1.28e-07 ***
#> p32    4.0018     1.0862   3.684 0.000239 ***
#> p33    6.1309     1.0819   5.667 1.79e-08 ***
#> p34    3.7866     1.0827   3.498 0.000485 ***
#> p35    6.0693     1.0879   5.579 2.94e-08 ***
#> p36    7.9278     1.0869   7.294 5.20e-13 ***
#> p37    4.1578     1.0928   3.805 0.000149 ***
#> p38    6.3038     1.0828   5.822 7.33e-09 ***
#> p39    5.5015     1.0783   5.102 3.86e-07 ***
#> p40    4.9836     1.0811   4.610 4.43e-06 ***
#> p41    5.0221     1.0799   4.650 3.65e-06 ***
#> p42    4.7914     1.0799   4.437 9.89e-06 ***
#> p43    4.3587     1.0774   4.046 5.53e-05 ***
#> p44    3.0150     1.0811   2.789 0.005368 **
#> p45    3.5245     1.0800   3.263 0.001129 **
#> p46    4.7406     1.0880   4.357 1.42e-05 ***
#> p47    6.0698     1.0853   5.593 2.72e-08 ***
#> p48    3.6449     1.0835   3.364 0.000791 ***
#> p49    5.2338     1.0776   4.857 1.34e-06 ***
#> p50    5.9314     1.0803   5.490 4.82e-08 ***
#> p51    3.5477     1.0890   3.258 0.001152 **
#> p52    5.7565     1.0948   5.258 1.70e-07 ***
#> p53    5.9139     1.0778   5.487 4.92e-08 ***
#> p54    4.7276     1.0821   4.369 1.35e-05 ***
#> p55    5.2633     1.0794   4.876 1.21e-06 ***
#> p56    2.1691     1.0804   2.008 0.044887 *
#> p57    5.6533     1.0881   5.196 2.36e-07 ***
#> p58    5.4465     1.0789   5.048 5.09e-07 ***
#> p59    4.0392     1.0815   3.735 0.000196 ***
#> p60    8.5768     1.0844   7.909 5.49e-15 ***
#> p61    7.7501     1.0796   7.179 1.18e-12 ***
```

```
#> p62      4.7129      1.0866    4.337 1.55e-05 ***
#> p63      7.6098      1.0792    7.051 2.88e-12 ***
#> p64      3.9122      1.0945    3.574 0.000364 ***
#> p65      4.1076      1.0792    3.806 0.000148 ***
#> p66      6.1803      1.0814    5.715 1.36e-08 ***
#> p67      3.1658      1.0857    2.916 0.003606 **
#> p68      6.6770      1.0845    6.157 9.89e-10 ***
#> p69      5.2784      1.0949    4.821 1.60e-06 ***
#> p70      3.7960      1.0825    3.507 0.000469 ***
#> p71      3.5271      1.0792    3.268 0.001110 **
#> p72      3.6436      1.0833    3.364 0.000792 ***
#> AV       2.1432      0.1506   14.233  < 2e-16 ***
#> ---
#> Signif. codes:  0 '***' 0.001 '**' 0.01 '*' 0.05 '.' 0.1 ' ' 1
#>
#> (Dispersion parameter for gaussian family taken to be 7.551858)
#>
#>     Null deviance: 71872.9  on 1374  degrees of freedom
#> Residual deviance:  9817.4  on 1300  degrees of freedom
#> AIC: 6751.1
#>
#> Number of Fisher Scoring iterations: 2
AIC(m3)
#> [1] 6751.12
```

## SI3.7 Visualising and comparing the three fitted DI models

We predicted from each of the three models and graphed the fitted models with the raw data (Figure SI3.3).

```r
# Single scatter plot with each predicted line included
## Group model objects in list
models <- list('Richness' = m1, 'AV (theta = 1)' = m2, 'AV (theta estimated)' = m3)

# Get predictions from each of the three models
predictions <- sapply(models, predict, simplify = T, USE.NAMES = T)

# Prepare data for plotting and include predictions from each of the models
plot_data <- bind_cols('response' = Bell$response, 'richness' = Bell$richness,
 ↪   predictions) %>%
              pivot_longer(cols = colnames(predictions), names_to = 'Model', values_to
               ↪ = 'Prediction') %>%
              mutate('Model' = fct_inorder(Model))

# Grouped data to find mean prediction at each level of richness for the three different
 ↪   models
group_data <- plot_data %>% group_by(richness, Model) %>% summarise('Prediction' =
 ↪   mean(Prediction), .groups = 'drop')

# Create plot
ggplot(data = plot_data, aes(x = richness, y = response))+
  # Raw data
  geom_point(size = 2)+
```

```
# Mean predictions at each level of richness
geom_line(data = group_data, aes(x = richness, y = Prediction, colour = Model), size =
    1.2)+
geom_point(data = group_data, aes(x = richness, y = Prediction, colour = Model), size =
    3)+
# Theme for plot
theme_bw()+
# Axis labels
labs(x = 'Richness', y = "Average daily respiration rate")+
# Change colours of the lines and points
scale_colour_manual(values =c('tomato', 'gold', 'steelblue'))+
# Change axis and legend text size
theme(axis.title = element_text(face = 'plain', size = 14),
      axis.text = element_text(face = 'plain', size = 12),
      legend.title = element_text(face = 'plain', size = 14),
      legend.text = element_text(face = 'plain', size = 12),
      legend.position = 'top')
```

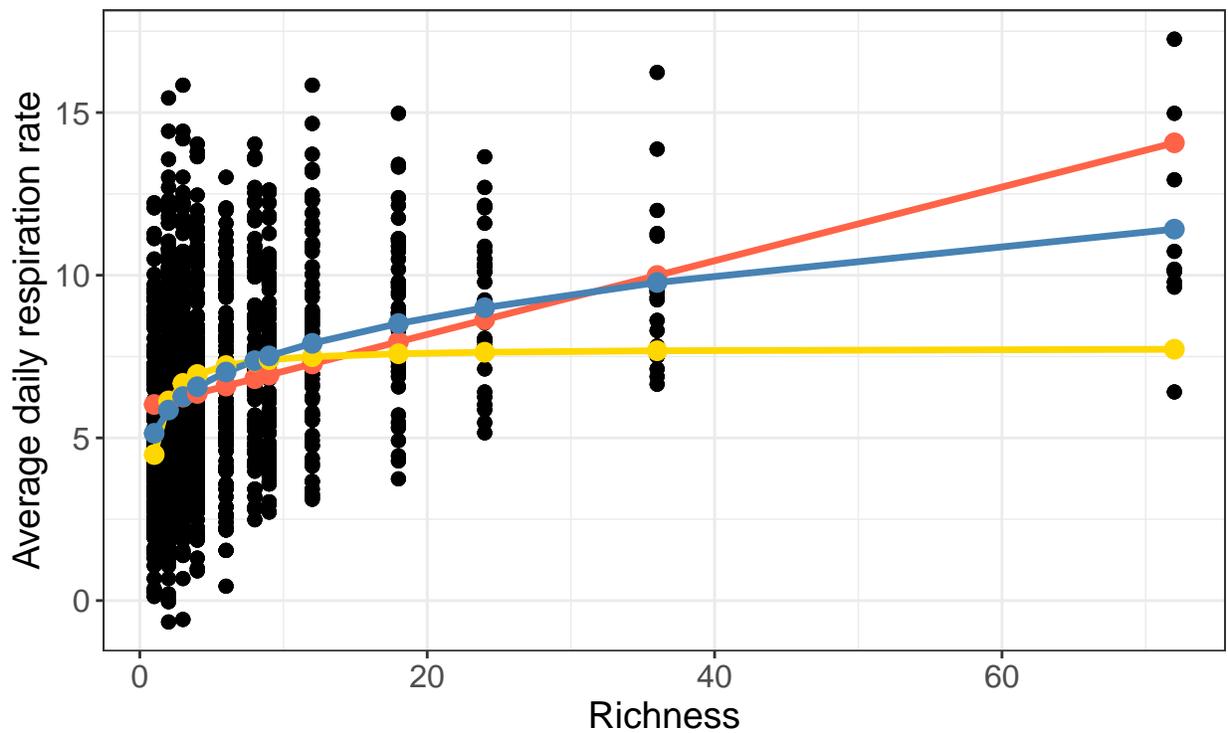

**Figure SI3.3**: *Scatter plot of the raw average daily respiration rate values versus richness, with the three fitted DI model regression lines; the red line is the richness model, the yellow model is the average pairwise model with theta set to 1, and the blue line is the average pairwise model with theta estimated (estimate = 0.79).*

The AIC values for the three models are richness model = 6789.95, average pairwise model (with $\theta = 1$) = 6814.61, and average pairwise model (with theta estimated) = 6751.12; the AIC value for the third model is considerably lower and according to AIC is the best of these three models.

## SI3.8 Predicting for specific communities

When predicting the response for specific communities, we can use existing design points within the dataset, as below where we selected two rows from the Bell dataset.

We predicted from one row containing the species proportions for a monoculture of species p5 and another row with an equal mixture of 18 species (p9, p10, p15, p16, p28, p29, p30, p33, p35, p39, p42, p47, p52, p61, p62, p63, p67, p72)

```r
predict(m3, newdata = Bell[c(5, 433),])
#> [1] 3.882534 9.099395
```

We can also predict from communities that aren't present in the original dataset, as in the example below where we predicted the response using a new 18 species mixture community with species p1 present at a proportion of 1/2 and species p2 through p18 are equally represented at 1/34.

```r
newCommunity <- data.frame(rbind(c(1/2, rep(1/34, 17), rep(0, 54))))
colnames(newCommunity) <- colnames(Bell)[4:75]

predict(m3, newdata = newCommunity)
#> [1] 7.942423
```

```r
# Create new data with communities of interest
communities <- data.frame(rbind(c(1/2, rep(1/34, 17), rep(0, 54)),
                                c(rep(0, 54), rep(1/34, 17), 1/2),
                                c(rep(0, 4), 1, rep(0, 69))))
colnames(communities) <- colnames(Bell)[4:75]
communities$id <-  c('1', '2', '3')
communities$richness <-  c(18, 18, 1)
communities$prediction <- predict(m3, newdata = communities)

# Create plot
ggplot(data = Bell, aes(x = richness, y = response)) +
  # plot raw responses
  geom_point(size = 1.75) +
  # Add predictions in specific communities of interest
  geom_point(data = communities,
             aes(x = richness, y = prediction, colour = id),
             size = 4)+
  # Theme for graph
  theme_bw()+
  # Axis labels
  labs(x = 'Richness', y = "Average daily respiration rate")+
  # Change colours of the lines and points
  scale_colour_manual(values =c("#cb66ff", "#ffcb66", "#669aff"))+
  # Change axis and legend text size
  theme(axis.title = element_text(face = 'plain', size = 14),
        axis.text = element_text(face = 'plain', size = 12),
        legend.position = 'none')
```

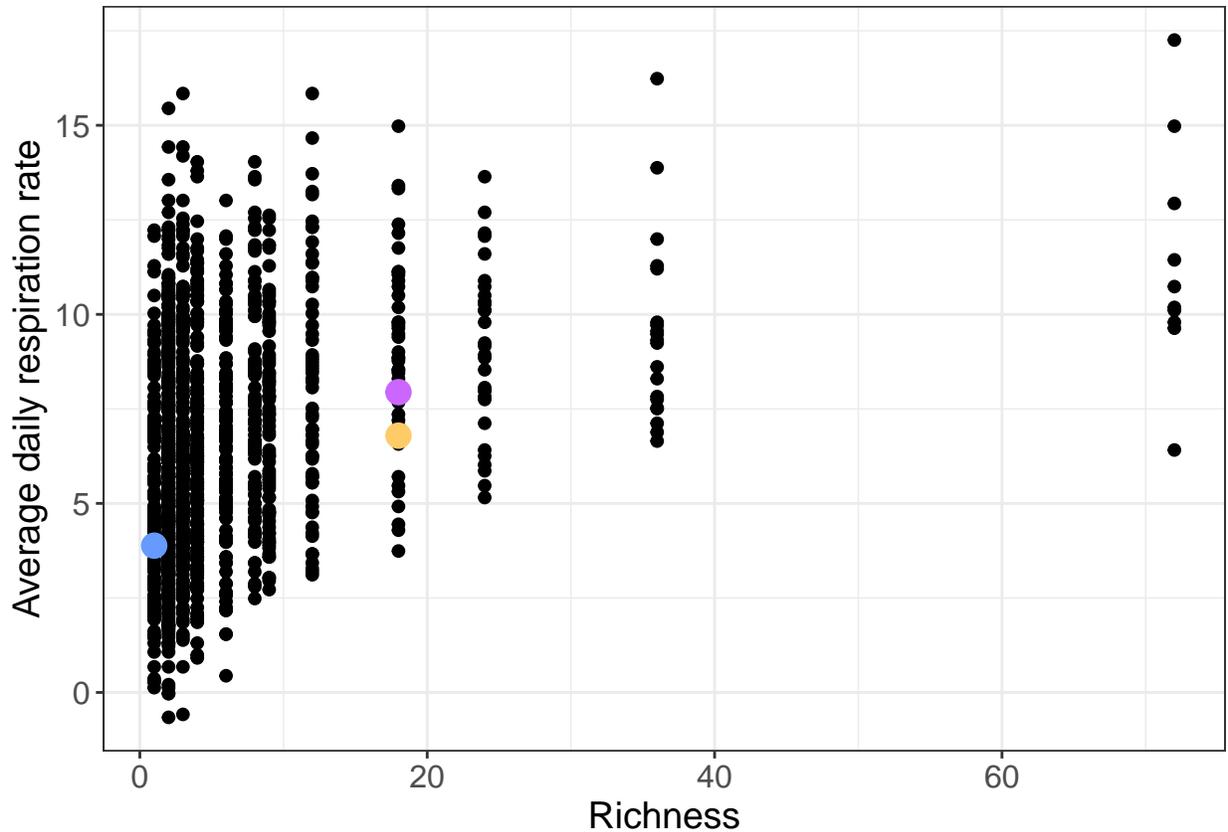

**Figure SI3.4**: *Scatter plot of the raw average daily respiration rate values versus richness, with predictions for three select communities, predicted from the selected DI model (the average pairwise model with theta estimated).*

## SI3.9    Using `autoDI` for the Bell dataset

When analysing the Bell dataset using `autoDI`, we get the same model as the selected model above, i.e., the average pairwise model with $\theta$ estimated.

Some points to note in the `autoDI()` output:

- In step 1, $\theta$ differs from 1 (p < 0.001) and is estimated as 0.7917.

- In step 2, the intercept only model (STR; but since there are no structures specified), identity effects (ID) model, average pairwise (AV) model, additive species (ADD) model and full (FULL) pairwise models are compared. The $\theta$ parameter is estimated for each of the models that include interactions. The selected model at this stage is the average pairwise model, with $\theta$ estimated. While the FULL model is included, it is not possible to fit the full pairwise interactions model and so this version has set many interaction terms to 0 automatically.

- In step 3, no treatment is tested as none was specified in the `treat` argument.

- In step 4, the selected DI model is compared to the 'reference' model, which is the one-way ANOVA model, i.e., a factor is specified for each unique community (set of proportions). There is no evidence of lack of fit (p = 0.4795).

```
autoDI(y = "response", prop = 4:75, data = Bell)
#> 
#> ---------------------------------------------------------------------------------
```

```
#> Step 1: Investigating whether theta is equal to 1 or not for the AV model, including
↪   all available structures
#>
#> Theta estimate: 0.7917
#> Selection using F tests
#>           Description
#> DI Model 1 Average interactions 'AV' DImodel
#> DI Model 2 Average interactions 'AV' DImodel, estimating theta
#>
#>           DI_model treat estimate_theta Resid. Df Resid. SSq Resid. MSq Df
#> DI Model 1       AV  none          FALSE      1301  10296.704     7.9145
#> DI Model 2       AV  none           TRUE      1300   9817.415     7.5519  1
#>              SSq      F   Pr(>F)
#> DI Model 1
#> DI Model 2 479.2882 63.4663 <0.0001
#>
#> The test concludes that theta is significantly different from 1.
#>
#> -----------------------------------------------------------------------------------
#> Step 2: Investigating the interactions
#> Selection using F tests
#>           Description
↪
#> DI Model 1 Structural 'STR' DImodel
↪
#> DI Model 2 Species identity 'ID' DImodel
↪
#> DI Model 3 Average interactions 'AV' DImodel, estimating theta
↪
#> DI Model 4 Additive species contributions to interactions 'ADD' DImodel, estimating
↪   theta
#> DI Model 5 Separate pairwise interactions 'FULL' DImodel, estimating theta
#>
#>           DI_model treat estimate_theta Resid. Df Resid. SSq Resid. MSq  Df
#> DI Model 1      STR  none          FALSE      1373  12401.456     9.0324
#> DI Model 2       ID  none          FALSE      1302  11347.193     8.7152  71
#> DI Model 3       AV  none           TRUE      1300   9817.415     7.5519   2
#> DI Model 4      ADD  none           TRUE      1229   9341.491     7.6009  71
#> DI Model 5     FULL  none           TRUE       694   5238.987     7.5490 535
#>              SSq       F   Pr(>F)
#> DI Model 1
#> DI Model 2 1054.263    1.967 <0.0001
#> DI Model 3 1529.7776 101.3236 <0.0001
#> DI Model 4  475.9244    0.888  0.7307
#> DI Model 5 4102.5037    1.0158  0.4223
#>
#> Functional groups (argument 'FG') were not specified, and therefore not investigated.
#>
#> Selected model: Average interactions 'AV' DImodel, estimating theta
#>
#> -----------------------------------------------------------------------------------
#> Step 3: No investigation of treatment effect included, since no treatment was
↪   specified
```

```
#>           (argument 'treat' omitted)
#> 
#> ----------------------------------------------------------------------------------
#> Step 4: Comparing the final selected model with the reference (community) model
#> 'community' is a factor with 679 levels, one for each unique set of proportions.
#> 
#>                 model Resid. Df Resid. SSq Resid. MSq  Df      SSq      F
#> DI Model 1  Selected       1300   9817.415     7.5519
#> DI Model 2 Reference        695   5238.987     7.5381 605 4578.4282 1.0039
#>            Pr(>F)
#> DI Model 1
#> DI Model 2 0.4795
#> 
#> ----------------------------------------------------------------------------------
#> autoDI is limited in terms of model selection. Exercise caution when choosing your
#> ↪  final model.
#> ----------------------------------------------------------------------------------
#> 
#> Call:  glm(formula = formula(obj), family = family, data = data_theta_E_AV)
#> 
#> Coefficients:
#>     p1      p2      p3      p4      p5      p6      p7      p8      p9     p10
#> 6.3646  6.2287  4.1466  2.9689  3.8825  6.1671  4.1201  2.2673  4.0724  7.7669
#>    p11     p12     p13     p14     p15     p16     p17     p18     p19     p20
#> 4.2444  5.6520  3.2082  4.2334  6.0592  5.0915  6.7652  3.8568  4.0477  8.5007
#>    p21     p22     p23     p24     p25     p26     p27     p28     p29     p30
#> 4.8163  7.1037  5.0405  5.4132  3.6078  8.0247  5.2696  6.7333  4.8043  7.3342
#>    p31     p32     p33     p34     p35     p36     p37     p38     p39     p40
#> 5.7395  4.0018  6.1309  3.7866  6.0693  7.9278  4.1578  6.3038  5.5015  4.9836
#>    p41     p42     p43     p44     p45     p46     p47     p48     p49     p50
#> 5.0221  4.7914  4.3587  3.0150  3.5245  4.7406  6.0698  3.6449  5.2338  5.9314
#>    p51     p52     p53     p54     p55     p56     p57     p58     p59     p60
#> 3.5477  5.7565  5.9139  4.7276  5.2633  2.1691  5.6533  5.4465  4.0392  8.5768
#>    p61     p62     p63     p64     p65     p66     p67     p68     p69     p70
#> 7.7501  4.7129  7.6098  3.9122  4.1076  6.1803  3.1658  6.6770  5.2784  3.7960
#>    p71     p72      AV   theta
#> 3.5271  3.6436  2.1432  0.7917
#> 
#> Degrees of Freedom: 1374 Total (i.e. Null);  1300 Residual
#> Null Deviance:        71870
#> Residual Deviance: 9817     AIC: 6751
```

# Supporting Information 4
# Vignette: Analysis of the `Switzerland` dataset.

This vignette provides an additional analysis to those provided in the Case Studies section in the main text. The `Switzerland` dataset is available in the `DImodels` package.

## SI4.1   The `Switzerland` dataset

The `DImodels` package is installed from CRAN and loaded in the typical way.

```
#install.packages("DImodels")
library(DImodels)
```

The `Switzerland` dataset may be loaded directly from the `DImodels` package by executing:

```
data(Switzerland)
```

This case study comes from a grassland biodiversity experiment that was conducted in Switzerland as part of the "Agrodiversity Experiment", detailed in Kirwan et al. (2014), henceforth referred to as the "Switzerland dataset". A total of 68 grassland plots were established across a gradient of species diversity, and two additional treatments (nitrogen fertiliser and total seed density) were also manipulated. The proportions of four species (two grasses and two legumes) were varied across the plots: there were plots with 100% of a single species, and 2- and 4-species mixtures with varying proportions (e.g., (0.5, 0.5, 0, 0) and (0.7, 0.1, 0.1, 0.1)). Nitrogen fertiliser was either 50 or 100 kg N per annum and total seed density was either low or high. The total annual yield per plot was recorded for the first year after experimental set up. See Kirwan et al. (2014) for more details. Here we present a brief analysis using `DImodels`.

We begin by producing an exploratory plot of richness versus yield, for each combination between density and nitrogen levels:

```
library(tidyverse)
#> -- Attaching core tidyverse packages ------------------------ tidyverse 2.0.0 --
#> v dplyr     1.1.1     v readr     2.1.4
#> v forcats   1.0.0     v stringr   1.5.0
#> v ggplot2   3.4.2     v tibble    3.2.1
#> v lubridate 1.9.2     v tidyr     1.3.0
#> v purrr     1.0.1
#> -- Conflicts ------------------------------------------ tidyverse_conflicts() --
#> x tidyr::extract() masks DImodels::extract()
#> x dplyr::filter()  masks stats::filter()
#> x dplyr::lag()     masks stats::lag()
#> i Use the conflicted package (<http://conflicted.r-lib.org/>) to force all conflicts
↪   to become errors

Switzerland %>%
  mutate(richness = apply(Switzerland[,4:7], 1, function(x) sum(x > 0))) %>%
  ggplot(aes(x = richness, y = yield, col = nitrogen)) +
    # Add theme
    theme_bw() +
```

```
# Create separate facets for the two seeding densities
facet_wrap(~ density, nrow = 2,
           labeller = as_labeller(c('low' = 'Seeding density: low',
                                    'high' = 'Seeding density: high'))) +
# Jitter points
geom_jitter(width = .05, alpha = .8, size = 2.5) +
# Add line showing mean yield for each richness level and nitrogen level
stat_summary(geom = "line", fun = "mean", size = 1.25) +
# Adjust labels of axes and legend
labs(x = 'Richness', y = 'Annual Yield (tons/ha)',
     colour = 'Nitrogen (kg/yr)')+
# Adjust size of axis text
theme(axis.title = element_text(face = 'plain', size = 14),
      axis.text = element_text(face = 'plain', size = 12),
      legend.title = element_text(face = 'plain', size = 14),
      legend.text = element_text(face = 'plain', size = 12),
      strip.text = element_text(face = 'plain', size= 12))
```

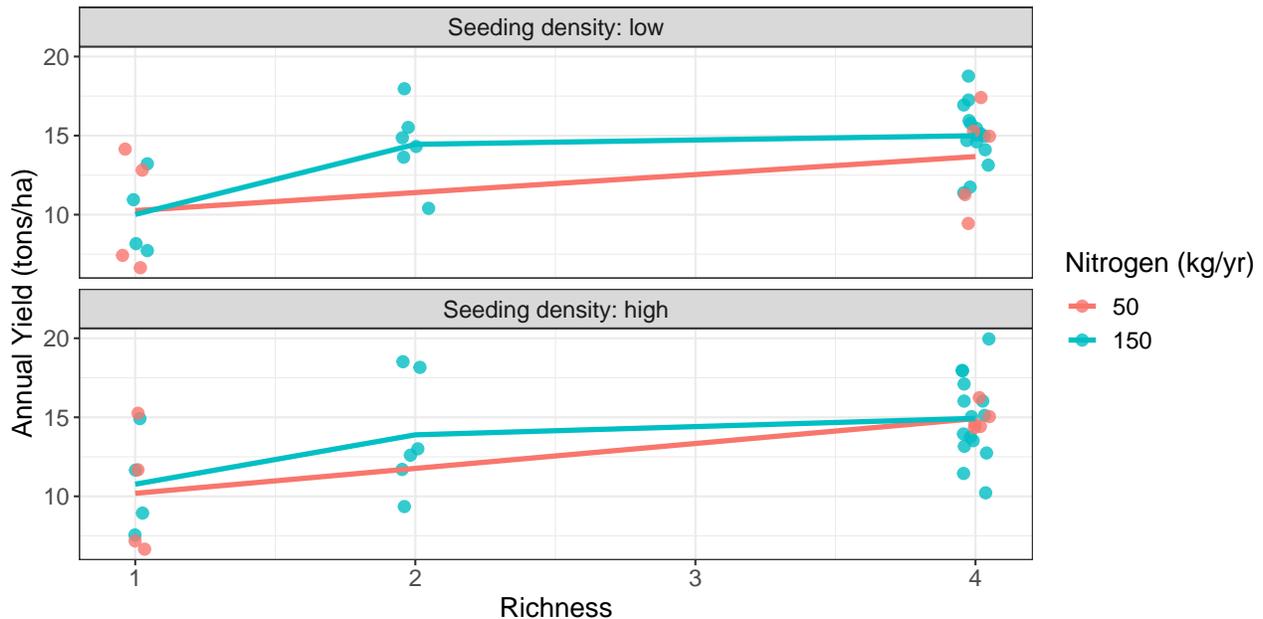

**Figure S3.1**: *Yield versus richness for low and high nitrogen management.*

There does not seem to be an interaction between density and nitrogen level, and it appears that nitrogen is not the main driver of yield increase, but richness does play an important role. By fitting DI models, we are able to decompose the effect of richness by looking at the *Identities* and different types of *Interactions* effects in the linear predictor. In the next section, we perform automatic model selection through `autoDI()`.

## SI4.2  Automatic model selection using `autoDI`

We begin by performing automatic model selection using the `autoDI` function. We need to specify the names of the variables representing the response variable, species proportions, treatment and density variables. Since we have two functional groups represented in this dataset (grasses and legumes), we also inform which species belong to each, through the argument `FG`. We use the default selection method using F-tests and save the final selected model as `auto_fit`. Here we are declaring that the first two species proportions variables refer to grasses and the last two to legumes, through the `FG` argument.

```r
auto_fit <- autoDI(y = "yield", density = "density", treat = "nitrogen",
                   prop = c("p1","p2","p3","p4"), FG =
                   ↪ c("Grass","Grass","Legume","Legume"),
                   data = Switzerland)
#>
#> --------------------------------------------------------------------------------
#> Step 1: Investigating whether theta is equal to 1 or not for the AV model, including
#> ↪  all available structures
#>
#> Theta estimate: 0.7603
#> Selection using F tests
#>            Description
#> DI Model 1 Average interactions 'AV' DImodel with treatment
#> DI Model 2 Average interactions 'AV' DImodel with treatment, estimating theta
#>
#>            DI_model     treat estimate_theta Resid. Df Resid. SSq Resid. MSq
#> DI Model 1       AV 'nitrogen'          FALSE        61   174.2937     2.8573
#> DI Model 2       AV 'nitrogen'           TRUE        60   164.8882     2.7481
#>            Df    SSq      F Pr(>F)
#> DI Model 1
#> DI Model 2  1 9.4055 3.4225 0.0692
#>
#> The test concludes that theta is not significantly different from 1.
#>
#> --------------------------------------------------------------------------------
#> Step 2: Investigating the interactions
#> All models include density
#> Since 'Ftest' was specified as selection criterion and functional groups were
#> ↪  specified, dropping the ADD model as it is not nested within the FG model.
#> Selection using F tests
#>            Description
#> DI Model 1 Structural 'STR' DImodel with treatment
#> DI Model 2 Species identity 'ID' DImodel with treatment
#> DI Model 3 Average interactions 'AV' DImodel with treatment
#> DI Model 4 Functional group effects 'FG' DImodel with treatment
#> DI Model 5 Separate pairwise interactions 'FULL' DImodel with treatment
#>
#>            DI_model     treat estimate_theta Resid. Df Resid. SSq Resid. MSq
#> DI Model 1      STR 'nitrogen'          FALSE        65   647.9784     9.9689
#> DI Model 2       ID 'nitrogen'          FALSE        62   374.8340     6.0457
#> DI Model 3       AV 'nitrogen'          FALSE        61   174.2937     2.8573
#> DI Model 4       FG 'nitrogen'          FALSE        59   136.9301     2.3208
#> DI Model 5     FULL 'nitrogen'          FALSE        56   124.1971     2.2178
#>            Df      SSq       F  Pr(>F)
#> DI Model 1
#> DI Model 2  3 273.1444 41.0533 <0.0001
#> DI Model 3  1 200.5404 90.4229 <0.0001
#> DI Model 4  2  37.3636  8.4235   6e-04
#> DI Model 5  3  12.7331  1.9138  0.1378
#>
#> Selected model: Functional group effects 'FG' DImodel with treatment
#>
```

```
#> --------------------------------------------------------------------------------
#> Step 3: Investigating the treatment effect
#> Selection using F tests
#>            Description
#> DI Model 1 Functional group effects 'FG' DImodel
#> DI Model 2 Functional group effects 'FG' DImodel with treatment
#>
#>            DI_model    treat estimate_theta Resid. Df Resid. SSq Resid. MSq
#> DI Model 1       FG     none          FALSE        60   143.8881     2.3981
#> DI Model 2       FG 'nitrogen'        FALSE        59   136.9301     2.3208
#>            Df    SSq     F Pr(>F)
#> DI Model 1
#> DI Model 2  1 6.9579 2.998 0.0886
#>
#> Selected model: Functional group effects 'FG' DImodel
#>
#> --------------------------------------------------------------------------------
#> Step 4: Comparing the final selected model with the reference (community) model
#> 'community' is a factor with 25 levels, one for each unique set of proportions.
#>
#>                 model Resid. Df Resid. SSq Resid. MSq Df     SSq      F Pr(>F)
#> DI Model 1   Selected        60   143.8881     2.3981
#> DI Model 2  Reference        42    84.6499     2.0155 18 59.2382 1.6329 0.0954
#>
#> --------------------------------------------------------------------------------
#> autoDI is limited in terms of model selection. Exercise caution when choosing your
#>   final model.
#> --------------------------------------------------------------------------------
```

The output from the `autoDI()` function is separated in four steps. In Step 1, it investigates whether $\theta$ is significantly different from 1 using the `AV` model. Then, in Step 2, it investigates the diversity effect by fitting models with different *Interactions* structures (see Table 1 in Moral et al.). By default, it performs F-tests to assess the significance of the effects in the linear predictor in a forward selection approach (Structural (`STR`), then Species Identity (`ID`), then Average interactions (`AV`), then Functional group effects (`FG`), then Separate pairwise interactions (`FULL`)).

From Step 1, we concluded that $\theta$ is not significantly different from 1 ($F_{1,60} = 3.42, p = 0.07$). Therefore, in Step 2, `autoDI()` investigated the Interactions component while assuming $\theta = 1$, fixed. From Step 2, the functional group effects model (`FG`) was selected (comparing the average interactions model `AV` with the `FG` model yielded $F_{1,61} = 90.42, p < 0.01$, while comparing the `FG` model with the separate pairwise interactions model `FULL` yielded $F_{3,56} = 1.91, p = 0.14$). Next, in Step 3, the additive treatment effect (nitrogen dose) was investigated, and we concluded that it was not significant ($F_{1,59} = 6.96, p = 0.09$). Finally, in Step 4, `autoDI()` performs a test for lack-of-fit, by comparing the `FG` model to the reference (community) model. Since the test was not significant ($F_{18,42} = 1.63, p = 0.10$), we may conclude that the functional group effects are suitable to explain the variability generated by the species interactions.

## SI4.3   Exploring fitted model

We may explore the coefficients of the final selected model by executing:

```
summary(auto_fit)
#>
#> Call:
#> glm(formula = new_fmla, family = family, data = new_data)
```

```
#> 
#> Deviance Residuals:
#>     Min       1Q   Median       3Q      Max
#> -3.3626  -0.9923  -0.1629   0.9537   3.9269
#> 
#> Coefficients:
#>                    Estimate Std. Error t value Pr(>|t|)
#> p1                   8.0540     0.6357  12.670  < 2e-16 ***
#> p2                   8.1107     0.6357  12.759  < 2e-16 ***
#> p3                  15.5788     0.6357  24.508  < 2e-16 ***
#> p4                  12.0943     0.6357  19.026  < 2e-16 ***
#> FG_bfg_Grass_Legume 18.6205     1.8923   9.840 3.92e-14 ***
#> FG_wfg_Grass        10.0412     4.1004   2.449   0.0173 *
#> FG_wfg_Legume        1.1087     4.1004   0.270   0.7878
#> densitylow          -0.1378     0.3756  -0.367   0.7151
#> ---
#> Signif. codes:  0 '***' 0.001 '**' 0.01 '*' 0.05 '.' 0.1 ' ' 1
#> 
#> (Dispersion parameter for gaussian family taken to be 2.398135)
#> 
#>     Null deviance: 13310.90  on 68  degrees of freedom
#> Residual deviance:   143.89  on 60  degrees of freedom
#> AIC: 261.94
#> 
#> Number of Fisher Scoring iterations: 2
```

The graph below shows species identity effects and net interactions effects for the high density treatment:

```r
## get model coefficients
swiModel <- DI(y = "yield", prop = 4:7, FG = c("G", "G", "L", "L"), DImodel = "FG", data
  = Switzerland)
#> Warning in DI_data_prepare(y = y, block = block, density = density, prop = prop, : One
#>  or more rows have species proportions that sum to approximately 1, but not exactly 1.
#>  This is typically a rounding issue, and has been corrected internally prior to
#>  analysis.
#> Fitted model: Functional group effects 'FG' DImodel
betas <- swiModel$coefficients

## Create desired dataset to predict from
swipred <- Switzerland[c(12, 13, 14, 15), 4:7] #Monos
swipred[5,] <- c(0, 0, 0.5, 0.5)
swipred[6,] <- c(0.5, 0, 0.5, 0)
swipred[7,] <- c(0.5, 0.5, 0, 0)
swipred[8,] <- c(0.25, 0.25, 0.25, 0.25)

## Add Richness level for grouping
swipred <- cbind(swipred, c("Richness = 1", "Richness = 1", "Richness = 1", "Richness =
  1", "Richness = 2", "Richness = 2", "Richness = 2", "Richness = 4"))

## Get predictions
swipred$response <- predict(swiModel, newdata = swipred)
```

```r
## Divide up predicted responses according to contributing species (reverse order for
↪    stacked bar order)
conts <- data.frame(1:8)
conts$p4Cont <- betas[4]*swipred$p4
conts$p3Cont <- betas[3]*swipred$p3
conts$p2Cont <- betas[2]*swipred$p2
conts$p1Cont <- betas[1]*swipred$p1

intCont <- swipred$response - conts$p1Cont - conts$p2Cont - conts$p3Cont - conts$p4Cont

swipred <- cbind(swipred, intCont, conts[, 2:5])

## Name communities/columns for easy labelling
communities <- c("1:0:0:0", "0:1:0:0", "0:0:1:0", "0:0:0:1", "0:0:0.50:0.50",
↪    "0.50:0:0.50:0", "0.50:0.50:0:0", "0.25:0.25:0.25:0.25")
swipred <- cbind(communities, swipred)
colnames(swipred) <- c("Community", "Sp1Sown", "Sp2Sown", "Sp3Sown", "Sp4Sown",
↪    "richness", "response", "Net interactions", "Sp4", "Sp3", "Sp2", "Sp1")

## Reformat into long data for stacking
library(reshape2)
#>
#> Attaching package: 'reshape2'
#> The following object is masked from 'package:tidyr':
#>
#>     smiths
swipred <- reshape2::melt(swipred,
                          id.vars = c("Community", "Sp1Sown", "Sp2Sown", "Sp3Sown",
↪                             "Sp4Sown", "response", "richness"),
                          variable.name = "Species",
                          value.name = "Predicted response")

swipred$`Community` <- factor(swipred$`Community`, levels = c("1:0:0:0", "0:1:0:0",
↪    "0:0:1:0", "0:0:0:1", "0:0:0.50:0.50", "0.50:0:0.50:0", "0.50:0.50:0:0",
↪    "0.25:0.25:0.25:0.25"))

## Create stacked bar chart
library(ggplot2)

ggplot(swipred, aes(fill = Species, y = `Predicted response`, x = `Community`)) +
  # Add bars showing the predicted response
  geom_bar(position = "stack", stat = "identity", width = 0.75)  +
  # Create a separate facet for each level of richness
  facet_grid(~richness, scales = "free_x", space = "free_x") +
  # Add white space at end of facet
  expand_limits(x = 1.1) +
  # Theme of plot
  theme_bw()+
```

```r
# Adjust specific elements of the theme
theme(axis.text = element_text(size=12),
      axis.title = element_text(size=14),
      legend.position="top",
      legend.text = element_text(size=12),
      legend.title = element_text(size=14),
      strip.text.x = element_text(size=12),
      panel.border = element_blank(),
      panel.spacing = unit(0.25, "cm")) +
# Choose colours for sections in the bar
scale_fill_manual(values=c( "black", "#cb66ff", "#669aff", "#ffcb66", "#9aff66")) +
# Ensure legend is in proper order
guides(fill = guide_legend(reverse = TRUE))
```

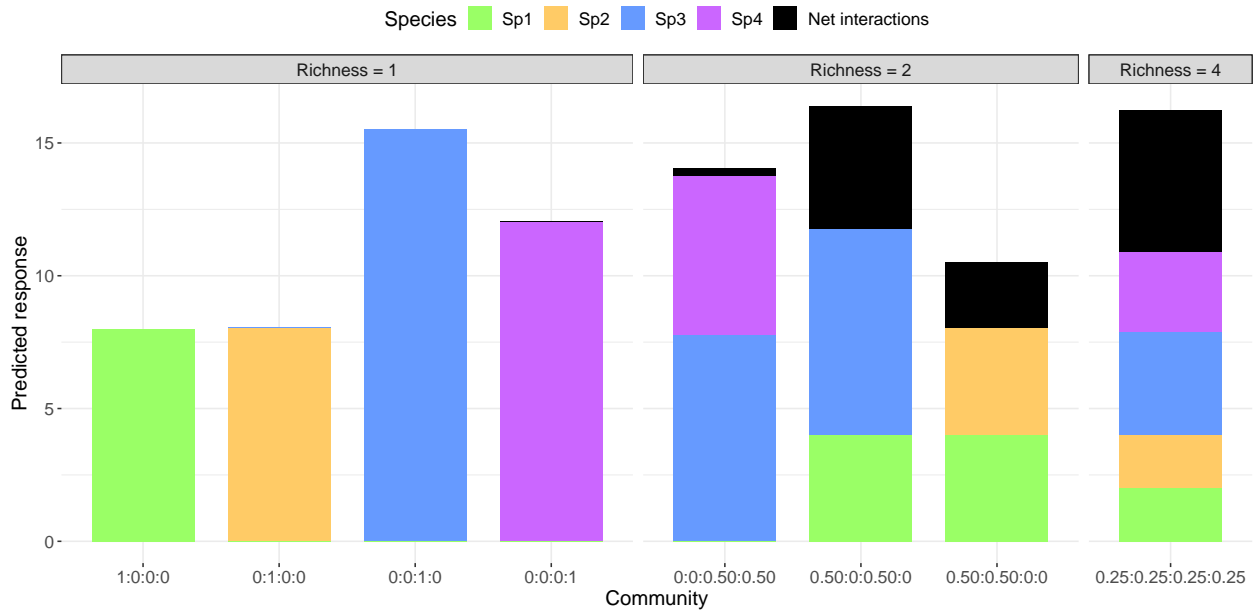

**Figure S3.2**: *Yield predictions for monocultures (richness = 1), two-, and four-species mixtures obtained from the FG model fitted to the Switzerland dataset. The predictions for the three two-species mixtures presented include the within- (0:0:0.5:0.5 and 0.5:0.5:0:0 communities) and between- (0.5:0:0.5:0) functional group effects estimated by the DI model. The prediction for the four-species mixture includes a combination of the within- and between-functional group interaction effects.*

We can estimate a contrast comparing a 4-species mixture with the highest-performing monoculture:

```r
## obtaining contrast vector
Switzerland_contrast <- data.frame(p1 = c(0,1/4),
                                   p2 = c(0,1/4),
                                   p3 = c(1,1/4),
                                   p4 = c(0,1/4))

## Create the interaction terms for the communities
FG_var <- DI_data(prop = c("p1","p2","p3","p4"),
                  FG = c("Grass","Grass","Legume","Legume"),
                  data = Switzerland_contrast,
                  what = "FG")

Switzerland_contrast <- data.frame(Switzerland_contrast, FG_var)
```

```r
contrast_vector <- as.numeric(Switzerland_contrast[1,] - Switzerland_contrast[2,])

# Calculate contrast
the_contrast <- contrasts_DI(object = auto_fit,
                             contrast = list(
                               "p3 vs 4sp mix Low density" = c(contrast_vector, 0),
                               "p3 vs 4sp mix High density" = c(contrast_vector, 1)))
#> Generated contrast matrix:
#>                               p1    p2   p3    p4 FG_bfg_Grass_Legume
#> p3 vs 4sp mix Low density  -0.25 -0.25 0.75 -0.25               -0.25
#> p3 vs 4sp mix High density -0.25 -0.25 0.75 -0.25               -0.25
#>                            FG_wfg_Grass FG_wfg_Legume densitylow
#> p3 vs 4sp mix Low density       -0.0625       -0.0625          0
#> p3 vs 4sp mix High density      -0.0625       -0.0625          1

summary(the_contrast)
#>
#>   Simultaneous Tests for General Linear Hypotheses
#>
#> Fit: glm(formula = new_fmla, family = family, data = new_data)
#>
#> Linear Hypotheses:
#>                                 Estimate Std. Error z value Pr(>|z|)
#> p3 vs 4sp mix Low density == 0   -0.7327     0.7472  -0.981    0.418
#> p3 vs 4sp mix High density == 0  -0.8704     0.8363  -1.041    0.383
#> (Adjusted p values reported -- single-step method)
```

We conclude that the highest performing monoculture performs as well as an equiproportional four-species mixture, for both low and high density.

We can also assess goodness-of-fit by looking, e.g., at plots using the studentised residuals as follows.

```r
res <- tibble(res = rstudent(auto_fit))
linetypes <- c("Density estimate" = 2,
               "N(0,1) curve" = 1)

res %>%
  ggplot(aes(x = res)) +
  # theme of plot
  theme_bw() +
  # Create histogram
  geom_histogram(aes(y = ..density..),
                 binwidth = .5, fill = "lightgrey",
                 col = 1) +
  # Add density curve
  geom_density(aes(lty = "Density estimate")) +
  # Add the N(0, 1) curve
  stat_function(aes(lty = "N(0,1) curve"),
                fun = dnorm) +
  # Labels for the axes and legend
  labs(x = "Studentised residuals",
       y = "Density",
       linetype = "") +
  # Adjust linetypes shown in legend
```

```r
  scale_linetype_manual(values = linetypes) +
  # Adjust specific element of theme
  theme(axis.text = element_text(size=12),
        axis.title = element_text(size=14),
        legend.text = element_text(size = 12))
```

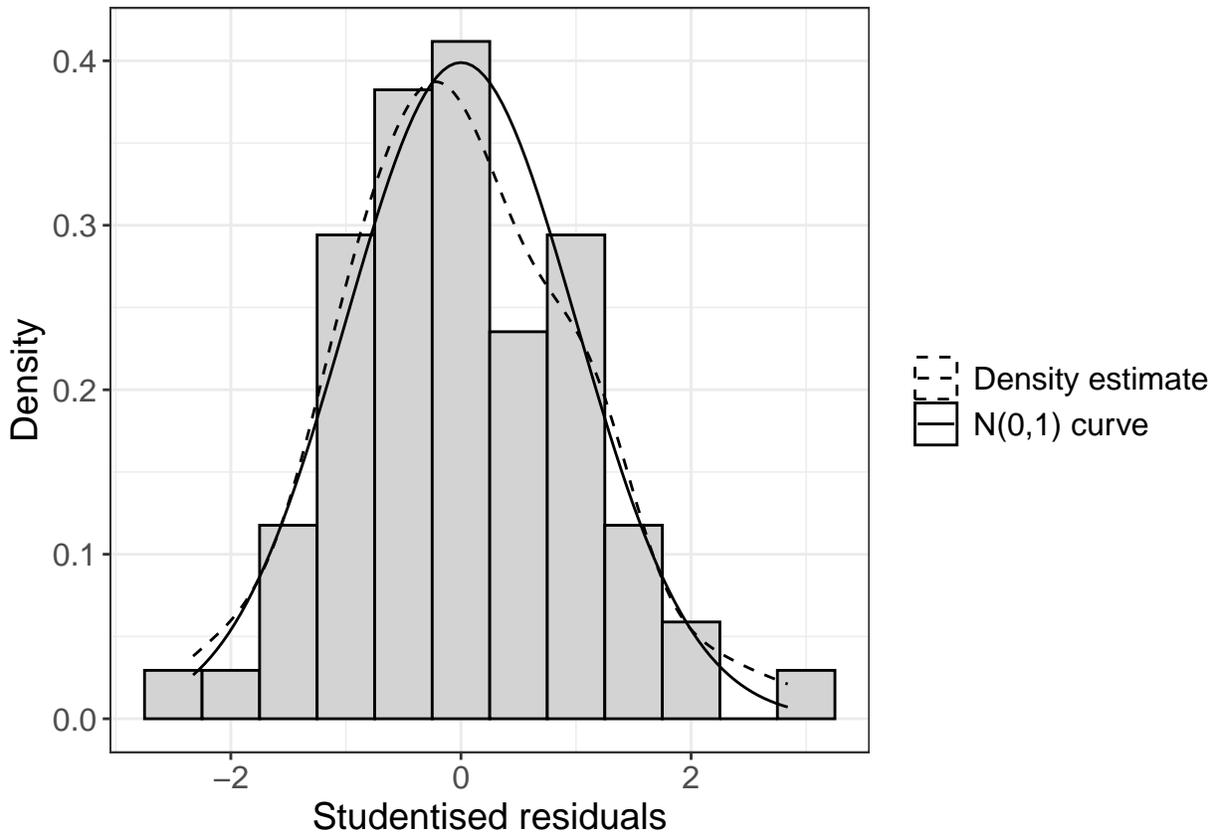

**Figure S3.3**: *Histogram of the studentised residuals from the selected model.*

The histogram of these residuals with overlaid non-parametric kernel density estimate and standard normal density curves indicates that their distribution does not seem to depart from the standard normal.

A half-normal plot with a simulated envelope (Moral et al., 2017) indicates that the observed yields are a plausible realisation of the fitted `FG` model:

```r
library(hnp)

set.seed(2020)
hnp_fit <- hnp(auto_fit, plot = FALSE)
#> Gaussian model (glm object)
hnp_ggplot <- tibble(residuals = hnp_fit$residuals,
                     lower = hnp_fit$lower,
                     median = hnp_fit$median,
                     upper = hnp_fit$upper,
                     x = hnp_fit$x)

hnp_ggplot %>%
  ggplot(aes(x = x, y = residuals)) +
  # theme of plot
```

```r
  theme_bw() +
  # Add points
  geom_point(cex = 1, pch = 16, alpha = .75) +
  # Add the median
  geom_line(aes(y = median),
            lty = 2, lwd = .2) +
  # Add the CI around the median
  geom_ribbon(aes(ymin = lower, ymax = upper),
              fill = "gray", alpha = .2,
              colour = 'black') +
  # Axis labels
  labs(x = "Half-normal scores",
       y = "Studentised residuals") +
  # Adjust axis label size
  theme(axis.text = element_text(size=12),
        axis.title = element_text(size=14))
```

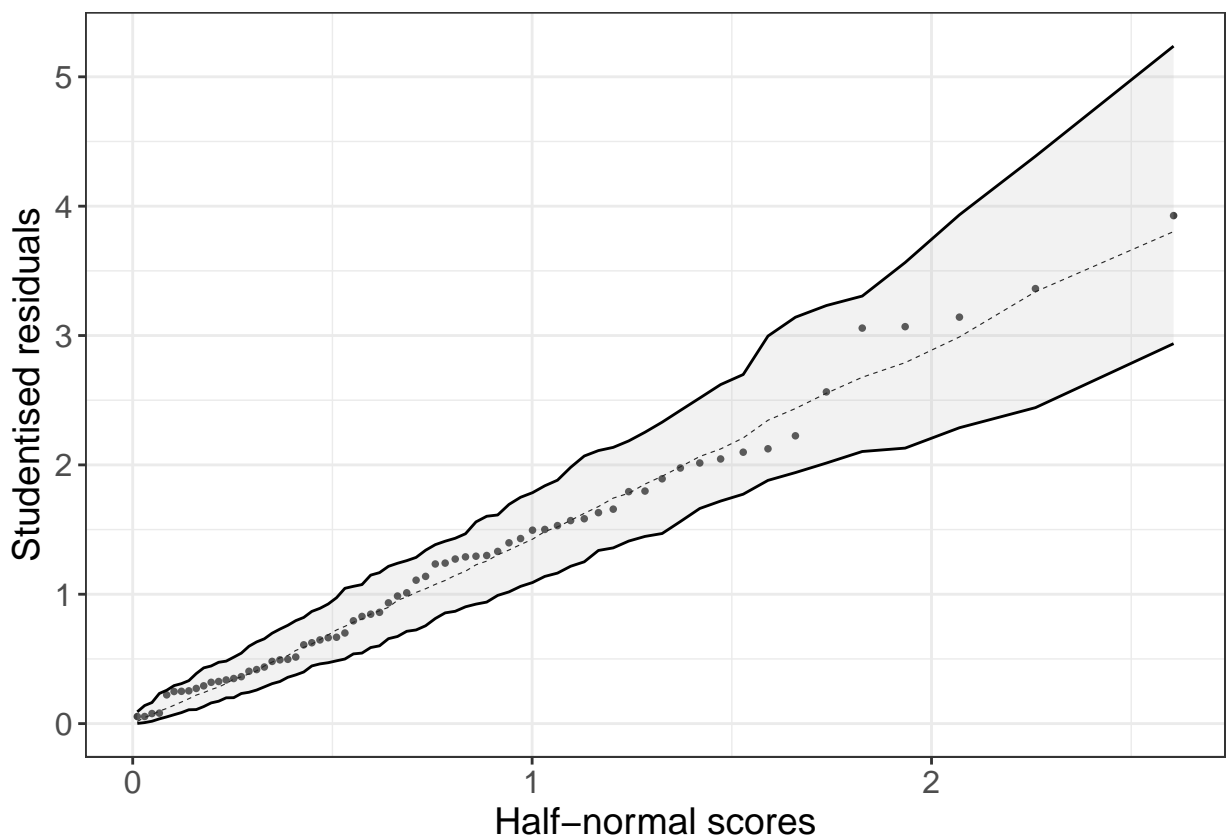

**Figure S3.3**: *Half normal plot to assess the fit of selected model.*

## SI4.4    Exploring advanced DI-modelling options:

We explore further options within the DI modelling framework by including, for example, the interactions between nitrogen level and the functional group effects. The rationale for this exploration is that even though the additive effects of nitrogen levels are not significant, there could be significant interactions to be explored. We make use of the `DI_data()` function to compute the *Interactions* variables needed to fit this model, and use the `extra_formula` argument within the `DI()` function to include the new effects in the model.

We start by showcasing how to use `DI()` to fit the `FG` model, which was selected by `autoDI()` in the previous section:

```
fit_FG <- DI(y = "yield", density = "density",
             prop = 4:7, FG = c("Grass","Grass","Legume","Legume"),
             data = Switzerland, DImodel = "FG")
#> Fitted model: Functional group effects 'FG' DImodel
```

We now create auxiliary variables using `DI_data`:

```
Switz_prep <- DI_data(prop = 4:7, FG = c("Grass","Grass","Legume","Legume"),
                      data = Switzerland, what = "FG")
Switzerland2 <- data.frame(Switzerland, Switz_prep)
```

And fit a custom DI model to test for interactions between functional groups and the nitrogen treatment:

```
fit_FG_treat <- DI(y = "yield", density = "density", treat = "nitrogen",
                   prop = 4:7, FG = c("Grass","Grass","Legume","Legume"),
                   data = Switzerland2, DImodel = FG,
                   extra_formula = ~ (wfg_Grass + wfg_Legume) * nitrogen)
#> Fitted model: Functional group effects 'FG' DImodel

anova(fit_FG, fit_FG_treat, test = "F")
#> Analysis of Deviance Table
#>
#> Model 1: yield ~ 0 + p1 + p2 + p3 + p4 + FG_bfg_Grass_Legume + FG_wfg_Grass +
#>     FG_wfg_Legume + densitylow
#> Model 2: yield ~ 0 + p1 + p2 + p3 + p4 + FG_bfg_Grass_Legume + FG_wfg_Grass +
#>     FG_wfg_Legume + nitrogen50 + densityhigh + `nitrogen150:wfg_Grass` +
#>     `nitrogen150:wfg_Legume`
#>   Resid. Df Resid. Dev Df Deviance      F Pr(>F)
#> 1        60     143.89
#> 2        57     135.74  3   8.1526 1.1412 0.3403
```

Here we conclude these interactions are not significant.

## SI4.5 Updating model call and obtaining confidence intervals for $\theta$

We may also explore the 95% CI for $\theta$ when estimating it for the final selected model. First we update the DI model call to estimate $\theta$, using `update_DI`:

```
fit_FG_theta <- update_DI(fit_FG, estimate_theta = TRUE)
#> Fitted model: Functional group effects 'FG' DImodel
#> Theta estimate: 0.8299
```

Computing the 95% CI for theta:

```
CI_95 <- theta_CI(fit_FG_theta, conf = .95)
CI_95
#>      lower     upper
#> 0.5868526 1.0338467
```

Plotting the profiled log-likelihood:

```
ggplot(fit_FG_theta$profile_loglik, aes(x = grid, y = prof)) +
  # theme for plot
```

```
  theme_bw() +
  # plotting likelihood values
  geom_line() +
  # Restricting x-axis
  xlim(0,1.5) +
  # CI for the theta estimate
  geom_vline(xintercept = CI_95, lty = 3) +
  # Estimate of theta
  geom_vline(xintercept = fit_FG_theta$coefficients['theta'])+
  # Labels for the axes and plot title
  labs(x = expression(theta), y = "Log-likelihood",
       title = "   Log-likelihood versus theta",
       caption = "The vertical line is the theta estimate and dotted vertical lines are
      ↪ upper and lower bounds of 95% CI for theta") +
  # Adjust size of axis labels
  theme(axis.text = element_text(size=12),
        axis.title = element_text(size=14))
```

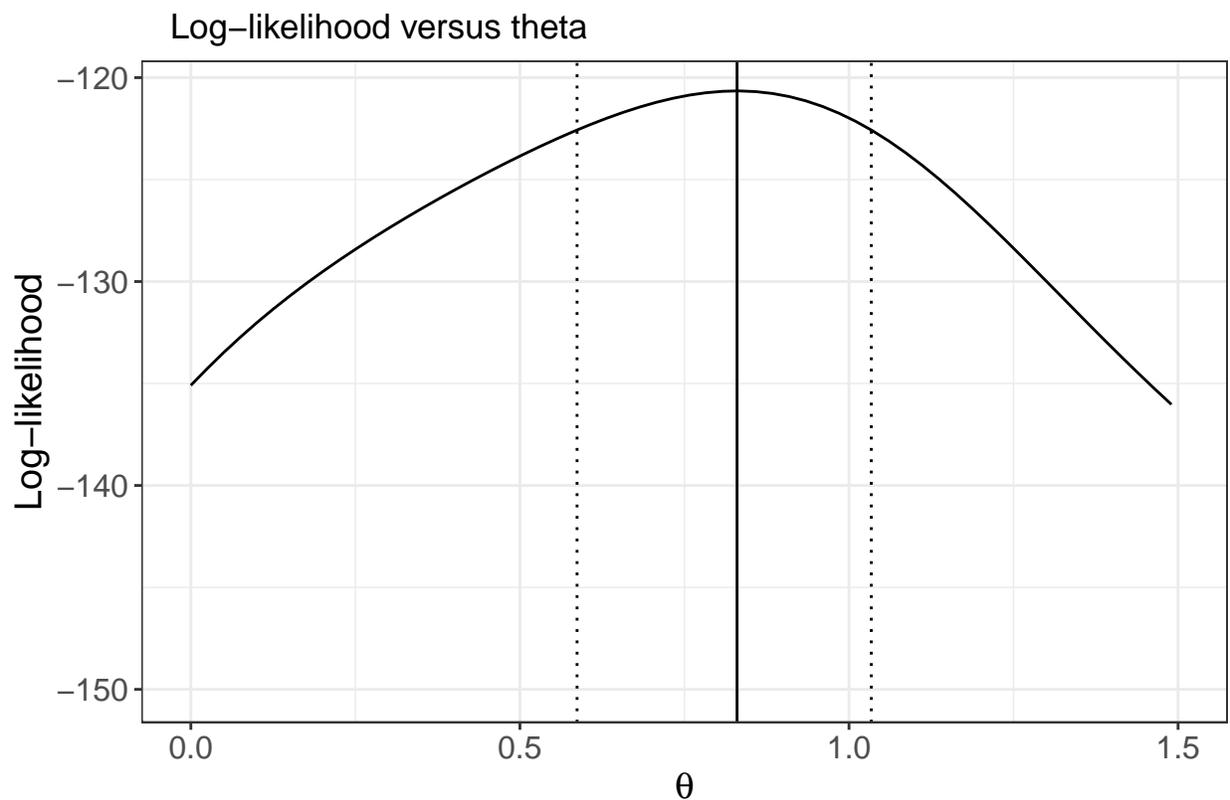

**Figure S3.3**: *Plot of the log-likelihood versus $\theta$*

The 95% confidence interval for $\theta$ includes 1, and is therefore indicative that we may assume $\theta = 1$ for this particular dataset.

## SI4.6 Comparing with the reference/community model – ANOVA approach

The reference model (also termed the 'community' model) estimates an average yield for each unique set of species proportions. In the Switzerland dataset, we have 25 unique sets of species proportions, or communities. We can use the internal function `get_community` to produce a factor where each level represents a unique set of species proportions in the data. This is helpful when fitting the reference model:

```
Switzerland$community <- DImodels:::get_community(prop = 4:7, data = Switzerland)
#> 'community' is a factor with 25 levels, one for each unique set of proportions.
fit_reference <- glm(yield ~ density + community, data = Switzerland)
summary(fit_reference)
#> 
#> Call:
#> glm(formula = yield ~ density + community, data = Switzerland)
#> 
#> Deviance Residuals:
#>     Min       1Q   Median       3Q      Max
#> -3.4269  -0.7702   0.0000   0.7382   2.4077
#> 
#> Coefficients:
#>              Estimate Std. Error t value Pr(>|t|)
#> (Intercept)  12.9969     0.7304  17.794  < 2e-16 ***
#> densityhigh   0.1378     0.3443   0.400 0.691126
#> community2   -0.1270     1.0039  -0.126 0.899952
#> community3    4.4144     1.0039   4.397 7.33e-05 ***
#> community4    2.5277     1.0039   2.518 0.015702 *
#> community5    2.0866     1.0039   2.079 0.043803 *
#> community6    1.9760     1.2295   1.607 0.115511
#> community7    2.3408     1.2295   1.904 0.063784 .
#> community8    1.2526     1.2295   1.019 0.314114
#> community9    4.1116     1.2295   3.344 0.001745 **
#> community10   2.5062     1.2295   2.038 0.047834 *
#> community11   2.6743     1.2295   2.175 0.035299 *
#> community12  -5.6150     1.0039  -5.593 1.52e-06 ***
#> community13  -5.4450     1.0039  -5.424 2.66e-06 ***
#> community14   1.3171     1.0039   1.312 0.196650
#> community15  -1.2919     1.0039  -1.287 0.205182
#> community16  -3.1898     1.2295  -2.594 0.012993 *
#> community17   5.1751     1.2295   4.209 0.000132 ***
#> community18   0.6682     1.2295   0.544 0.589650
#> community19   3.7711     1.2295   3.067 0.003772 **
#> community20  -0.3989     1.2295  -0.324 0.747216
#> community21   0.5933     1.2295   0.483 0.631898
#> community22  -2.0847     1.2295  -1.696 0.097359 .
#> community23  -1.6510     1.2295  -1.343 0.186541
#> community24   3.8912     1.2295   3.165 0.002884 **
#> community25   1.7725     1.2295   1.442 0.156799
#> ---
#> Signif. codes:  0 '***' 0.001 '**' 0.01 '*' 0.05 '.' 0.1 ' ' 1
#> 
#> (Dispersion parameter for gaussian family taken to be 2.015473)
#> 
#>     Null deviance: 680.29  on 67  degrees of freedom
```

```
#> Residual deviance:  84.65  on 42  degrees of freedom
#> AIC: 261.87
#> 
#> Number of Fisher Scoring iterations: 2
```

`DImodels` allows for a comparison between any fitted DI model and the reference model that includes automatic fitting of the reference model through `DI_compare`. For instance, we may compare the FG model fitted in the previous section with the reference model by executing:

```
DI_compare(model = fit_FG, test = "F")
#> 'community' is a factor with 25 levels, one for each unique set of proportions.
#> Fitted model: Functional group effects 'FG' DImodel
#> Analysis of Deviance Table
#> 
#> Model 1: yield ~ 0 + p1 + p2 + p3 + p4 + FG_bfg_Grass_Legume + FG_wfg_Grass +
#>     FG_wfg_Legume + densitylow
#> Model 2: yield ~ 0 + p1 + p2 + p3 + p4 + FG_bfg_Grass_Legume + FG_wfg_Grass +
#>     FG_wfg_Legume + densitylow + community2 + community3 + community4 +
#>     community5 + community6 + community7 + community8 + community9 +
#>     community10 + community11 + community12 + community13 + community14 +
#>     community15 + community16 + community17 + community18 + community19
#>   Resid. Df Resid. Dev Df Deviance      F Pr(>F)
#> 1        60     143.89                            
#> 2        42      84.65 18   59.238 1.6329 0.09536 .
#> ---
#> Signif. codes:  0 '***' 0.001 '**' 0.01 '*' 0.05 '.' 0.1 ' ' 1
```

This is equivalent to:

```
anova(fit_FG, fit_reference, test = "F")
#> Analysis of Deviance Table
#> 
#> Model 1: yield ~ 0 + p1 + p2 + p3 + p4 + FG_bfg_Grass_Legume + FG_wfg_Grass +
#>     FG_wfg_Legume + densitylow
#> Model 2: yield ~ density + community
#>   Resid. Df Resid. Dev Df Deviance      F Pr(>F)
#> 1        60     143.89                            
#> 2        42      84.65 18   59.238 1.6329 0.09536 .
#> ---
#> Signif. codes:  0 '***' 0.001 '**' 0.01 '*' 0.05 '.' 0.1 ' ' 1
```